\documentclass[12pt,a4paper]{article}
\usepackage{latexsym}	
\usepackage{amsmath}
\usepackage{amsthm}
\usepackage{amssymb}
\usepackage{graphicx}

\usepackage{amsmath}
\usepackage{amsthm}
\usepackage{amsmath,amssymb}
\usepackage{graphicx}
\def\ba{\begin{eqnarray}}
\def\ea{\end{eqnarray}}

\def\ba{\begin{eqnarray}}
\def\ea{\end{eqnarray}}

\def\be{\begin{equation}}
\def\ee{\end{equation}}

\theoremstyle{plain}

\begin{document}

\title{Corrections of the GR eikonal limit by a class of renormalizable gravity models}

\author{Leandro Lanosa$^1$ and Osvaldo P. Santill\'an$^1$
\thanks{Electronic addresses:lanosalf@hotmail.com and firenzecita@hotmail.com}\\
\textit{\small{$^1$Instituto de Matem\'atica Luis Santaló (IMAS) and CONICET,}}\\
\textit{\small{Ciudad Universitaria, 1428 Buenos Aires, Argentina.}}}
\date{}

\maketitle

\begin{abstract}
In the present work, the scattering between a light scalar particle $\phi$ and a heavy scalar $\sigma$  in the eikonal limit is considered, for  gravity scenarios containing  higher order derivatives, such as the ones studied in \cite{stelle1}-\cite{modesto4}.  It is suggested that if one of the new gravity scales introduced in the higher order action is smaller than the Planck mass, for instance of the order of $M_{GUT}\sim 10^{15}$ GeV, the functional form of  the GR eikonal formulas appears changed by a factor. However, in this situation, the conditions for the eikonal approximation to hold has to be revised, this issue is analyzed in the text. The statements of the present work should be taken with a grain of salt, as the Schwarzschild radius for these polynomials theories is not yet established. The results presented here, in our opinion, are in agreement with the suppression of corrections to GR pointed out in \cite{brandhuber}, \cite{fradkin} and \cite{deser} for the Stelle gravity. The next to leading order approximation and part of the seagull diagrams are estimated. Different to the GR case, this order generically is non vanishing. An explicit regularization scheme is presented, based on Riesz and Hadamard procedures. The need of a regularization  is partially expected, as the inclusion of small  energy fluctuations  may spoil the eikonal approximation.
\end{abstract}

\section{Introduction}
The study of effective field quantum theories of gravity beyond GR is one of the main topics of theoretical physics. There exist renormalizable theories of gravity, such as the Stelle ones \cite{stelle1}-\cite{stelle2}
or the super-renormalizable ones presented in \cite{modesto1}-\cite{modesto4} and \cite{giacchini}, which improve the bad perturbative behavior of GR, at the price of accepting apparent instabilities \cite{dewitt}. The main argument of \cite{dewitt} is that a theory with classical fourth order equations of motion may improve the bad  power counting that GR has at perturbative level. The propagator of such fourth order theory will be given in a Pauli Villars form as
$$
D\sim \frac{1}{k^2}-\frac{1}{k^2+M^2}.
$$
It is a simple exercise to see that for large values of $k$ this propagator will go as $M^2 k^{-4}$. In QFT
without gravity, the mass $M$ is interpreted as a cutoff, which tends $M\to\infty$ at the end of the calculations. In the gravity approach, this cutoff  is rendered  finite, and the minus sign
suggests the presence of a state with negative kinetic energy, which may indicate a potential instability. 

However, there exist several references suggesting that these instabilities may be avoided. An incomplete list is \cite{mannheim1}-\cite{donogue} and the references therein.  These works suggest that the problem of the apparent instabilities may be solved by employing variants of the quantization method, some of them with roots at the Pauli and Dirac days. Those approaches are of clear interest.  A first question that arises is that, if one starts with a theory whose classical formulation is unstable, and becomes stable after quantization, which type of classical limit it leads to. 

Clearly, the study of classical or semi-classical  limits of renormalizable theories of gravity, with or without the prescriptions given in \cite{mannheim1}-\cite{donogue}, is of interest in this context. After the appearance of pioneer works \cite{levisucher}-\cite{barbashov} about the eikonal approximation 
in QFT, besides the fundamental works about QCD reviewed in \cite{lipatovlibro}-\cite{barone}, a large number of applications in gravity theories were found, even beyond the Planck scale \cite{verlinde}-\cite{lipatov}. The work in \cite{pionero0} presents a link between  classical  and quantum aspects for horizon formation in black holes, and this line was further studied in several works such as \cite{duff}-\cite{aplicacion3}. In addition, in recent years, new semi classical methods for effective field theories for studying compact binaries were reported in several papers \cite{porto1}-\cite{porto6}, detailed reviews can be found in  \cite{review1}-\cite{review5} and references therein. Semiclassical methods for modified gravity theories have been considered, for instance, in \cite{black hole}-\cite{ciafa10} and also in \cite{pionero1}-\cite{pionero2}.  More recently, the study of particle creations inside the so named Page time interval \cite{gaddam1}-\cite{gaddam3} near a black hole horizon, which may throw light on the information paradox problem and the nature of gravitational collapse, was studied by the use of these tools. Furthermore, eikonal methods were applied recently in order to characterize the gravitational wave spectrum that is obtained after a collision between two particles takes place \cite{veneziano}. Modern references covering these aspects are \cite{veneziano2}-\cite{veneziano10}. The eikonal limit was employed also in \cite{edelstein}-\cite{zhiboedov} to the study of causality issues in gravity theories. 

The present work deals with the eikonal limit of the gravitational models of \cite{modesto1}-\cite{modesto4}. 
There exist several types of eikonal limits in the literature, some are related to scattering of massless particles  with high center of mass energy, and others study  scattering between a extremely heavy and a extremely light particle. This second situation will be considered here, and the corrections to the Shapiro time delay and angle deflection will be partially estimated. In addition, the validity of the eikonal limit is analyzed and the lower bounds for the value of the  energy of the center of mass $\sqrt{s}$ are discussed. The exact bounds are based on the dependence of the generalized Schwarzschild radius for the theory. The explicit black hole solutions are not known for  gravity theories with higher order derivatives, even though some hints of them can be seen in \cite{schwarz1}-\cite{schwarz3}. 

The present work is organized as follows. In section 2 the general situation to be analyzed is described, and some eikonal identities that are useful are summarized. In section 3 the gravity models under consideration are described in some detail, and the effective coupling constants of the problem are identified.  These couplings describe the strength of the gravitational interaction for these models. In section 4 the eikonal approximation for the above mentioned scattering is worked out, together with a description of the validity of such approximation. Section 5 contains some considerations for specific gravity models. Section  6 describes the next to leading order and seagull approximations for these models. Section 7 contains a discussion about the obtained results and  new perspectives related to them. 

\section{Generalities about the problem}
     The aim of the present work  is the study of gravitational scattering between an ultra energetic light (almost massless) scalar particle $\phi$ of energy $E_\phi$ and an extremely heavy, almost static, scalar particle $\sigma$ of mass $M_\sigma$. It is assumed that the scattering involves  small angles, however, the conditions for this to be the case has to and will be worked out in a more precise form.  The situation above describes the same scenario than the one in reference \cite{fazio}, with the difference that the gravity model considered here contains higher derivatives.  In fact, this problem was studied  in \cite{nexteikonal}-\cite{nexteikonal2} as well. Both references study the same physical situation, the difference is that  the ones \cite{nexteikonal}-\cite{nexteikonal2}  employ the standard techniques of Feynman diagrams while \cite{fazio} approaches the problem by use of Fradkin modified perturbation theory,  which is a less known tool.  Even taking into account that this second approach is more complicated or less known, the present work will follow the methods of \cite{fazio} which, although being more sophisticated, were for us simpler  to generalize. 
     
The ingoing and outgoing momenta of the light particle $\phi$ will be denoted by $p$ and $p^\prime$, while for $\sigma$ the analogous quantities will be denoted as $q$ and $q^\prime$. The elastic property and the mass shell conditions for the particles participating in the scattering are standard
\begin{equation}\label{cinema}
p+q= p’+q’\qquad p^2=p’^2=0\, \qquad q^2=q’^2=M_\sigma^2,
\end{equation}
while that the transferred impulse in the eikonal limit, assumed to hold here,  is required to fulfill the following inequality $$\Delta\equiv\sqrt{-(p-p’)^2}<<E_\phi=p^0<< M_\sigma.$$ The  incoming and outgoing momenta of the scalar light particle, denoted above as $p$ and $p’$, are much larger than the transferred momentum $\Delta^\mu=p^{\prime\mu}-p^{\mu}$.  If in addition, the recoil of the heavy particle is neglected, which is justified by the fact that $M_\sigma$ is very large,  it follows that $p^0\sim p’^0 \sim E_\phi \sim |\vec{p}|$. The three dimensional components $\vec{p}$  of the momentum will be taken to lay almost along the $z$-axis. The following notation will be employed along the text
\begin{equation}\label{frames}
p^\mu=\left(E_\phi, p^z, -\frac{\vec{\Delta}}{2}\right)\,\,\,\,\,\,\,\, p^{\prime\mu}=\left(E_\phi, p^z,\frac{\vec{\Delta}}{2}\right).
\end{equation}
At first order $p^z\simeq E_\phi$,  while for higher orders the expansion parameter is taken as $\frac{\Delta^2}{E_\phi^2}$. It follows that  $p^z=\sqrt{E_\phi^2-\frac{\Delta^2}{4}}=p’^z$. The four momentum of the heavy particle is $q=(M_\sigma, 0, 0, 0)$ and $q^{\prime}\sim q$, as its recoil will be neglected at first order, due to its massive nature. 

The eikonal limit is customarily studied in terms of the Mandelstam kinematic variables. These variables are defined by
\begin{equation}
	\label{mandelstamo}
	s=(p+q)^2,\qquad t=(p-p^{\prime})^2,\qquad u=(p-q^{\prime})^2.
\end{equation}
Note that here a plus sign convention is employed, while in several works it is standard to include a minus sign in the definition of these quantities. The energy of the center of mass of the system is $E=\sqrt s$, while $t$ is minus the square of the momentum transfer, that is, $t=-\Delta^2$. The variable $u$ can be written in terms of $s$ and $t$ using momentum conservation and the mass-shell relation
\begin{equation}
	\label{eq:stum}
	s+t+u =2M_\sigma^2\,,
\end{equation}
which shows that $u$  is not an independent quantity. 

The previous approximation about the kinematics of the process, of course, is not generically satisfied for every $2\to 2$ process. The regime of validity will be specified below. The resulting constraints put limits on the possible values of the impact parameter, $\vec{b}$ of the collision process.

Following the approach of \cite{kabatortiz} and \cite{itzykson},  the study of the above described scattering process requires the calculation of the  full four-point connected Green function between four scalars. Depending on the model of quantum gravity in consideration, this quantity is given by
\begin{equation}\label{green}
G(x_1,x’_1, x_2, x’_2)=\int\limits_{\substack{\text {Connected}\\ \text{diagrams}}} [Dh][D\phi_1] [D\phi_2] \phi_1(x_1)\phi_1(x’_1)\phi_2(x_2)\phi_2(x’_2)
\end{equation}
$$
\exp\bigg[{i\int[\text{L}_{g} +\frac{1}{2}g^{\mu\nu}\partial_\mu\phi_1\partial_\nu\phi_1+\frac{1}{2}g^{\mu\nu}\partial_\mu\phi_2\partial_\nu\phi_2-\frac{1}{2}M_\sigma^2\phi_2^2]\sqrt{-g} d^4 x }\bigg],
$$
where  $\text{L}_g$ is the part of the lagrangian describing the graviton. It may be metric dependent or metric and connection dependent. In the present approach, only metric dependent schemes will be considered. The metric  $g_{\mu\nu}$ is defined as the sum of a flat Minkowski component $\eta_{\mu\nu}$ plus an small perturbation $ h_{\mu\nu}$.  The scattering amplitude for the process  is given by 
$$
i(2\pi)^4\delta^4(p+q-p’-q’) T(p,p’, q,q’)=e^{ i\int d^4x d^4y \left(\frac{\delta}{\delta h^{\alpha\beta}(x)}D^{\alpha\beta,\gamma\delta}(x-y)\frac{\delta}{\delta h^{\prime \gamma\delta}(y)}\right)}$$
\begin{equation}\label{rodnis}
\left.<p’|G^c(x_1,x’_1|h)|p><q’|G^c(x_2,x’_2|h’)|q>\right |_{h,h’=0 }.
\end{equation}
Here $D^{\mu\nu,\alpha\beta}$ is the graviton propagator, defined as follows. First write the lowest order terms of the graviton action  as 
\begin{equation}\label{cuadratizante2}
S_g\sim \frac{1}{2}\int h^{\alpha\beta}(D^{-1})_{\alpha\beta,\mu\nu} h^{\mu\nu} d^4x.
\end{equation}
where $D^{-1}$ is some operator, depending on the choice of the gravity theory.  The graviton propagator is the inverse of this operator, that is
$$
2\int D_{\alpha\beta, \gamma \delta}(x-y)(D^{-1})^{\gamma\delta}_{,\mu \nu}(y-z)d^4y=\delta^4(x-z)\eta_{\mu(\alpha}\eta_{\beta)\nu}.
$$
In addition, in \eqref{rodnis},  the following matrix element
\begin{equation}
<p’|G^c(x,y|h_{\mu\nu})|p>=\lim_{p^2,p’^2\rightarrow 0} \int d^4x  \int d^4 y e^{-ip\cdot x}e^{ip’\cdot y}\square_x \square_y (G(x,y|h)-G_0(x,y)),
\label{green0}
\end{equation}
has been introduced. It represents the Fourier transform of the  full  amputated Green function in the perturbed background. The notation $G_0$ refers to the free propagation for the scalar field without any external background.  Analogously, for the heavy scalar of mass $M_\sigma$, the following element
\begin{equation}\label{green1}
<q’|G^c(x,y|h_{\mu\nu})|q>=\lim_{q^2,q’^2\rightarrow M_\sigma^2} \int d^4x  \int d^4 y e^{-iq\cdot x}e^{iq’\cdot y}(\square_x+M_\sigma^2) (\square_y+M_\sigma^2) (G(x,y|h)-G_0(x,y)),
\end{equation}
has been defined in \eqref{rodnis}. 

An important role in the subsequent calculations will be played by the so called eikonal identities and, for this reason, they will be stated here.  The first of these identities, proved in \cite{nexteikonal}, is given by
\begin{equation}
\int_0^{+\infty}d\nu e^{i\nu x}\prod_{m=1}^n \frac{1-e^{i\nu a_m}}{a_m+i\epsilon}=i\sum_\pi \frac{1}{x+i\epsilon}\frac{1}{x+a_{\pi(1)}+i\epsilon}\dots \frac{1}{x+a_{\pi(1)}+\dots a_{\pi(n)}+i\epsilon},
\label{idento}
\end{equation}
where $\pi$ belongs to the set of permutations of $n$ indices and $a_m$ are $n$ independent parameters.  The sum is understood to cover all the possible permutations of the $n$-parameters $a_m$. In addition, the sum of the permutations is done by the use of the summation formula proved in \cite{nexteikonal2}
\begin{equation}\label{idento2}
\delta(x_1+\dots x_n)\sum_{\pi} \frac{1}{x_{\pi(1)}+i\epsilon}\dots \frac{1}{x_{\pi(1)}+\dots +x_{\pi(n-1)}+i\epsilon}=(-2\pi i)^{n-1}\delta(x_1)\dots\delta(x_n).
\end{equation}
The last identity that concerns the present work  can be found in \cite{levisucher}, \cite{cardy}, and can be expressed as
$$
\int_0^\nu d\xi_1\dots\int_0^\nu d\xi_n \sum_{r=1}^n \vec{k}_r\cdot \vec{k}_r (\nu-\xi_r)\exp\left[-\sum_{s=1}^n 2i\vec{p^{\prime}}\cdot \vec{k}_s(\nu-\xi_s)\right]$$
\begin{equation}\label{cardio}
=-\sum_{r=1}^n  \vec{k}_r\cdot \vec{k}_r  \frac{\partial}{\partial (2i\vec{p^{\prime}}\cdot \vec{k}_r)}\prod_{s=1}^n \frac{1}{2i\vec{p^{\prime}}\cdot \vec{k}_s}[1-e^{-2i\nu\vec{p^\prime}\cdot \vec{k}_s}]
\end{equation}
$$
=\sum_{r=1}^n   \frac{\vec{k}_r\cdot \vec{k}_r}{2i\vec{p^{\prime}}\cdot \vec{k}_r}\prod_{s=1}^n \frac{1}{2i\vec{p^{\prime}}\cdot \vec{k}_s}[1-e^{-2i\nu\vec{p^\prime}\cdot \vec{k}_s}]-\nu\sum_{r=1}^n  \vec{k}_r\cdot \vec{k}_r\prod_{s=1}^n \frac{1}{2i\vec{p^{\prime}}\cdot \vec{k}_s}e^{-2i\nu\vec{p^\prime}\cdot \vec{k}_s}.
$$
The three identities \eqref{idento}-\eqref{cardio} are very important for obtaining the results presented along the text.

The description given above, in particular formula \eqref{rodnis}, shows that for the determination of the propagator $D_{\alpha\beta,\gamma\delta}$ of the gravitational model in consideration, and to describe the elements \eqref{green0}-\eqref{green1} is essential for describing the desired scattering process.  Fortunately the last elements have been characterized in detail in \cite{fazio}. The results of that reference will be described below. However, at this point, it will be convenient  to specify the gravity model to be  concerned with.

\section{The gravitational model and the validity of the eikonal approximation}
\subsection{The defining action and the graviton  propagator}
The gravitational models to be considered here, for which the eikonal limit will be partially characterized, are given by the generic action
\begin{equation}\label{tat}
S_g=\frac{1}{16\pi G_N}\int \bigg[R+R F_1(\square) R+R_{\mu\nu}F_2(\square)R^{\mu\nu}+R_{\mu\nu\alpha\beta}F_3(\square)R_{\mu\nu\alpha\beta}\bigg]\sqrt{-g}d^4x +S_{gf}.
\end{equation}
Here $S_{gf}$ is a gauge fixing term,  $\square$ denotes the wave D'Alambert operator on the background, and $F_i(\square)$ are functions of this operator. The motivation for this graviton action is that it leads to the renormalizable model of Stelle \cite{stelle1}-\cite{stelle2} or the super-renormalizable models introduced in \cite{modesto1}-\cite{modesto4} as particular cases. Furthermore, in the momentum space, for  small perturbations of the Minkowski metric, these functions $F_i(-k^2)$ run with the momentum, and this may imitate the behavior of the gravitational coupling described by some renormalization group approaches. In several, but not in all,  references, these functions are taken as polynomials.

The propagator $D_{\mu\nu,\alpha\beta}$ corresponding to a perturbation  $g_{\mu\nu}=\eta_{\mu\nu}+h_{\mu\nu}$ around flat space can be calculated by retaining the quadratic terms in $h_{\mu\nu}$ in the gravitational action. After some standard calculation, it follows that the quadratic terms for the model given above are collected as \cite{biswas}
$$
S^2_g=\frac{1}{32\pi G_N}\int \bigg[\frac{1}{2}h^{\alpha\beta}\eta_{\alpha\mu}\eta_{\beta\nu} \square a(\square) h^{\mu\nu}+h^{\alpha\beta}\eta_{\beta\mu}  b(\square)\partial_\alpha\partial_\nu h^{\mu\nu}+h   c(\square)\partial_\mu\partial_\nu  h^{\mu\nu}
$$
\begin{equation}\label{cuadrtizada}
+\frac{1}{2}h   \square d(\square)  h+h^{\alpha\beta}   \frac{f(\square)}{\square}\partial_\alpha\partial_\beta\partial_\mu\partial_\nu  h^{\mu\nu}\bigg] d^4x+S_{gf}.
\end{equation}
In the last expressions, the quantities
$$
a(\square)=1-\frac{1}{2} F_2(\square)\square-2F_3(\square)\square+a_{gf},\qquad 
b(\square)=-1+\frac{1}{2} F_2(\square)\square+2F_3(\square)\square+b_{gf},
$$
$$
c(\square)=1+2 F_1(\square)\square+\frac{1}{2}F_2(\square)\square+c_{gf},\qquad
d(\square)=-1-2 F_1(\square)\square-\frac{1}{2}F_2(\square)\square+d_{gf},
$$
$$
f(\square)=-2 F_1(\square)\square-F_2(\square)\square-2F_3(\square)\square+f_{gf},
$$
were introduced, where subscript ``$gf$" indicates gauge fixing terms. In order to generalize the results of \cite{fazio}  to the present situation, the gravitational model has to be worked in the De Donder gauge, which constraints the perturbation to satisfy
$$
\partial_\mu h^\mu_\nu=\frac{1}{2}\partial_\nu h^\mu_\mu.
$$
This is the gauge employed in \cite{fazio}. The  GR limit  corresponds to $F_i(\square)\to 0$. This limit, for the De Donder gauge, after the redefinition $h_{\mu\nu}\to \sqrt{16\pi G_N}h_{\mu\nu}$ should lead to the well known GR linearized gauge fixed action
$$
S_{GR}=\int \bigg[\frac{1}{4}h^{\alpha\beta}\eta_{\alpha\mu}\eta_{\beta\nu} \square h^{\mu\nu}
-\frac{1}{8}h   \square  h\bigg] d^4x.
$$
This leads to the conclusion that the gauge fixed coefficients are
$$
a_d(\square)=1-\frac{1}{2} F_2(\square)\square-2F_3(\square)\square,\qquad 
b_d(\square)=\frac{1}{2} F_2(\square)\square+2F_3(\square)\square,
$$
$$
c_d(\square)=2 F_1(\square)\square+\frac{1}{2}F_2(\square)\square,\qquad
d_d(\square)=-\frac{1}{2}-2 F_1(\square)\square-\frac{1}{2}F_2(\square)\square,
$$
\begin{equation}\label{donde}
f_d(\square)=-2 F_1(\square)\square-F_2(\square)\square-2F_3(\square)\square.
\end{equation}
It should be remarked that the work \cite{fazio} employs the redefinition $h_{\mu\nu}\to \sqrt{32\pi G_N}h_{\mu\nu}$ while
the redefinition here is $h_{\mu\nu}\to \sqrt{16\pi G_N}h_{\mu\nu}$, thus there will be an apparent factor $2$ discrepancy in some results when going to the GR limit. However, when the final results are written in terms of the Newton constant $G_N$, this apparent discrepancy disappears.

The calculation of the propagator of the model requires to express the action \eqref{cuadrtizada} as in \eqref{cuadratizante2}. This task may be achieved with the help of the Barnes -Rivers operators, which  are given in terms of the elementary tensors,
\begin{equation}\label{ripol2}
\omega_{\mu\nu}=\frac{k_\mu k_\nu}{k^2}, \qquad \theta_{\mu\nu}=\eta_{\mu\nu}-\frac{k_\mu k_\nu}{k^2},
\end{equation}
in the following way
$$
P^2_{\alpha \beta,\mu\nu}=\frac{1}{2}(\theta_{\beta \mu}\theta_{\alpha\nu}+\theta_{\beta\nu}\theta_{\alpha\mu})-\frac{1}{3}\theta_{\beta\alpha}\theta_{\mu\nu}
$$
$$
P^1_{\alpha \beta,\mu\nu}=\frac{1}{2}(\theta_{\beta\mu}\omega_{\alpha\nu}+\theta_{\beta\nu}\omega_{\alpha\mu}+\theta_{\alpha\mu}\omega_{\beta\nu}+\theta_{\alpha\nu}\omega_{\beta\mu})
$$
$$
P^{0-s}_{\alpha \beta,\mu\nu}=\frac{1}{3}\theta_{\beta\alpha}\theta_{\mu\nu},\qquad
P^{0-w}_{\alpha \beta,\mu\nu}=\omega_{\beta\alpha}\omega_{\mu\nu},
$$
\begin{equation}\label{ripol}
P^{0-sw}_{\alpha \beta,\mu\nu}=\frac{1}{\sqrt{3}}\theta_{\beta\alpha}\omega_{\mu\nu},\qquad
P^{0-ws}_{\alpha \beta,\mu\nu}=\frac{1}{\sqrt{3}}\omega_{\beta\alpha}\theta_{\mu\nu}.
\end{equation}
These formulas may be expressed in the coordinate space by making the change
$$
\omega_{\mu\nu}=\frac{\partial_\mu \partial_\nu}{\square}, \qquad \theta_{\mu\nu}=\eta_{\mu\nu}-\frac{\partial_\mu \partial_\nu}{\square}.
$$
There are several identities satisfied for these operators, which have been worked out in the literature, for instance in \cite{barnes1}. For the present purposes, it is enough to remark that, given an expression of the form
$$
M=a_2 P^2+a_1 P^1+a_s P^{0-s}+a_w P^{0-w}+a_{sw}\sqrt{3}(P^{0-sw}+P^{0-ws}),
$$
its inverse is 
\begin{equation}\label{inverso}
M^{-1}=\frac{1}{a_2} P^2+\frac{1}{a_1} P^1+\frac{1}{a_s a_w-3 a_{sw}^2}\bigg[a_w P^{0-s}+a_s P^{0-w}-a_{sw}\sqrt{3}(P^{0-sw}+P^{0-ws})\bigg].
\end{equation}
The last formula is useful for calculating the desired propagator. In order to see how it should be applied, note that by going to the momentum space, the first term of the action \eqref{cuadratizante2} is mapped to
$$
h^{\alpha\beta}\eta_{\alpha\mu}\eta_{\beta\nu} \square a(\square) h^{\mu\nu}\to -k^2a(-k^2)h^{\alpha\beta}\eta_{\alpha\mu}\eta_{\beta\nu}  h^{\mu\nu}.
$$
By employing the definition \eqref{ripol2}, it is evident that $\eta_{\mu\nu}=\theta_{\mu\nu}+\omega_{\mu\nu}$, and with this simple identity the last term in momentum space may be worked out as
$$
-k^2a(-k^2)h^{\alpha\beta}\eta_{\alpha\mu}\eta_{\beta\nu}  h^{\mu\nu}=-k^2a(-k^2)h^{\alpha\beta}(\theta_{\alpha\mu}\theta_{\beta\nu}+\theta_{\alpha\mu}\omega_{\beta\nu}+\omega_{\alpha\mu}\theta_{\beta\nu}+\omega_{\alpha\mu}\omega_{\beta\nu})  h^{\mu\nu}
$$
$$
=-k^2a(-k^2)h^{\alpha\beta}(P^{2}_{\alpha\beta, \mu\nu}+P^{0}_{\alpha\beta, \mu\nu}+P^{1}_{\alpha\beta, \mu\nu}+P^{0-w}_{\alpha\beta, \mu\nu})  h^{\mu\nu}
$$
In the last identity, the definition of the Barnes Rivers operators \eqref{ripol} was employed. The analogous procedure may be applied for the remaining terms of the action. The result is
$$
S^2_{gf}=\int h^{\alpha\beta} D^{-1}_{\alpha\beta,\mu\nu} h^{\mu\nu} d^4x,
$$
where the quantity $D^{-1}_{\alpha\beta,\mu\nu}$ in momentum space reads in terms of the gauge fixed coefficients \eqref{donde} as follows
$$
D^{-1}_{\alpha\beta, \mu\nu}=-k^2\bigg[\frac{a}{2} P^2+\frac{a+b}{2} P^1+\frac{a+3d}{2} P^{0-s}
$$
$$
+\frac{a+b+2c+d+2f}{2} P^{0-w}+\frac{\sqrt{3}(c+d)}{2}(P^{0-sw}+P^{0-ws})\bigg]_{\alpha\beta,\mu\nu}.
$$
 The propagator in De Donder gauge is the inverse of the last expression, which can be found easily by use of \eqref{inverso}. The result is
$$
D_{\alpha\beta, \mu\nu}=-\frac{1}{k^2}\bigg[\frac{2}{a} P^2+\frac{2}{a+b} P^1+\frac{a+b+2c+d+2f}{2D} P^{0-s}
$$
$$
+\frac{a+3d}{2D} P^{0-w}-\frac{\sqrt{3}(c+d)}{2D}(P^{0-sw}+P^{0-ws})\bigg]_{\alpha\beta,\mu\nu}.
$$
Here, the following quantity $$
4D=(a+3d)(a+b+2c+d+2f)-(c+d)^2,$$
has been introduced.

The expressions given above can be simplified further by using \eqref{donde}, which implies that
$a+b=1$,  $c+d=-\frac{1}{2}$  and $b+c+f=0$. These relations allows to express the quantities of interest in terms of two of the functions, which may be  chosen as $a$ and $d$, resulting in
$$
Q_{\alpha\beta, \mu\nu}=-k^2\bigg[\frac{a}{2} P^2+\frac{1}{2} P^1+\frac{a+3d}{2} P^{0-s}
+\frac{2a+d-1}{2} P^{0-w}-\frac{\sqrt{3}}{4}(P^{0-sw}+P^{0-ws})\bigg]_{\alpha\beta,\mu\nu},
$$
\begin{equation}\label{propagador}
D^{\alpha\beta, \mu\nu}=-\frac{1}{k^2}\bigg[\frac{2}{a} P^2+ 2P^1+\frac{2a+d-1}{2D} P^{0-s}
+\frac{a+3d}{2D} P^{0-w}+\frac{\sqrt{3}}{4D}(P^{0-sw}+P^{0-ws})\bigg]_{\alpha\beta,\mu\nu}.
\end{equation}
with
$$D=\frac{1}{4}(a+3d)(2a+d-1)-\frac{3}{16}.$$
The quantity \eqref{propagador} will be fundamental for studying the desired process in the eikonal limit.  As a consistency test, it may be noticed that the GR limit, which follows by imposing $F_i(\square)=0$, implies by \eqref{donde} that $a\to 1$ and $2d\to -1$. In this limit, the last expressions reduce to
\begin{equation}\label{relativiza}
Q_{\alpha\beta,\mu\nu}=-\frac{k^2}{2}\eta_{\alpha\mu}\eta_{\beta\nu}+\frac{k^2}{4}\eta_{\alpha\beta}\eta_{\mu\nu},\qquad 
D_{\alpha\beta,\mu\nu}=-\frac{1}{2k^2}\bigg[\eta_{\alpha\mu}\eta_{\beta\nu}+\eta_{\alpha\nu}\eta_{\beta\mu}-\eta_{\alpha\beta}\eta_{\mu\nu}\bigg],
\end{equation}
which are the known quantities  corresponding to GR in the De Donder gauge. 

For the computational calculation of the vertices of the theory, modern references are \cite{dubna1}-\cite{prinz}.
\subsection{The strength of the gravitational interaction}

For the minimally coupled scalar field model in consideration, whose lagrangian is given in \eqref{green}
as follows
$$
L=L_{g} +\frac{1}{2}g^{\mu\nu}\partial_\mu\phi_1\partial_\nu\phi_1+\frac{1}{2}g^{\mu\nu}\partial_\mu\phi_2\partial_\nu\phi_2-\frac{1}{2}M_\sigma^2\phi_2^2,
$$the leading scalar-graviton scalar vertex is well known. It is given by
\begin{equation}\label{vertices}
	\tau_a^{\mu\nu}(p,p') =- i\kappa
	\Big[
	p^\mu p'^\nu + p^\nu p'^\mu
	- \eta^{\mu\nu} (p\cdot p'-m_\phi^2)
	\Big],
\end{equation}
where $\kappa=\sqrt{16\pi G_N}$.  The task is now to determine the tree level amplitude for the $2\to 2$ scattering between $\phi$ and $\sigma$. As the transfer  $\Delta$ is small, the term $p\cdot p^{\prime}\sim p\cdot p\sim 0$, the last condition follows from the fact that the mass  $m_\phi$ of the light particle can be neglected. These considerations lead to a simplified vertex
\begin{equation}\label{bicho}
	\tau^{\mu\nu}_1(p,p^{\prime})  \simeq 2i\kappa p_1^\mu p_1^\nu
\end{equation}
By making an analogy with QED or QCD, the quantity $g_g=2\kappa p_1^\mu$ can be interpreted as a dimensionless coupling constant.  Furthermore, for the massive particle,  as $q\sim q^{\prime}\sim (M_\sigma, 0, 0, 0)$ the corresponding vertex expression simplifies to
\begin{equation}\label{vertices}
\tau_{a}^{\mu\nu}(q,q^{\prime}) =- 2i\kappa M_{\sigma}^2 \delta^{\mu 0}\delta^{\nu 0},
\end{equation}
In terms of the above approximations,  the scattering amplitude at tree level becomes
\begin{equation}\label{amplitudes}
	A^T(p, p', q, q')=\tau^{\mu\nu}_1(p, p') 
	D_{\mu\nu,\rho\sigma}(\vec{\Delta})
	\tau^{\rho\sigma}_2(q, q')=-4k^2 p^\mu p^{\prime \nu} D_{\mu \nu, 00}(\vec{\Delta}) M_\sigma^2. 
\end{equation}
Here the propagator has to be evaluated at transferred momentum $\vec{\Delta}$, which has only transversal components since, in the present approximation, $p^{\prime}_z\sim p_z$. In order to evaluate the last scattering amplitude  the following formulas  
$$
\omega_{00}=\frac{k_0^2}{k^2}, \qquad \theta_{00}=\eta_{00}-\frac{k_0^2}{k^2}, \qquad
p^\alpha\omega_{\alpha 0}=\frac{(p\cdot k)k_0}{k^2},\qquad p^\alpha\theta_{\alpha 0}=E_0-\frac{(p\cdot k)k_0}{k^2},
$$
$$
p^\alpha p^\beta \omega_{\alpha \beta}=\frac{(p\cdot k)^2}{k^2},\qquad p^\alpha\theta_{\alpha 0}=E^2_0-\frac{(p\cdot k)^2}{k^2},
$$
are useful. They follow directly from the definition \eqref{ripol2} and \eqref{mandelstamo}. With these identities at hand, it is not difficult to see that
$$
 p^{\prime\mu} p^{\prime\nu} P^{0-w}_{\mu \nu,00}=\frac{k^2_{0}(k\cdot p^{\prime})^2}{k^4},\qquad 
p^{\prime\mu} p^{\prime\nu}P^{0-sw}_{\mu\nu,00}=\frac{1}{\sqrt{3}}\bigg[E^2_0-\frac{(p\cdot k)^2}{k^2}\bigg]\frac{k_0^2}{k^2},
$$
$$
p^{\prime\mu}p^{\prime\nu}P^{0-ws}_{\mu \nu,00}=\frac{1}{\sqrt{3}}\bigg[\eta_{00}-\frac{k_0^2}{k^2}\bigg]\frac{(p^{\prime}\cdot k)^2}{k^2},\qquad
p^{\prime\mu}p^{\prime\nu}P^{0-s}_{\mu\nu,00}=\frac{1}{3}\bigg[E^2_0-\frac{(p^{\prime}\cdot k)^2}{k^2}\bigg]\bigg[\eta_{00}-\frac{k_0^2}{k^2}\bigg],
$$
$$
p^{\prime\mu}p^{\prime\nu}P^1_{\mu \nu,00}=2\bigg[E_0-\frac{(p^{\prime}\cdot k)k_0}{k^2}\bigg]\frac{(p^{\prime}\cdot k)k_0}{k^2},
$$
\begin{equation}\label{evaluando1}
p^{\prime \mu}p^{\prime \nu}P^{2}_{\mu \nu,00}=\bigg[E_0-\frac{(p^{\prime}\cdot k)k_0}{k^2}\bigg]^2-\frac{1}{3}\bigg[E^2_0-\frac{(p^{\prime}\cdot k)^2}{k^2}\bigg]\bigg[\eta_{00}-\frac{k_0^2}{k^2}\bigg],
\end{equation}
which follows directly from the definition of the Barnes-Rivers operators given in  \eqref{ripol}.  In addition, from \eqref{ripol2} it can be deduced that
$$
\omega_\alpha^{\;\alpha}=1, \qquad \theta_\alpha^{\;\alpha}=3,\qquad 
\omega_{\mu\alpha}\omega_{\nu}^{\;\alpha}=\omega_{\mu\nu},\qquad \omega_{\mu}^{\;\alpha}\theta_{\alpha\nu}=0,\qquad \theta_{\mu\alpha}\theta_{\nu}^{\;\alpha}=\theta_{\mu\nu}.
$$Under the assumption that  $\vec{k}=(0, \vec{\Delta}, 0)$, as $p^{\prime}$ is almost directed along the $z$ axis, it is seen from the last expressions that everything in \eqref{evaluando1} can be neglected, except for the term proportional to $E_0^2$. This implies that \eqref{amplitudes} is expressed in a remarkably simple form
\begin{equation}\label{amplitudes2}
	A^T(p, p', q, q')\sim \frac{2\kappa^2E_0^2 M_\sigma^2}{ \Delta^2 }\bigg[\frac{4}{3a(- \Delta^{2})}+\frac{2a(-\Delta^{2})+d( -\Delta^{2})-1}{6D ( -\Delta^{2})}\bigg].
\end{equation}
In the present approximation $\Delta<<E_0\sim p_z<< M_\sigma$ the definition of the $s$ Mandelstam variable \eqref{mandelstamo} reduces in this case to
$$
s=(E_0+M_\sigma)^2-p_z^2-\Delta^2\sim 2 E_0M_\sigma,
$$
and therefore the amplitude can be written as
\begin{equation}\label{amplitudes2}
A^T(s, \Delta)\sim \frac{\kappa^2 s^2}{ 2\Delta^2 }\bigg[\frac{4}{3a(- \Delta^{2})}+\frac{2a(-\Delta^{2})+d( -\Delta^{2})-1}{6D ( -\Delta^{2})}\bigg].
\end{equation}
By making an analogy with QCD, where the eikonal amplitude is given by $A(s, \Delta)\sim g_s^2 s \Delta^{-2}$ up to color factors, one may define the dimensionless quantity
\begin{equation}\label{fuerza2}
\alpha_e(s, \Delta)=\frac{\kappa^2s}{ 2}\bigg[\frac{4}{3a(- \Delta^{2})}+\frac{2a(-\Delta^{2})+d( -\Delta^{2})-1}{6D ( -\Delta^{2})}\bigg],
\end{equation}
which may be interpreted as the strength of the gravitational interaction in this approximation. This quantity will be useful for describing the limits of the eikonal approximation, this will be emphasized below.

\section{The scattering matrix in the eikonal limit}

After giving a description of  the gravity models in consideration,  the problem of scattering between a massless and an extremely massive particle interchanging gravitons described by such models is reduced to the estimation of  the matrix elements $<p|G^{c}(x,y|h)|p>$.  These elements were defined in \eqref{green0}.  These quantities are fundamental for calculating the scattering amplitude through \eqref{green}.   In the forward limit, $p\sim p’$, the matrix elements  were calculated  in \cite{fazio} by use of Fradkin  modified perturbation theory, an scheme which generalizes the Schwinger  proper time method \cite{fradkin}- \cite{fried}.  The master formula for the light particle is given by
$$
<p|G^{c}(x,y|h)|p> =\lim\limits _{\substack{p^{2}\rightarrow0\\
p^{\prime{2}}\rightarrow 0}
}\int d^{4}xd^{4}ye^{-ip\cdot x}e^{ip\cdot y}\square_{x}\square_{y}\int\frac{d^{4}l}{(2\pi)^{4}}e^{-il\cdot(x-y)}
$$
\begin{equation} \label{forum}
 \int_{0}^{+\infty}d\nu e^{i\nu(\ell^{2}+i\epsilon)}(Y(x,\nu)-1),
\end{equation}
where the following Fradkin nucleus 
$$
Y(x,\nu)= 1+\sum_{n=1}^{+\infty}\frac{(-i)^{n}\kappa^{n}}{n!}\left(\prod\limits _{m=1}^{n}\int_{0}^{\nu}d\xi_{m}\int\frac{d^{4}k_{m}}{(2\pi)^{4}}l^{\mu}l^{\beta}\hat{h}_{\mu\beta}(k_{m})e^{-ik_{m}\cdot x}\right)$$
\begin{equation}\label{funcionar} \exp\left[2i\sum_{m=1}^{n}l\cdot k_{m}(\nu-\xi_{m})\right] \exp\left[i\sum_{m,m_{1}}^n k_{m}\cdot k_{m_{1}}\left(\frac{\xi_{m}+\xi_{m_{1}}}{2}+\frac{1}{2}|\xi_{m}-\xi_{m_{1}}|-\nu\right)\right],\end{equation}
has been introduced. In the eikonal limit $p$ and $p^\prime$ are $M_\sigma$ are taken to be very large, and the terms quadratic in $k_m$ may be neglected, which is equivalent to say that the last exponential in the last formula can be set equal to one. This simplification leads to the approximated matrix element
$$
<p^\prime |G^{c}(x,y|h)|p>=(2\pi)^{4}\lim\limits _{\substack{p^{2}\rightarrow0\\
{p^{\prime}}^{2}\rightarrow0
}
}p^{2}{p^{\prime}}^{2}\sum_{n=1}^{\infty}\frac{(-i)^{n}\kappa^{n}}{n!}\int_{0}^{+\infty}d\nu e^{i\nu(p^{2}+i\epsilon)}
\int\frac{d^{4}k_{1}\dots d^{4}k_{n}}{(2\pi)^{4n}}
$$
\begin{equation}\label{lighparticle}
\delta^{4}(p^{\prime}-p+k_{1}+..+k_{n})
p^{\prime\mu_{1}}p^{\prime\beta_{1}}\hat{h}_{\mu_{1}\beta_{1}}(k_{1})\dots p^{\prime\mu_{n}}p^{\prime\beta_{n}}\hat{h}_{\mu_{n}\beta_{n}}(k_{n})\prod_{m=1}^n \frac{1-e^{-2i\nu p’\cdot k_m}}{2ip’\cdot k_m}.
\end{equation}
In similar fashion, the matrix elements for the heavy particle has been calculated as \cite{fazio}
$$
<q’|G^c(x,y|h_{\mu\nu})|q>=  (2\pi)^{4}\lim\limits_{\substack {q^2\rightarrow M_\sigma^2\\ {q^{\prime}}^2\rightarrow M_\sigma^2}}(q^2-M_\sigma^2)({q^{\prime}}^{2}-M_\sigma^2)\sum_{r=1}^{+\infty}\frac{(-i)^r\kappa^r}{r!}
$$
$$
\int_0^{+\infty}d\nu_1 e^{i\nu_1(q’^2-M_\sigma^2+i\epsilon)}\int\frac{d^4 \tilde{k}_1\dots d^4\tilde{k}_r}{(2\pi)^{4r}}\delta^4(q’-q+\tilde{k}_1+\dots \tilde{k}_r)
$$
\begin{equation}\label{heavyparticle}
 q’^{\mu_1}q’^{\beta_1}\hat{h}_{\mu_1\beta_1}(\tilde{k}_1)\dots q’^{\mu_r}q’^{\beta_r}\hat{h}_{\mu_r\beta_r}(\tilde{k}_r)\prod_{m=1}^r \frac{1-e^{-2i \nu_1 q’\cdot \tilde{k}_m }}{2iq’\cdot \tilde{k}_m}. \end{equation}

With the help of the last two expressions, the scattering amplitude \eqref{green} is written as
$$
i(2\pi)^4\delta^8(p+q-p^{\prime}-q^{\prime}) T(p,p^{\prime}, q,q^{\prime})=(2\pi)^{8}\lim\limits _{\substack{p^{2}\rightarrow0\\
p^{\prime2}\rightarrow0
}
}\lim\limits_{\substack {q^2\rightarrow M_\sigma^2\\ q^{\prime2}\rightarrow M_\sigma^2}}(q^2-M_\sigma^2)(q^{\prime2}-M_\sigma^2)p^{2}p^{\prime 2}
$$
$$
\sum_{n=1}^{\infty}\sum_{r=1}^{+\infty}\frac{(-i)^{n+r}\kappa^{n+r}}{n!r!}\int_{0}^{+\infty}d\nu e^{i\nu(p^{2}+i\epsilon)}\int_0^{+\infty}d\nu_1 e^{i\nu_1(q’^2-M_\sigma^2+i\epsilon)}
$$
$$
\int\frac{d^{4}k_{1}\dots d^{4}k_{n}d^4 \tilde{k}_1\dots d^4\tilde{k}_r}{(2\pi)^{4n+4r}}\delta^{4}(p^{\prime}-p+k_{1}+..+k_{n})\delta^4(q’-q+\tilde{k}_1+\dots \tilde{k}_r)
$$
$$
\prod_{m=1}^n \frac{1-e^{-2i\nu p’\cdot k_m}}{2ip’\cdot k_m}\prod_{p=1}^r \frac{1-e^{-2i\nu_1 q’\cdot \tilde{k}_p}}{2iq’\cdot \tilde{k}_p}
$$
$$\exp{ \bigg[i\int d^4x d^4y \left(\frac{\delta}{\delta h^{\alpha\beta}(x)}D^{\alpha\beta,\gamma\delta}(x-y)\frac{\delta}{\delta h^{\prime \gamma\delta}(y)}\right)\bigg]}
$$
$$
p^{\prime\mu_{1}}p^{\prime\beta_{1}}\hat{h}_{\mu_{1}\beta_{1}}(k_{1})\dots p^{\prime\mu_{n}}p^{\prime\beta_{n}}\hat{h}_{\mu_{n}\beta_{n}}(k_{n})q’^{\mu_1}q’^{\beta_1}\hat{h}^{\prime}_{\mu_1\beta_1}(\tilde{k}_1)\dots q’^{\mu_r}q’^{\beta_r}\hat{h}^{\prime}_{\mu_r\beta_r}(\tilde{k}_r)\bigg|_{h, h^{\prime}=0}
$$
By taking into account the following characterization of the Fourier components of the metric
$$
h_{\alpha\beta}(x)=\int e^{ip\cdot x} h_{\alpha\beta}(p) d^4p,\qquad h_{\alpha\beta}(p)=\int e^{-ip\cdot y} h_{\alpha\beta}(y) d^4y.
$$
$$
\frac{\delta}{\delta h^{\prime \gamma\delta}(x)}= \int d^4 p\; e^{-ip\cdot x} \frac{\delta}{\delta h^{\prime \gamma\delta}(p)},
$$
together with 
$$
\exp{ \bigg[i\int d^4k\;\left(\frac{\delta}{\delta h^{\alpha\beta}(-k)}D^{\alpha\beta,\gamma\delta}(-k)\frac{\delta}{\delta h^{\prime \gamma\delta}(k)}\right)\bigg]}
$$
$$
=\sum_{l=0}^{\infty}\frac{i^l}{l!}\prod_{j=1}^{l}\int d^4k_j\left(\frac{\delta}{\delta h^{\alpha\beta}(-k_j)}D^{\alpha\beta,\gamma\delta}(-k_j)\frac{\delta}{\delta h^{\prime \gamma\delta}(k_j)}\right),
$$
 it is found that
$$
i(2\pi)^4\delta^4(p+q-p’-q’)T(p,p^{\prime}, q,q^{\prime})=(2\pi)^{8}\lim\limits _{\substack{p^{2}\rightarrow0\\
p^{\prime2}\rightarrow0
}
}\lim\limits_{\substack {q^2\rightarrow M_\sigma^2\\ q^{\prime2}\rightarrow M_\sigma^2}}(q^2-M_\sigma^2)(q^{\prime2}-M_\sigma^2)p^{2}p^{\prime{2}}
$$
$$
\sum_{n=1}^{\infty}\sum_{r=1}^{+\infty}\frac{(-i)^{n+r}\kappa^{n+r}}{n!r!}\int_{0}^{+\infty}d\nu e^{i\nu(p^{2}+i\epsilon)}\int_0^{+\infty}d\nu_1 e^{i\nu_1(q’^2-M_\sigma^2+i\epsilon)}
$$
$$
\int\frac{d^{4}k_{1}\dots d^{4}k_{n}d^4 \tilde{k}_1\dots d^4\tilde{k}_r}{(2\pi)^{4n+4r}}\delta^{4}(p^{\prime}-p+k_{1}+..+k_{n})\delta^4(q’-q+\tilde{k}_1+\dots \tilde{k}_r)
$$
$$
\prod_{m=1}^n \frac{1- e^{-2i\nu p’\cdot k_m}}{2ip’\cdot k_m}\prod_{p=1}^r \frac{1-e^{-2i\nu_1 q’\cdot \tilde{k}_p}}{2iq’\cdot \tilde{k}_p}
$$
$$\sum_{l=0}^{\infty}\frac{1}{l!}\prod_{j=1}^{l}i\int d^4K_j\;\left(\frac{\delta}{\delta h^{\alpha\beta}(-K_j)}D^{\alpha\beta,\gamma\delta}(-K_j)\frac{\delta}{\delta h^{\prime \gamma\delta}(K_j)}\right)
$$
\begin{equation}\label{fisura1}
p^{\prime\mu_{1}}p^{\prime\beta_{1}}\hat{h}_{\mu_{1}\beta_{1}}(k_{1})\dots p^{\prime\mu_{n}}p^{\prime\beta_{n}}\hat{h}_{\mu_{n}\beta_{n}}(k_{n})q’^{\mu_1}q’^{\beta_1}\hat{h}^{\prime}_{\mu_1\beta_1}(\tilde{k}_1)\dots q’^{\mu_r}q’^{\beta_r}\hat{h}^{\prime}_{\mu_r\beta_r}(\tilde{k}_r)\bigg|_{h, h^{\prime}=0}
\end{equation}
The product of $l$ functional derivatives acting on $n+r$ terms, if $2l\neq n+r$,  gives vanishing contribution due to the prescription  $h=h^{\prime}=0$ .  Therefore, only the situation $2l=n+r$ has to be considered and furthermore, as the Green function is the connected one, the pairs have to be chosen corresponding to different particles, i.e, one to the heavy scalar and one to the light one.   Therefore, the action of the operator $\frac{\delta}{\delta h^{\alpha\beta}(-K_j)}D^{\alpha\beta,\gamma\delta}(-K_j)\frac{\delta}{\delta h^{\prime \gamma\delta}(K_j)}$ is only non trivial a pair of different metrics $h_{\alpha\beta}$ and $h^{\prime}_{\alpha\beta}$, and it follows  that $n=r=l$ .  As the functional derivatives of the metric $h^{\alpha\beta}$ correspond to opposite momentum values $K_j$ and $-K_j$, the action of the operator $\frac{\delta}{\delta h^{\alpha\beta}(-K_j)}D^{\alpha\beta,\gamma\delta}(-K_j)\frac{\delta}{\delta h^{\prime \gamma\delta}(K_j)}$ on an product of the form $\hat{h}_{\mu_{i}\beta_{i}}(k_{n})\hat{h}^{\prime}_{\mu_j\beta_j}(\tilde{k}_j)$, in particular,  fixes $k_i=-\tilde{k}_j$ , and this follows for all the possible choices of pairs. As a consequence, the products of deltas become
$$
\delta^{4}(p^{\prime}-p+k_{1}+..+k_{n})\delta^4(q’-q+\tilde{k}_1+\dots \tilde{k}_n)\longrightarrow \delta^{4}(p^{\prime}-p+k_{1}+..+k_{n})\delta^4(q’-q-k_1-\dots -k_n)
$$
$$
\longrightarrow \delta^{4}(p^{\prime}-p+k_{1}+..+k_{n})\delta^4(p^{\prime}+q’-p-q)
$$
After some simple calculation and by deleting a common $\delta^4(p^{\prime}+q’-p-q)$ factor, it is obtained from \eqref{fisura1} that
$$
 iT(p,p^{\prime},q,q^{\prime})=(2\pi)^{4}\lim\limits _{\substack{p^{2}\rightarrow0\\
p^{\prime2}\rightarrow0
}
}\lim\limits_{\substack {q^2\rightarrow M_\sigma^2\\ q^{\prime2}\rightarrow M_\sigma^2}}(q^2-M_\sigma^2)(q^{\prime2}-M_\sigma^2)p^{2}p^{\prime{2}}
 \sum_{n=1}^{+\infty}\frac{i^n\kappa^{2n}M_{\sigma}^{2n}}{n!(2\pi)^{4n}}\int_{0}^{+\infty}d\nu\,e^{i\nu(p^{2}+i\epsilon)}
$$
$$
 \int_{0}^{+\infty}d\nu_{1}\,e^{i\nu_{1}[(q^{2}-M_{\sigma}^{2})+i\epsilon]}\prod_{m=1}^{n}\int d^4 k_ m p^{\prime\mu_{m}}p^{\prime\beta_{m}}D_{\mu_m \beta_m,00}
 $$
 \begin{equation}\label{matrix}
 \frac{(1-e^{-2i\nu p\cdot k_m})}{2ip\cdot k_m}\frac{(1-e^{-2i\nu_{1}q\cdot k_{m}})}{2iq\cdot k_m}\delta^{4}(q^{\prime}-q-k_{1}-k_{2}-\dots k_{n}).
 \end{equation}
These expressions will be simplified further by use of two of the eikonal identities \eqref{idento}-\eqref{idento2}.  
These identities are employed as follows. From \eqref{idento}, the integrals in $\nu_1$ may be worked as 
$$
\delta^4(q’-q-k_1-k_2-\dots k_n)\lim_{\substack {q^2\rightarrow M_\sigma^2\\ q’^2\rightarrow M_\sigma^2}}(q^2-M_\sigma^2)(q^{\prime 2}-M_\sigma^2)
\int_0^{+\infty} d\nu_1 e^{i\nu_1[q^{\prime 2}-M_\sigma^2+i\epsilon]}\prod_{m=1}^n \frac{1-e^{-2i\nu_1q’\cdot k_m}}{2iq’\cdot k_m}
$$
$$
=-i\delta^4(q’-q-k_1-k_2-\dots -k_n)\lim\limits_{\substack {q^2\rightarrow M_\sigma^2\\ q’^2\rightarrow M_\sigma^2}}(q^2-M_\sigma^2)(q^{\prime 2}-M_\sigma^2) 
\sum_{\pi}\bigg(\frac{1}{q^{\prime 2} -M_\sigma^2+i\epsilon}\bigg)
$$
$$
\bigg(\frac{1}{q^{\prime 2} -M_\sigma^2+2q’\cdot k_{\pi(1)}+i\epsilon}\bigg)
\dots \bigg(\frac{1}{q^{\prime 2} -M_\sigma^2+2q’\cdot (k_{\pi(1)}+\dots k_{\pi(n)})+i\epsilon}\bigg)
$$
$$
=-i\delta^4(q’-q-k_1-k_2-\dots -k_n) \lim_{q^2-M_\sigma^2} (q^2-M_\sigma^2)
\sum_{\pi}\
\bigg(\frac{1}{q^{\prime 2} -M_\sigma^2+2q’\cdot k_{\pi(1)}+i\epsilon}\bigg)
$$
\begin{equation}\label{bose}
\dots \bigg(\frac{1}{q^{\prime 2} -M_\sigma^2+2q’\cdot (k_{\pi(1)}+\dots k_{\pi(n)})+i\epsilon}\bigg)
\end{equation}
The delta enforces momentum conservation, which implies  $q’-q=k_{\pi(1)}+k_{\pi(2)}\dots+k_{\pi(n)}$. This implies that the last term of the product in the above expression may be worked out as
\begin{equation}
 \frac{1}{q^{\prime 2} -M_\sigma^2+2q’\cdot (k_{\pi(1)}+\dots k_{\pi(n)})+i\epsilon}=\frac{1}{q^{\prime 2} -M_\sigma^2+2q’\cdot (q’-q)}\sim \frac{1}{q^2-M_\sigma^2}.
\label{eikonalizando}
\end{equation}
Here, the fact that $q=(M_\sigma, 0, 0, 0)$ and that $q^{\prime} \sim q$ in the quasi static limit in consideration was taken into account. By employing the identity  \eqref{idento2} the last two lines  in \eqref{bose} become
$$
i\delta^4(q’-q-k_1-k_2-\dots -k_n) \lim_{q^2-M_\sigma^2} (q^2-M_\sigma^2)
\sum_{\pi}\
\bigg(\frac{1}{q^{\prime 2} -M_\sigma^2+2q’\cdot k_{\pi(1)}+i\epsilon}\bigg)
$$
$$
\dots \bigg(\frac{1}{q^{\prime 2} -M_\sigma^2+2q’\cdot (k_{\pi(1)}+\dots k_{\pi(n)})+i\epsilon}\bigg)
$$
$$
=
-\frac{i(-2\pi i)^{n-1}}{(2M_\sigma)^{n-1}}\delta(k^0_1)\dots\delta(k^0_n)
 \delta^3(\vec{\Delta}+\vec{k}_1+\vec{k}_2+\dots +\vec{k}_n),
$$
the last step follows by recognizing that  $q\cdot k=M_\sigma k^0$. By collecting all the results from \eqref{bose} till the last formula, it follows that the identity that has been proved is simply
$$
\delta^4(q’-q-k_1-k_2-\dots k_n)\lim\limits_{\substack {q^2\rightarrow M_\sigma^2\\ q’^2\rightarrow M_\sigma^2}}(q^2-M_\sigma^2)(q^{\prime 2}-M_\sigma^2)
$$
$$
\int_0^{+\infty} d\nu_1 e^{i\nu_1[q^{\prime 2}-M_\sigma^2+i\epsilon]}\prod_{m=1}^n \frac{1-e^{-2i\nu_1q’\cdot k_m}}{2iq’\cdot k_m}=-\frac{i(-i\pi)^{n-1}}{(M_\sigma)^{n-1}}\delta(k^0_1)\dots\delta(k^0_n)\delta^3(\vec{\Delta}+\vec{k}_1+\vec{k}_2+\dots +\vec{k}_n).
$$
This implies that the time components $k_{i0}$ can be selected to be zero. By use of the last formula, the scattering matrix  \eqref{matrix} may be worked out to the more simple form
$$
 iT(p,p^{\prime}, q,q^{\prime})=-i(2\pi)^{4}\lim\limits _{\substack{p^{2}\rightarrow0\\
p^{\prime2}\rightarrow0}
}p^{2}p^{\prime{2}}
 \sum_{n=1}^{+\infty}\frac{i^n\kappa^{2n} M_\sigma^{2n}}{n!(2\pi)^{4n}}\int_{0}^{+\infty}d\nu\,e^{i\nu(p^{2}+i\epsilon)}
 $$
 $$\frac{(-i\pi)^{n-1}}{(M_\sigma)^{n-1}}\prod_{m=1}^{n}\int d^{3}\vec{k}_{m}p^{\prime\mu_{m}}p^{\prime\beta_{m}}D_{\mu_m \beta_m,00}\frac{1-e^{-2i\nu \vec{p}\cdot \vec{k}_{m}}}{2i\vec{p}\cdot \vec{k}_{m}}
 \delta^3(\vec{\Delta}+\vec{k}_1+\vec{k}_2+\dots +\vec{k}_n).
 $$
Still, the constraint that $p^{\prime}$ is almost directed in the $\hat{z}$ direction gives further simplifications. First, note that
\begin{equation}
\vec{p}\,\,’\cdot \vec{k}_m= p^{\prime z}k_m^z+\frac{\vec{\Delta}}{2}\cdot \vec{k}_m^\perp\sim 
 E_\phi k_m^z +\frac{\vec{\Delta}}{2}\cdot \vec{k}_m^\perp.
\end{equation}
In these terms it can be shown by a first order Taylor expansion that 
\begin{equation}
\prod_{m=1}^n  \frac{1-e^{-2i\nu \vec{p}\,\,’\cdot \vec{k}_m}}{2i\vec{p}\,’\cdot \vec{k}_m}\sim\prod_{m=1}^n \frac{1-e^{-2i\nu E_\phi k_m^z}}{2i E_\phi k_m^z},
\end{equation}
up to a term  proportional to $\sum\limits_{m=1}^n \vec{\Delta}\cdot \vec{k}_m^\perp$ , which is of order ${\Delta}^2$ due to the presence of  $\delta^3(\vec{\Delta}+\vec{k}_1+\vec{k}_2+\dots \vec{k}_n)$. Therefore this term  can be  neglected at first order.  The amplitude is now
$$
 iT(p,p^{\prime}, q,q^{\prime})=-i(2\pi)^{4}\lim\limits _{\substack{p^{2}\rightarrow0\\
p^{\prime2}\rightarrow0}
}p^{2}p^{\prime{2}}
 \sum_{n=1}^{+\infty}\frac{i^n\kappa^{2n} M_\sigma^{2n}}{n!(2\pi)^{4n}}\int_{0}^{+\infty}d\nu\,e^{i\nu(p^{2}+i\epsilon)}
 $$
 \begin{equation}\label{rus}
 \frac{(-i\pi)^{n-1}}{(M_\sigma)^{n-1}}\prod_{m=1}^{n}i\int d^{3}\vec{k}_{m}p^{\prime\mu_{m}}p^{\prime\beta_{m}}D_{\mu_m \beta_m,00}\frac{1-e^{2i\nu E_\phi k_{mz}}}{2i E_\phi k_{mz}}\delta^3(\vec{\Delta}+\vec{k}_1+\vec{k}_2+\dots +\vec{k}_n).
 \end{equation}
Furthermore, by use of \eqref{idento} it is easily seen that
 $$
\int_{0}^{+\infty}d\nu\,e^{i\nu(p^{2}+i\epsilon)} \prod_{m=1}^n \frac{1-e^{-2i\nu E_\phi k_{mz}}}{2i E_\phi k_{mz}}=-i\sum_\pi \frac{1}{p^2+i\epsilon}\frac{1}{p^2+2E_\phi k_{\pi(1)z} +i\epsilon}
$$
$$
\dots \frac{1}{p^2+2E_\phi k_{\pi(1)z}+\dots +2E_\phi k_{\pi(n)z}+i\epsilon},
 $$
and from \eqref{idento2}  and the last formula, it is inferred that
$$
\lim\limits _{\substack{p^{2}\rightarrow0\\
p^{\prime2}\rightarrow0}
}p^{2}p^{\prime{2}}\delta^3(\vec{\Delta}+\vec{k}_1+\vec{k}_2+\dots +\vec{k}_n)
\sum_\pi \frac{1}{p^2+i\epsilon}\frac{1}{p^2+2E_\phi k_{\pi(1)z} +i\epsilon}
$$
$$
\dots \frac{1}{p^2+2E_\phi k_{\pi(1)z}+\dots +2E_\phi k_{\pi(n)z}+i\epsilon}
=\frac{i(-\pi i)^{n-1}}{(E_\phi)^{n-1}}\delta(k_{1z})\dots\delta(k_{nz})\delta^{2}(\vec{k}_{1}^{\perp}+\dots\vec{k}_{n}^{\perp}+\vec{\Delta}).
$$
The last formula shows that longitudinal components $k_{iz}$ can effectively be set to zero, due to the presence of the terms $\delta(k_{iz})$. By use of the last two formulas, the amplitude \eqref{rus} is simplified to
$$
 iT(p,p^{\prime}, q,q^{\prime})=-4iM_\sigma E_\phi (2\pi)^2
 \sum_{n=1}^{+\infty}\frac{\kappa^{2n}}{n!}
 \frac{(-iM_{\sigma})^{n}}{(16\pi^2 E_\phi)^{n}}
$$
 $$
 \prod_{m=1}^{n}
 \int d^{2}k^{\perp}_{m}p^{\prime\mu_{m}}p^{\prime\beta_{m}}D_{\mu_m \beta_m,00}\delta^{2}(\vec{k}_{1}^{\perp}+\dots\vec{k}_{n}^{\perp}+\vec{\Delta}).
 $$
The final task is to evaluate the propagator terms $p^{\prime\mu_{m}}p^{\prime\beta_{m}}D_{\beta_m \beta_n,00}$. This can be done following the procedure described in \eqref{evaluando1} and with the propagator definition \eqref{propagador}. As the longitudinal momentum $k_{iz}$
and  the energy $k_{i0}$ both effectively vanishes due to the Dirac delta terms in the integrand, the procedure in \eqref{evaluando1} boils down to a single term
 \begin{equation}\label{false3}
p^{\prime\mu_{m}}p^{\prime\beta_{m}}D_{\beta_m \beta_n,00}=-\bigg[\frac{4}{3a(- k_m^{\perp 2})}+\frac{2a( -k_m^{\perp 2})+d( -k_m^{\perp 2})-1}{6D( -k_m^{\perp 2})}\bigg]E_{\phi}^2.
\end{equation}
This reduces the amplitude to a manageable form
$$
 iT(p,p^{\prime}, q,q^{\prime})=-4iM_\sigma E_\phi(2\pi)^{2}
 \sum_{n=1}^{+\infty}\frac{1}{n!}
 \frac{(i\kappa^2 E_\phi M_{\sigma})^{n}}{(16\pi^2)^{n}}
 \prod_{m=1}^{n}
$$
$$
 \int \frac{d^{2}k^{\perp}_{m}}{k^{\perp 2}_m}\bigg[\frac{4}{3a(- k_m^{\perp 2})}+\frac{2a(- k_m^{\perp 2})+d( -k_m^{\perp 2})-1}{6D( -k_m^{\perp 2})}\bigg]\delta^{2}(\vec{k}_{1}^{\perp}+\dots\vec{k}_{n}^{\perp}+\vec{\Delta}).
 $$
 A convenient trick in order to eliminate the complications due to the Dirac delta inside the integral is to take the Fourier transform with respect to the transferred momentum $\vec{\Delta}=\vec{k}^{\perp}$. This transformation is simply
 $$
iT(\vec{b}^{\perp})=\frac{1}{(2\pi)^2} \int d^2 k^\perp e^{-i\vec{b}^{\perp}\cdot \vec{k}^\perp} iT(p, p^{\prime}, q, q^{\prime}).
 $$
and translate the amplitude into the transverse impact parameter space, described by a two dimensional vector   $\vec{b}^{\perp}$. The last integral can be calculated directly, and the resulting transformed amplitude can be written in terms of an eikonal phase $\chi_0$ as
 \begin{equation}
iT(\vec{b}^{\perp})= -4iE_\phi M_\sigma (e^{i\chi_0}-1)
\end{equation}
with 
\begin{equation}
\chi_0 ({\vec{b}}^\perp)= \frac{\kappa^2 M_\sigma E_\phi}{8\pi^2}\int\frac{d^2 k^{\perp}}{k^{\perp 2}}\bigg[\frac{4}{3a(- k_m^{\perp 2})}+\frac{2a( -k_m^{\perp 2})+d( -k_m^{\perp 2})-1}{6D( -k_m^{\perp 2})}\bigg]e^{-i\vec{b}^{\perp}\cdot \vec{k}^{\perp}}.
\label{fases}
\end{equation}
This phase depends on the choice of the function $a(-k^{\perp 2})$ and $d(-k^{\perp2})$, which of course is a characteristic quantity for the higher derivative scenario under study.  

Note that in the GR limit $a\to 1$ and $2d\to -1$ discussed in \eqref{donde}, for which the determinant $D$ defined below \eqref{propagador}  is  $D=-1/4$, the quantity in brackets in \eqref{fases} tends to $1$ and the GR phase in \cite{fazio}
is recovered. This is an important test for the obtained formulas.

.

\subsection{ General remarks about the eikonal approximation}
At this point, it should be remarked that the previous calculation was done in blind form, as the small impulse transfer approximation was employed without full justification.  For this reason, before going about the application of the eikonal formalism, it is convenient to discuss its range of applicability.

The classical arguments about the validity of the eikonal approximations may be summarized as follows.  The Compton length of the process is defined as  $\lambda_c\sim 1/\sqrt{s}$ and, if the process has large center of mass energy, this length is rather small. In GR, the strength \eqref{fuerza2}  of the gravitational interaction is related to the effective Schwarzschild radius by the relation $R_s\sqrt{s}\sim \alpha_e$. The problem is that, for the present theories, the exact black hole solutions are not known.  This complicates the subject. Despite this drawback, some possibilities will be discussed now.

As a first prejudice, it may be assumed that the radius $R_c=\frac{\alpha_e(s, \Delta)}{\sqrt{s}}$ is close to the characteristic scale of those unknown black hole solutions for the present models, where the dimensionless coupling constant $\alpha_e$ was already introduced in \eqref{fuerza2}. This assumption will be relaxed or taken with care latter on. It is instructive to analyze its consequences  in some detail. 

If the Compton length $\lambda_c$ is larger than the characteristic radius $R_c$, then a quantum description is required and the assumption that $h_{\mu\nu}$ is small may be dubious. Therefore, the description given above shows  that $\lambda_c=1/\sqrt{s}<<R_c=\alpha/\sqrt{s}$, which requires the strong coupling regime $\alpha>>1$,  a condition  which, by use of the definition \eqref{fuerza2}, is translated into
\begin{equation}\label{a}
2<<\kappa^2 s\bigg[\frac{4}{3a(- \Delta^{2})}+\frac{2a(-\Delta^{2})+d( -\Delta^{2})-1}{6D ( -\Delta^{2})}\bigg].
\end{equation}
Here the determinant $D(-\Delta^2)$ was introduced in \eqref{donde}. As it was shown in \eqref{relativiza}, in the limit  $\Delta\to 0$ it is true that $a(-\Delta^2)\to 1$ and $2d(-\Delta^2)\to -1$.  In this case, the quantity in brackets tends to one, as $4D\to-1$. It follows that if $\Delta$ is small enough, the last requirement implies that $s>>M_p$, that is, the energies involved are transplanckian. However, there can be exceptions. First, it may happen that the quantities in brackets have a real pole at some value of $\Delta$.  In this case $\kappa^2 s$ does not necessarily need to have large values, and therefore the transplanckian limit is not required if the transferred momentum $\Delta$ takes values near the pole.  In fact, this situation is not exceptional. An example is Stelle gravity. 

As is well known, the Stelle lagrangian is given by
$$L_S=(16\pi G_N)^{-1}( R+\alpha R^2+\beta R_{\mu\nu}R^{\mu\nu})\sqrt{-g}.$$
It is important not to confuse the Stelle coupling constant $\alpha$, which multiplies the $R^2$ term in the action, with the effective coupling constant $\alpha_e$ defined in \eqref{fuerza2}.

At this point, it is important to describe the stability region or allowed parameter values $\alpha$ and $\beta$ for this model.  It is convenient to write the Stelle action as
$$
S=-\frac{1}{2\kappa}\int \sqrt{-g} d^4x\bigg[R+\frac{\gamma}{2} R^2-\frac{\delta}{2}W_{\mu\nu\alpha\beta}W^{\mu\nu\alpha\beta}\bigg].
$$
Here  $W_{\mu\nu\alpha\beta}$  represents the Weyl tensor, which is constructed in terms of the scalar curvature  $R$,  the Ricci tensor $R_{\mu\nu}$ and the curvature tensor  $R_{\mu\nu\alpha\beta}$.  The standard expression of the lagrangian employed in the literature is given in terms of  $R$, $R^2$ and $R_{\mu\nu}R^{\mu\nu}$, and the difference with the above lagragian is a total derivative in four dimensions. In fact, in four dimensions, one may replace $W_{\mu\nu\alpha\beta}W^{\mu\nu\alpha\beta}\to R_{\mu\nu}R^{\mu\nu}-\frac{1}{3}R^2$. This leads to the identification
\begin{equation}\label{inaguro}
\beta=-\delta,\qquad \gamma=2\alpha-\frac{2\beta }{3}.
\end{equation}
The advantage of the action written in terms of the Weyl tensor is that the mass of the modes will be expressed in terms of the parameters $\gamma$ and $\delta$ in simple manner. After that, one may go back and obtain constraints  for $\alpha$ and $\beta$. 

Given a small perturbation around the Minkowski metric
$$
g_{\mu\nu}=\eta_{\mu\nu}+(\widetilde{h}_{\mu\nu}-\frac{1}{2}\eta_{\mu\nu}\widetilde{h}),
$$
it may  be decomposed as follows \cite{medeiros1}-\cite{medeiros2}
$$
\widetilde{h}_{\mu\nu}=h_{\mu\nu}+\Psi_{\mu\nu}+\frac{\eta_{\mu\nu}}{2}\bigg(\phi+\frac{\Psi}{3}\bigg).
$$
Here $h_{\alpha\beta}$ is the solution of the GR equations of motion in the traceless transverse gauge, which will be specified below. The component $\Psi_{\mu\nu}$ is of spin  two and the  field $\phi$ is a scalar.  The last expression for the perturbation assumes that the  trace of the perturbation is not necessarily zero, and its value is encoded in $\phi$. The curvature tensor for a generic perturbation can be expanded by the formulas
$$
\Gamma_{\rho\mu\nu}^{\alpha}=\frac{1}{2}\eta^{\alpha\beta}(\partial_{\mu} \widetilde{h}_{\beta\nu}+\partial_{\nu}\widetilde{h}_{\beta\mu}-\partial_{\beta}\widetilde{h}_{\mu\nu})+O(h^2),
$$
$$
R^\alpha_{\rho\mu\nu}=\frac{1}{2}\eta^{\alpha\beta}(\partial_\rho\partial_{\mu}\widetilde{h}_{\beta\nu}-\partial_\rho\partial_{\nu}\widetilde{h}_{\beta\mu}-\partial_\nu\partial_{\beta}\widetilde{h}_{\rho\mu}+\partial_\mu\partial_{\beta}\widetilde{h}_{\rho\nu})+O(h^2),
$$
and $R_{\mu\nu}=\eta^{\alpha\beta}R_{\mu\alpha\nu\beta}$, $R=\eta^{\alpha\beta}R_{\alpha\beta}$ up to higher orders in  $\widetilde{h}^{\mu\nu}$.  The determinant can be expanded as
$$
\sqrt{-g}=\sqrt{\det(-\eta- \widetilde{h})}=\exp{\frac{1}{2}\log(\det(-\eta-\widetilde{h}))}=\exp{\frac{1}{2}\log(\det(-1-\eta \widetilde{h}))},
$$
the last step employs that $\eta_{\mu\nu}=\eta^{-1}_{\mu\nu}$. By taking into account the identity "$\log \det=\text{Tr}\log$" it follows that
$$
\sqrt{-g}=\exp{\frac{1}{2}\text{Tr}(\log(-1-\eta \widetilde{h}))},
$$
and by a simple Taylor expansion of the exponential  and the logarithm it is found that
$$
\sqrt{-g}\simeq 1+\frac{1}{2} \widetilde{h}_\alpha^\alpha+\frac{1}{8} \widetilde{h}_\alpha^\alpha \widetilde{h}_\beta^\beta-\frac{1}{4} \widetilde{h}_\alpha^\beta  \widetilde{h}_\beta^\alpha.
$$
By taking  these expansions into account,  the lagrangian may be linealized as follows
$$
L=\frac{1}{2\kappa}\bigg[\frac{1}{4}\partial _{\mu}h^{\alpha \beta }\partial ^{\mu }h_{\alpha \beta }+\frac{1}{4}\partial _{\mu }\Psi^{\alpha \beta }\partial ^{\mu }\Psi_{\alpha \beta}-\frac{3}{2}\partial _{\mu} \phi\partial ^{\mu }\phi+\frac{1}{2}\partial _{\mu}h^{\alpha \beta }\partial ^{\mu }\Psi_{\alpha \beta }
$$
$$
+\partial _{\beta }h^{\alpha \beta }\partial _{\alpha }\phi
+\partial _{\beta }\Psi ^{\alpha \beta }\partial _{\alpha }\phi
. +\frac{1}{2}\partial _{\mu }h\partial ^{\mu }\phi 
\frac{1}{2}\partial _{\alpha }\phi \partial ^{\alpha }\Psi \bigg]  
$$
$$
+\frac{\gamma }{2\kappa }\bigg[-\frac{9}{2}\left( \square \phi \right)
^{2}+3\partial _{\alpha }\partial _{\beta }h^{\alpha \beta }\square
\phi +3\partial _{\alpha }\partial _{\beta }\Psi ^{\alpha \beta }\square \phi 
+\frac{3}{2}\square h\square \phi   -\frac{3}{2}\square \Psi \square \phi \bigg]  
$$
$$
+\frac{\delta }{2\kappa }\left[ \frac{1}{4}\square h_{\alpha \beta
}\square h^{\alpha \beta }+\frac{1}{2}\square h_{\alpha
\beta }\square \Psi ^{\alpha \beta } +\frac{1}{4}\square \Psi _{\mu \beta }\square \Psi ^{\mu \beta }%
\right]+L_{int},
$$
up to interaction terms collected in $L_{int}$ that appear due to the higher order expansion of the curvature and the metric determinant.

The  resulting free equation of motion for $\phi$  follows from this linearization  by ignoring the term $L_{int}$. The  Euler-Lagrange equation of motion is then
\begin{equation}\label{phi}
3\square\phi- \partial_{\alpha\beta} h^{\alpha\beta}- \partial_{\alpha\beta} \Psi^{\alpha\beta}- \frac{1}{2}\square h+\frac{1}{2}\square \Psi=\frac{\gamma}{2}\bigg[18\square^2 \phi-6\square \partial_\mu \partial_\nu h^{\mu\nu}-6\square \partial_\mu \partial_\nu \Psi^{\mu\nu}-3\square^2 h+3\square^2 \Psi\bigg],
\end{equation}
while for the spin two perturbation $\Psi_{\mu\nu}$ the equations become
\begin{equation}\label{psi}
\square \Psi_{\mu\nu}+\square  h_{\mu\nu}+2\square \phi-\eta_{\mu\nu}\square \phi-6\gamma \square \partial_\mu\partial_\nu \phi+3\gamma\eta_{\mu\nu}\square^2\phi-\delta \square^2 h_{\mu\nu}-\delta \square^2 \Psi_{\mu\nu}=0.
\end{equation}
The transverse-traceless gauge choice mentioned above is not only applied for $h_{\mu\nu}$ but for $\Psi_{\mu\nu}$ as well. This gauge is defined by 
$$
\partial_\nu h^{\mu\nu}=\partial_\nu \Psi^{\mu\nu}=0,\qquad  h=\Psi=0,
$$
and it reduces the equation for the scalar perturbation  to
$$
\square^2 \phi-\frac{1}{3\gamma}\square\phi=0,
$$
which shows that $\phi=\phi_1+\phi_2$ is a free real scalar field with two modes $\phi_1$ and $\phi_2$ with masses $m^2_1=0$ and $m^2_2=m^2=\frac{1}{3\gamma}$. On the other hand, in the equation of motion of $\Psi_{\mu\nu}$ given by \eqref{psi}, the terms proportional to $\phi$ in \eqref{psi} cancel in pairs due the last equation. Furthermore, as $h_{\mu\nu}$ was chosen as the massless GR solution in the transverse-traceless gauge, it follows that $\square h_{\mu\nu}=0$. These simplifications reduce the equations of motion \eqref{psi} to
$$
 \square^2 \Psi_{\mu\nu}-\frac{1}{\delta}\square \Psi_{\mu\nu}=0,
$$
which shows that $\Psi_{\mu\nu}=\Psi^1_{\mu\nu}+\Psi_{\mu\nu}^2$ is a massive spin two field with two excitations $\Psi^1_{\mu\nu}$ and $\Psi^2_{\mu\nu}$, one massless and one with mass $M^2=\frac{1}{\delta}$.

Clearly, the values of $\delta$ or $\gamma$ has to be positive, otherwise these excitations will have imagniary masses. By taking the identification \eqref{inaguro} into account, if follows that $\alpha>0$ and $\beta<0$, and furthermore $3\alpha+\beta>0$. This is to avoid tachyons in the spectrum. 

For this lagrangian, the resulting quantities which follows from \eqref{donde} are 
$$
a(-\Delta^2)=1+\frac{\beta}{2} \Delta^2,\qquad d(-\Delta^2)=-\frac{1}{2}+2\alpha \Delta^2,
$$
\begin{equation}\label{uh}
D(-\Delta^2)=\frac{1}{8}\bigg[-1+(12\alpha+\beta) \Delta^2\bigg]\bigg[\frac{1}{2}+(2\alpha+\beta) \Delta^2\bigg]-\frac{3}{16}.
\end{equation}
Clearly, the quantities $\alpha$ and $\beta$ have mass dimensions mass$^{-2}$.  The determinant $D(-\Delta^2)\sim-1/4$
for small values of $\Delta^2$ while it is positive for $\Delta^2$ very large. Asymptotically it goes like
$$
D\sim (12\alpha+\beta)(2\alpha+\beta)\Delta^2.
$$
This means that if $(12\alpha+\beta)$ and $2\alpha+\beta$ have different sign, then a zero for $D$ may be avoided, which is equivalent to say that the effective coupling constant $\alpha_e$ is never divergent. But this is not always guaranteed. 

For the stability region $3\alpha+\beta>0$ is clear that $12\alpha+\beta>0$. So, the only other needed bound to avoid poles for $\alpha_e$ is $2\alpha+\beta<0$. It is interesting to note that the bound $12\alpha+\beta>0$ does not forbid the condition $3\alpha+\beta<0$, so there are tachyonic situations in which poles are avoided anyway if $2\alpha+\beta<0$. On the other hand, there are regimes without tachyons, for instance $2\alpha+\beta>0$ and $\alpha>0$, which allow poles. So, it seems that there is not a perfect correlation between absence of poles and absence of tachyons.

A natural question is how the presence of tachyons may affect the eikonal approximation. First, a flaw may be our assumptions about the Schwarzschild radius, which in presence of tachyons may be debatable. In addition, in presence of tachyons, there may be time advance instead of time delay for a perturbation in Minkowski space.  These points will be discussed in the next sections. 

 The situation without tachyons and poles is physically attractive. However, for completeness, consider the possibility that $D(-\Delta^2)$ reaches a zero at some finite value of $-\Delta^2$. The concept of $\Delta^2$ being large depends on the value of $\alpha$ and $\beta$. Therefore, if one of this scales is related to some large scale such as $M_{GUT}$, still smaller than the Planck scale $M_{pl}$, there may be no need to employ transplanckian $s$  values for the  condition \eqref{a} to be satisfied. However, the energy range for which the determinant takes those small values is very narrow, as the energies are localized in a small region near the pole. 
 
For the reasons described in the previous paragraph, in a generic situation, consider the possibility that $D(-\Delta^2)$ is not close to the zero value. As this quantity is assumed  to be a polynomial expression, it is seen that $a(-\Delta^2)$ and $D(-\Delta^2)$ grows for $\Delta$ taking large values, which forces $\kappa^2 s$ to grow to adjust the condition $\alpha_e>>1$.  This is because the above expression has a denominator that grows faster than the numerators. The values of $\sqrt{s}$ then are expected to be transplanckian.

Furthermore, the fact that the effective gravitational coupling constant $\alpha_e$ is large, implies that perturbative techniques may not be adequate.  Instead of employing the standard Feynman diagrams calculation, the regimes of interest in this case are obtained  by varying the impact parameter $b$, as a large separation may soften the effect of large $\alpha_e$. Heuristically the force between the particles goes as  $F\sim \alpha_e/b^2$ and the torque is $F b$. This represents the variation of the angular momentum and if $Fb/p$ is small, then the deviation angle $\Theta\sim Fb/p\sim R_s/b$ should also be small, which restricts the impact parameter to values $b>>R_c$.  If the deviation angle is small, then obviously $\Delta^2<<s$ since 
$$
\tan\frac{\Theta}{2}\sim \frac{\Theta}{2}=\frac{\Delta}{\sqrt{s}}.
$$
 For a given impact parameter $b$  if the two scalar particles $\sigma$ and $\phi$ interact each other by interchanging one graviton, the transferred momentum is $q\sim b^{-1}$. As $b$ is large, there is no reason to restrict the attention to a single exchange process since the interchange of a macroscopically large number $N$ of gravitons gives a kick  $Nb^{-1}\sim \Delta$ whose square still can be smaller than $s$ if $b$ is large enough. 
 
 The calculation given below will be based on semi-classical or saddle point methods, which implies that the eikonal phase $\chi_0(b^{\perp})$ is  large in comparison with the Planck constant $\hbar$ which, in the unities employed here, has the value $\hbar=1$. These assumptions have to be checked after the calculation of the  deviation angle and time delay of the process.

\subsection{The explicit calculation of the deviation angle and time delay}

In previous sections, the eikonal phase \eqref{fases} have been found in terms of the characteristic functions $a$ and $d$ of the problem. By use of this formula, the deviation angle and the time delay for the scattering process in consideration will be estimated, by assuming that the eikonal limit takes place. The assumptions for this to hold will be checked later on.  Consider now the (not generic) situation in which there are no real roots for the Fourier transform of the function $a(\square)$. By factorization into its roots $m_i$, assuming that the zeros of $a(\square)$ are simple, it follows that
\begin{eqnarray} \label{potencial}
a(\square) = \frac{(M_{1}^2 - \square) (M_{2}^2 - \square) \cdots (M_{n}^2 - \square)}{M_{1}^2 M_{2}^2\cdots M_{n}^2},
\end{eqnarray}
which is translated in momentum space into
\begin{eqnarray} \label{potencialo}
a(-k^{\perp 2}) = \frac{(M_{1}^2 +k^{\perp 2}) (M_{2}^2 +k^{\perp 2}) \cdots (M_{n}^2 +k^{\perp 2})}{M_{1}^2 M_{2}^2\cdots M_{n}^2},
\end{eqnarray}
The same type of considerations hold for $d(\square)$ or  $d(-k^{\perp 2})$. On the other hand, the phase $\chi_0$ in \eqref{fases} is related to the quantity
$$
\bigg[\frac{4}{3a(- k_m^{\perp 2})}+\frac{2a(- k_m^{\perp 2})+d( -k_m^{\perp 2})-1}{6D( -k_m^{\perp 2})}\bigg]
$$
This quantity is related to a division of polynomials, the denominator has higher degree than the numerator.
At first sight, partial fractions are available for this type of situations.  As discussed above, the function $a(- k_m^{\perp 2})$ or $D(- k_m^{\perp 2})$ may vanish at some impulse values. In fact, it was argued in the previous sections that for the Stelle gravity there are choices for $\alpha$ and $\beta$ coupling constants for which the determinant $D(- k_m^{\perp 2})$ has effectively a zero. 

Starting first the situation in which none of these quantities has a zero, the integral to be calculated will be expressed as
\begin{equation}
\chi_0 ({\vec{b}}^\perp)= \frac{\kappa^2 M_\sigma E_\phi m_{1}^2 m_{2}^2\cdots m_{n}^2}{16\pi^2}\int\frac{d^2 k^{\perp}}{(m_{1}^2 + k^{\perp 2}) (m_{2}^2 + k^{\perp 2}) \cdots (m_{n}^2+ k^{\perp 2})k^{\perp 2}}e^{-i\vec{b}^{\perp}\cdot k^{\perp}}.
\label{fase}
\end{equation}
where the over arrow was deleted for simplicity of notation.  This not the most general choice,  as there can be a lower order polynomial in the numerator. However, both cases can be tackled by partial fractions and we chose this case for simplicity, as it is enough for illustrative purposes.

By writing $d^2k^\perp=k dk d\theta$ the solution of this integral is  simple exercise of partial fractions. First, one should employ  the integral formulas
$$
J_n(x)=\frac{1}{2\pi}\int_{-\pi}^{\pi}e^{in\theta-i x \sin(\theta)}d\theta,
$$
\begin{equation}\label{u}
\int_0^\infty \frac{J_0(bx)x dx}{(x^2+a^2)}= K_{0}(ab),\qquad a>0, \qquad b>0,
\end{equation}
the second is a particular case of formula 6.565.5 of \cite{gradshtein}. After partial fractions, the integral of $\chi_0(\overline{b}^{\perp })$ reduces to a sum of integrals  of the form
$$
I=\int_0^{\infty}\int_0^{2\pi}  \frac{e^{-ibk\cos\theta}kdkd\theta }{(k^2+a^2)}.
$$
By first integrating the phase \eqref{fases} with respect to $\theta$ and then with respect to $k$ with the above given integral formulas, it is arrived to
\begin{equation}
\chi_0 ({\vec{b}}^\perp)=\chi^{gr}_0(\vec{b})+ \frac{ \kappa^2 M_\sigma E_\phi}{16\pi^2}\bigg[2\pi\sum_{j=1}^{n} K_0(b  m_j)\prod_{k\neq j} \frac{m_{k}^2}{m_{k}^2 - m_{j}^2}\bigg].
\label{faser}
\end{equation}
The validity of the eikonal limit requires that the phase is very large $1<<\chi_0(b^{\perp})$, in this case the following semi classical approximations will be justified since the amplitude oscillates very violently.  The GR contribution is given by 
$$
\chi^{gr}_0(\vec{b})\sim -\frac{ \kappa^2 M_\sigma E_\phi}{8\pi}\log(\mu b)
$$
where $\mu$ is a renormalization scale, which can be interpreted as the regularized mass of the graviton. The formula given above is valid if $\mu b<<1$, which means that as this mass is taken to zero, larger volumes in the universe are allowed to be inspected. Under the assumption that  $M_\sigma>>E_\phi>>\Delta$  it was  found in previous subsections that  $s\sim 2 M_\sigma E_\phi$. The modified Bessel  function 
$$
 K_0(b  m_j)=-I_0(b m_i)\log(b m_i) +\sum_{k=0}^{\infty}\frac{1}{k!^2} \bigg(\frac{b m_i}{2}\bigg)^{2k}\psi(k+1),\qquad \psi(x)=\frac{d\Gamma(x)}{dx},
$$
has been introduced in the previous formula, with $\Gamma(x)$ the gamma function. This Bessel function decays exponentially as $K_0(x)\sim e^{-x}/\sqrt{x}$ for $x>>1$ while it goes as $K_0(x)\sim -\log(x)$ for small $x$.

Consider now the situation where the denominators have at least one root. An illustrative situation may be
\begin{equation}
\chi_0 ({\vec{b}}^\perp)=\frac{\kappa^2 M_\sigma E_\phi m_{1}^2 m_{2}^2\cdots m_{n}^2}{16\pi^2}\int\frac{d^2 k^{\perp}}{(m_{1}^2 - k^{\perp 2}) (m_{2}^2 - k^{\perp 2}) \cdots (m_{n}^2- k^{\perp 2})k^{\perp 2}}e^{-i\vec{b}^{\perp}\cdot k^{\perp}}.
\label{faso}
\end{equation}
In this case, the partial fraction procedure reduces the integrals analogous to \eqref{u}, except that the integral of the Bessel function is of the form
$$
I=\int_0^\infty \frac{J_0(bx)x dx}{(x^2-a^2)}.
$$
In other words, this integral is analogous to the one in \eqref{u} but with the replacement $a^2\to -a^2$. This integral can be calculated with the help of formula 6.562 of \cite{gradshtein}, namely
\begin{equation}\label{gaynor}
\int_0^\infty \frac{J_0(bx) dx}{x+a}=\frac{\pi}{2}\bigg[H_0(ba)-Y_0(ba)\bigg].\qquad b>0,\qquad -\pi<\text{arg}(a)<\pi.
\end{equation}
Here $H_\nu(x)$ is the Struve function
$$
H_\nu(x)=\sum_{m=0}^\infty \frac{x^{2m+\nu+1}}{2^{2m+\nu+1}\Gamma(m+\frac{3}{2})\Gamma(m+\frac{3}{2}+\nu)}.
$$
while the modified Bessel function $Y_n(x)$ is defined through the ordinary one $J_n(x)$ by the formula
$$
+\sum_{k=0}^{\infty}\frac{(-1)^k}{k!(k+n)!} \bigg(\frac{x}{2}\bigg)^{2k+n}[\psi(k+1)+\psi(k+n+1)],\qquad \psi(x)=\frac{d\Gamma(x)}{dx}.
$$
By use of partial fractions 
$$
I=\int_0^\infty \frac{J_0(bx)x dx}{(x^2-a^2)}=\frac{1}{2}\int_0^\infty \frac{J_0(bx) dx}{(x+a)}+\frac{1}{2}\text{P.V}\int_0^\infty \frac{J_0(bx) dx}{(x-a)}
$$
$$
=\frac{1}{2}\int_0^\infty \frac{J_0(bx) dx}{(x+a)}+\frac{1}{2}\lim_{\epsilon\to 0^+}\int_0^\infty \frac{J_0(bx) dx}{(x-a+i\epsilon)}+i\pi \lim_{\epsilon\to 0^+}J_0(b(-a+i\epsilon))
$$
The reason for including a positive imaginary part  $i\epsilon$ for $a$ is that formula \eqref{gaynor} is not valid for $a$ real and negative. However, note that $J_0(x)$ can be defined to negative values of $x$, and the same holds for the Struve function $H_0(x)$, as they contain even powers of $x$. Instead, the function $Y_0(x)$ has a logarithm not defined on the negative axis. The last formula then results into
$$
I=\frac{\pi}{4}\bigg[H_0(ba)-Y_0(ba)\bigg]+\frac{\pi}{4}\lim_{\epsilon\to 0^+}\bigg[H_0(ba)-Y_0(b(-a+i\epsilon))\bigg]+\frac{i\pi}{2} J_0(ba).
$$
Since $\pi Y_0(x)=2J_0(x)\log(x)$ plus powers of $x$, it is clear that the term $i\pi J_0(ba)$ cancels the imaginary part $i\pi$ of the logarithm when approaching the negative axis.  The integral is then 
$$
I=\frac{\pi}{2}\bigg[H_0(ba)-Y_0(ba)\bigg],
$$
where the logarithm in $Y_0(ba)$ is evaluated at the positive real value $ba$. By use of this formula. a procedure analogous to the above leads to
\begin{equation}
\chi_0 ({\vec{b}}^\perp)=\chi^{gr}_0(\vec{b})+ \frac{ \kappa^2 M_\sigma E_\phi}{32\pi^2}\bigg[\pi\sum_{j=1}^{n} [H_0(b  m_j)-Y_0(b  m_j)]\prod_{k\neq j} \frac{m_{k}^2}{m_{k}^2 - m_{j}^2}\bigg].
\label{fasero}
\end{equation}
Here, for small values of the variable $x$  the function $Y_0(x)$ diverges logarithmically, as $K_0(x)$. However, for large $x$
values $Y_0(x)\sim cos(x+\alpha)/\sqrt{x}$.  In fact, the formula 8-554 of \cite{gradshtein} shows that $H_0(x)-Y_0(x)\sim 1/\sqrt{x}$. This decay is much slower than for $K_0(x)\sim e^{-x}/\sqrt{x}$. This suggests that the eikonal modifications are more pronounced when poles appear in the integrand of the phase, for instance, for the Stelle gravity. 

In order to take care the small graviton mass $\mu$, which produces a logarithmic divergence, it is  customary to introduce the inverse Fourier transform
 $$
iA(t, s)=2is\int d^2b^{\perp} e^{-i\vec{b}^{\perp}\cdot \Delta+i\chi_0},
 $$
 where  $t=\Delta^2$ is the other independent Mandelstam variable in addition to $s$. If the value of the phase is large, then the semi-classical limit is justified, and the saddle point approximation 
$$
\Delta^\mu=\frac{\partial \chi_0}{\partial b^\mu},
$$
leads to
$$
\Delta^\mu=\frac{\kappa^2 s}{16\pi}\bigg[-1+ \sum_{j=1}^{n} K^{\prime}_0(b m_j)\prod_{k\neq j} \frac{bm_j m_{k}^2}{m_{k}^2 - m_{j}^2}\bigg]\frac{b^\mu}{b^2}. 
$$
when no poles are present or to 
$$
\Delta^\mu=\frac{\kappa^2 s}{16\pi}\bigg[-1+ \frac{1}{2}\sum_{j=1}^{n} [H^{\prime}_0(b m_j)-Y_o^{\prime}(bm_i)]\prod_{k\neq j} \frac{bm_j m_{k}^2}{m_{k}^2 - m_{j}^2}\bigg]\frac{b^\mu}{b^2}. 
$$
when poles appear. 

The angle of deflection is obtained as
$$
\tan(\frac{\Theta}{2})=\frac{\Delta}{\sqrt{s}},
$$
and since this angle is small, it follows that
 \begin{equation}\label{fasendo}
\Theta_{np}\sim\frac{\kappa^2\sqrt{s}}{8\pi b}\bigg[1- \sum_{j=1}^{n} K^{\prime}_0(b m_j)\prod_{k\neq j} \frac{bm_j m_{k}^2}{m_{k}^2 - m_{j}^2}\bigg].
\end{equation}
if no poles are present, or
\begin{equation}\label{fasendos}
\Theta_p\sim\frac{\kappa^2\sqrt{s}}{8\pi b}\bigg[1- \frac{1}{2}\sum_{j=1}^{n} [H^{\prime}_0(b m_j)-Y_o^{\prime}(bm_j)]\prod_{k\neq j} \frac{bm_j m_{k}^2}{m_{k}^2 - m_{j}^2}\bigg].
\end{equation}
when they are. It should be noticed that 
$$
bm_j K^{\prime}_0(b m_j)\sim \sqrt{bm_j}e^{-bm_j}\bigg[1-\frac{1}{2bm_j}\bigg],\qquad bm_j[H^{\prime}_0(b m_j)-Y_o^{\prime}(bm_j)]\sim \frac{1}{\sqrt{bm_j}}.
$$
which shows that the corrections for large $b$ in the second case are more pronounced than the first.

The next task is to characterize the time delay corresponding to the process. Following \cite{wigner1}-\cite{wigner2}, the following estimation 
$$
\Delta T=2\frac{\delta\chi_0}{\delta E},\qquad E=\sqrt{s},
$$
may be applied. The explicit expression is 
\begin{equation}
\Delta T=\Delta T^{GR}- \frac{ \kappa^2\sqrt{E_\phi M_\sigma}}{16\pi^2}\bigg[2\pi\sum_{j=1}^{n} K_0(b  m_j)\prod_{k\neq j} \frac{m_{k}^2}{m_{k}^2 - m_{j}^2}\bigg],
\label{faser2}
\end{equation}
where $\Delta T^{GR}$ corresponds to the GR term. This expression is valid if no poles are present, otherwise
\begin{equation}
\Delta T=\Delta T^{GR}- \frac{ \kappa^2\sqrt{E_\phi M_\sigma}}{16\pi^2}\bigg[\pi\sum_{j=1}^{n} [H_0(b  m_j)-Y_0(b  m_j)]\prod_{k\neq j} \frac{m_{k}^2}{m_{k}^2 - m_{j}^2}\bigg].
\label{faser22}
\end{equation}
is the formula to be employed.

\section{About the range of applicability of the eikonal approximation}

After deriving the formulas \eqref{fasendo}-\eqref{faser22}, the last point to be discussed is their range of applicability.  An instructive example is the Stelle gravity, which was found in  \eqref{tat} to be described by the following factors
$$
a(-k^2)=1+\frac{\beta}{2} k^2,\qquad d(-k^2)=-\frac{1}{2}+2\alpha k^2
$$
$$
D(-k^2)=\frac{1}{8}\bigg[-1+(12\alpha+\beta) k^2\bigg]\bigg[\frac{1}{2}+(2\alpha+\beta) k^2\bigg]-\frac{3}{16}.
$$
The integrand in the definition of the phase \eqref{fases} is related to the quantity
\begin{equation}\label{fracciones}
\bigg[\frac{4}{3a(- \Delta^{2})}+\frac{2a(-\Delta^{2})+d( -\Delta^{2})-1}{6D ( -\Delta^{2})}\bigg]=\frac{4m_1^2}{3(m_1^2+\Delta^2)}-\frac{M_2^2}{m_2^2\pm \Delta^2}+\frac{M_3^2}{\Delta^2+m_3^2}.
\end{equation}
This decomposition is a simple exercise of partial fractions,  and the  new mass scales introduced above are found in the appendix and are functions of $\alpha$ and $\beta$.  The $\pm$ sign is to indicate the possible presence of poles. As the expressions in brackets have to tend to $1$ when $\Delta\to 0$ in order to recover GR, it follows that
\begin{equation}\label{ceros}
\frac{M_2^2}{m_2^2}-\frac{M_3^2}{m_3^2}=\frac{1}{3}.
\end{equation}
In the opposite case, namely, when the value of $\Delta$ is much larger than the largest mass scale present $m_i$ this quantity is suppressed. Note that the five the mass scales depend only on two namely $1/\sqrt{\alpha}$ and $1/\sqrt{\beta}$, therefore they are not completely independent.

The requirement that $h_{\mu\nu}$ is a small perturbation for the Minkowski space, as stated in the previous section, is translated into $R_c>\lambda_c$. As shown  in \eqref{a}, this implies that the gravitational strength $\alpha_e$ is large and therefore
$$
2<<\kappa^2 s\bigg[\frac{4}{3a(- \Delta^{2})}+\frac{2a(-\Delta^{2})+d( -\Delta^{2})-1}{6D ( -\Delta^{2})}\bigg].
$$ This condition, specified for the Stelle gravity, becomes
\begin{equation}\label{comptono}
2<<\kappa^2 s\bigg[\frac{m_1^2}{m_1^2+\Delta^2}-\frac{M_2^2}{m_2^2\pm \Delta^2}+\frac{M_3^2}{\Delta^2+m_3^2}\bigg].
\end{equation}
On the other hand, the assumption that the deviation angle is small, leads to a large impact parameter $b>>R_c$. The quantity $q=b^{-1}$
represents the kick between the two particles when a single graviton is interchanged.   For a macroscopically large number $N$ of gravitons, the condition  $Nb^{-1}\sim \Delta$ shows that $\Delta b=\Delta q^{-1}>>1$.  In other words,  the transferred impulse $\Delta$ should be considerably larger than $q$. 

A first possibility is that $bm_i>>1$ for all the mass scales of the problem. However, consider the situation in which one of the denominator masses $m_i$ of the problem is such that  $bm_i<<1$.  In this case, the corrections in \eqref{fasendo}-\eqref{faser2} are logarithmic due to the behavior of $K_0(z)$ at small angles, and may be noticeable.  Suppose that this small mass is $m_1$. Then the previous paragraph  implies that $m_1<<q=b^{-1}<<\Delta$, which forces $m_1$ to be the slightest mass scale in the problem (up to the regularized mass $\mu$ of the graviton).   If the remaining masses $m_j$ are larger than $\Delta$ it follows from \eqref{comptono}  and \eqref{ceros}  that 
  \begin{equation}\label{uno}
2<<\kappa^2 s\bigg|\frac{4m_1^2}{3\Delta^2}-\frac{1}{3}\bigg|.
\end{equation}
It is interesting that the sign of the effective constant $\alpha_e$ becomes negative, that is, attraction becomes repulsion, as the quotient $m_1^2/\Delta^2$ is small by hypothesis, and can be neglected. For the moment, this fact will be overlooked and be reconsidered latter on. This change of sign is the reason for which the modulus has been introduced in the last expression.

The last formula alone implies that $\kappa^2 s>>1$, in other words, $\sqrt{s}$ takes transplanckian values if the eikonal approximation takes place.  In addition, the assumption of a large impact parameter $b>R_s\sim |\alpha_e|/\sqrt{s}$ becomes
 \begin{equation}\label{dos}
\frac{\kappa^2 s}{3}<<b\sqrt{s}.
\end{equation}
The estimation  of the deviation angle, which is  similar to the one leading to \eqref{fasendo}, shows that for the Stelle gravity when poles are present
\begin{equation}
\Theta\sim\frac{\Delta}{\sqrt{s}}=\frac{\kappa^2\sqrt{s}}{8\pi b}\bigg[\frac{4}{3}-\frac{4}{3}K^{\prime}_0(b m_1) bm_1- \frac{M_2^2}{2m_2^2}[H^{\prime}_0(b m_2)-Y_o^{\prime}(bm_2)]bm_2+\frac{M_3^2}{m_3^2}K^{\prime}_0(b m_3)bm_3\bigg]
\end{equation}
 \begin{equation}\label{tres}
\sim\frac{\kappa^2\sqrt{s}}{3\pi b}.
\end{equation}
In the last expression, the fact that $K_0(x)\sim -J_0(x)\log(x)$  for small $x$ was employed, together with the assumption $bm_1<<1$. The resulting angle is $8/3$ of the GR one, at least in functional form. This may indicate that these corrections could  be  detected at the leading eikonal order. However, this should not be taken in literal form, since in a new model the conditions for the validity of the eikonal approximation should be reviewed. 

At this point, it is  important to study the quotients $M_i^2/m_i^2$. The masses of these quotients are not independent, in fact $M_i=M_i(\alpha,\beta)$  and $m_i=m_i(\alpha,\beta)$. There are two parameters $\alpha$ and $\beta$ which parameterize five masses. This may signal that they values are not completely uncorrelated. The decomposition \eqref{fracciones} can be worked out as in the appendix, the result is
$$
\frac{M_2^2}{m_2^2}=\frac{1-(4\alpha+\beta) \widetilde{x}_1}{12 c\widetilde{x}_1(\widetilde{x}_2-\widetilde{x}_1)},\qquad \frac{M_3^2}{m_3^2}=\frac{M_2^2}{m_2^2}-\frac{1}{3},
$$
$$
m^2_1=\frac{1}{\beta},\qquad m_2^2=\frac{1}{\widetilde{x}_1},\qquad m_3^2=\frac{1}{\widetilde{x}_2},
$$
with $c =2 (12\alpha + \beta)(2\alpha + \beta)$, and
$$
\tilde{x}_{1,2} = \frac{x_1 + x_2}{2} \pm \sqrt{\frac{3}{16c} + \left(\frac{x_2 - x_1}{2}\right)^2}, \quad x_1 = \frac{-1}{(12\alpha + \beta)}, \quad x_2 = \frac{1}{2(2\alpha + \beta)}.
$$
If either of the  $\widetilde{x}_i$ tend to infinite, the values of the quotients $\frac{M_i^2}{m_i^2}$ remain finite. If some of the $\widetilde{x}_i$ vanishes, these quotients diverge. However, this is equivalent of taking divergent values of $\alpha$ or $\beta$, that is, enormous corrections to GR. The eikonal description in this case may be non adequate. For moderate or even large values of $\widetilde{x}_i$ the quotient is also moderate, this can be seen by writing everything in terms of $\alpha$ and $\beta$. The expression is a bit cumbersome, but it is seen that it is the quotient of two functions which go roughly as the inverse of a linear combination of $\alpha$ and $\beta$. The expected values of the quotient are then moderate. 

Now, going back to the situation in which $m_1^2$ is small, as $m_1\sim \beta^{-1}$ implies that $\beta$ is quite large. If $\alpha $ is small, then the quantities $\widetilde{x}_i$ above are large and $m_i$ and consequently $M_i$, are all small. If $\alpha$ is large, the same conclusion follows by inspecting the expressions of $x_i$ and $\widetilde{x}_i$. Therefore, the fact that $m_1$ is small implies that all the other masses of the problem are also small. 

The above facts force to reconsider the deduction made in \eqref{uno}. This formula suggested a change of sign of $\alpha_e$. However, this perhaps disturbing conclusion was due to the fact that the other masses of the problem are small as well, and this was not taken into account.  The corrected equation will be
\begin{equation}\label{unocorr}
2<<\kappa^2 s\bigg[\frac{4m_1^2}{3\Delta^2}-\frac{M_2^2+M_3^2}{\Delta^2}\bigg].
\end{equation}
 Then a large $\alpha_e$ value is obtained if $\kappa^2 s$ is highly transplanckian, as $\Delta^2$ is larger than $m_1^2$ and consequently, considerably larger than  the other masses appearing in the last expression (since they are correlated and arguably of the same order). The large impact parameter $b>R_s\sim \alpha/\sqrt{s}$ request, in these terms, becomes
$$
\frac{4\kappa^2 s}{3\Delta^2}(4m_1^2-3M_2^2+3M_3^2)<<b\sqrt{s}.
$$
In addition, in analogous fashion as before
$$
m_i\leq q=\frac{1}{b}<<\Delta<<\sqrt{s}.
$$
The small angle condition \eqref{tres}  can be worked out again by taking the small argument expansion 
$$
zY'_0(z)\sim \frac{2}{\pi }(1-\frac{z^2}{4})+\frac{z^2}{\pi}-\frac{2}{\pi}(\log (z)+\gamma)\frac{z^2}{2},
$$
$$
zH'_0(z)\sim\frac{z}{2\Gamma(\frac{3}{2})\Gamma(\frac{3}{2})}+\frac{3z^3}{8\Gamma(\frac{5}{2})\Gamma(\frac{5}{2})},\qquad z<<1,
$$
into account, leading to
$$
\Theta\sim\frac{\Delta}{\sqrt{s}}= \frac{\kappa^2 s}{8\pi }(2-\frac{M_3^2}{m_3^2}+\frac{2M_2^2}{\pi m^2_2}).
$$
This results again, deviates from GR. Still, a final test that all the conditions derived can be satisfied may be desirable. All the conditions obtained above for the validity  of the eikonal approximation  are
$$
m_i\leq q=\frac{1}{b}<<\Delta<<\sqrt{s},\qquad \kappa^2 \sqrt{s}\sim \frac{4\pi \Delta}{q\sqrt{s}}<<\frac{4\pi}{q},\qquad 
$$
$$
2q<<\frac{\kappa^2 s}{\Delta^2}(4m_1^2-3M_2^2+3M_3^2)q<<\sqrt{s}.
$$
The last two are simply the smallness of the angle \eqref{tres} and the conditions \eqref{uno}-\eqref{dos}.  The last two conditions are hard to satisfy. The problem is that $m_i\sim \Delta  N_i$, which  roughly implies that  $\Delta^2\sim \overline{N}^2 [4m_1^2-3M_2^2+3M_3^2]$ with $\overline{N}$ an average number of interchanged gravitons, which has to be macroscopically large for the eikonal approximation to apply. This and the third condition imply that  $\kappa^2 s\geq\overline{N}^2$. On the other hand, the second condition implies
$$
2\kappa  s\sim 4\pi\overline{N}.
$$
This forces $\overline{N}^2\leq 4\pi \overline{N}$, that is $\overline{N}\leq 12$. Thus, the number of interchanged gravitons seems not so large. This compromises the eikonal approximation. 

In the remaining case, in which all the masses of the problem are equal or  larger than the Planck mass and the impact parameter $b$ is large, the GR corrections are of the order $1/\sqrt{bm_2}$  if there are poles in the effective constant, due to the terms proportional to $[H^{\prime}_0(b m_2)-Y_o^{\prime}(bm_2)]bm_2$ in \eqref{uno}. Note that the corrections related to $m_1$ and $m_3$ turn out to be exponentially suppressed due to the asymptotic behavior of the functions $K^{\prime}_0(z)z$ in \eqref{uno}. These corrections of the form $1/\sqrt{bm_2}$ are expected to be small. However that for a  value  $bm_2\sim 10^{2}$ the result may  still be noticeable, although present as a subleading contribution.

The above analysis is suggesting that the introduction of a scale $m_i$ considerably smaller than the Planck mass forces the other mass scales to the same behavior. The resulting angle is different from GR. However, the eikonal approximation is likely to break down.  For the other case namely, where the additional masses are above the Planck scale, the leading eikonal angle is the same of GR.  In this case, if no poles are present in the effective constant $\alpha_e$ then the corrections are exponentially suppressed and may not even be resolvable below the quantum uncertainty limit.  The presence of poles may give noticeable corrections of the form $1/\sqrt{bm_2}$, however, the leading order is still GR.

We believe that the above results are in harmony with the findings of \cite{brandhuber}, \cite{tseytlin} and \cite{deser}. Those references conclude that, for Stelle gravity, the eikonal approximation for the scattering of two scalars is equivalent to the one in GR.  This is justified by a metric transformation that converts the model into GR plus two scalar fields with non standard couplings. The resulting interaction vertices apparently do not affect the leading eikonal order. These findings seem to be confirmed by sophisticated methods in \cite{javier}.  


As a final comment, we would like to add that in presence of tachyons, the mass factors $m_i^2$ change the sign and, depending on the other parameters, there may appear a time advance effect due to \eqref{faser22} which, as a perturbations on Minkowski space, may suggest a conflict with causality.  The validity of the eikonal approximation in this case is to be taken with care and requires further analysis.  If there are no tachyons, the results are those of \cite{salvio}-\cite{salvio2}.

\section{Higher order  estimations}
The results presented above can be generalized to higher orders, by use of the methods of \cite{fazio}. The purpose of this section is to work out these contributions partially.  These orders correspond to the quadratic (or even higher) terms of the expansion of the exponential in \eqref{funcionar}.  In the terminology of \cite{nexteikonal}-\cite{nexteikonal2} this corresponds to corrections of the leading order of the scalar propagator, and results into further terms suppressed by powers of $E_\phi/\Delta$.  This contribution vanishes in GR, but not in the models presented here. In the terminology of those references, this corresponds to the  "next to eikonal" approximation. 

In addition, there are corrections corresponding  to graviton trees with loops diagrams, such as the ones of figures 4 and 5 of \cite{nexteikonal2}.  One of these diagrams involves a simple triangle with two gravitons and one scalar internal lines, this is known as the seagull graph. The other involves the same triangle, but a further three graviton interaction vertex. The second contribution is harder to estimate, since the three graviton interaction depends on the lagrangian under consideration. For GR the contribution is considered in those references. For generic lagrangians such as \eqref{tat} the vertex interaction will involve the contributions of higher curvature terms such as $R^2$ or $R_{\mu\nu}R^{\mu\nu}$ and the effect of the functions $F_i(\square)$ on them. The resulting contributions for such vertices are not easy to be estimated and will be considered in a separate publication. Only the seagull approximation will be considered here, at least partially, since it do not involve graviton-graviton interaction and is easier to deal with.

The fact that contributions corresponding to the quadratic terms of the expansion of the exponential in \eqref{funcionar} are zero in GR, have been shown in \cite{nexteikonal}-\cite{nexteikonal2} by standard  Feynman diagram expansions and in \cite{fazio} by use of Schwinger-Fradkin techniques. For generalizations of GR, this is not the case, and we turn the attention to this point.

\subsection{Next to eikonal approximation}

The next eikonal approximation is obtained by taking the next orders in the Fradkin kernel \eqref{funcionar}. The next to leading order is in this case
$$
iT^1(p,p’,q,q’)_{NE}=\lim\limits _{\substack{p^{2}\rightarrow0\\
p^{\prime2}\rightarrow0
}
}\lim\limits_{\substack {q^2\rightarrow M_\sigma^2\\ q^{\prime2}\rightarrow M_\sigma^2}}\sum_{n=1}^{+\infty}\frac{(-1)^n\kappa^{2n}}{n!}$$
$$
\int_0^{+\infty}d\nu e^{i\nu(p’^2+i\epsilon)}\int_0^{+\infty}d\nu_1 e^{i\nu_1[(q’^2-M_\sigma^2)+i\epsilon]}\prod_{m=1}^{n}\int\frac{d^4 k_m}{(2\pi)^4} iM_\sigma^2 p^{\prime\mu_{m}}p^{\prime\beta_{m}}D_{\mu_m \beta_m,00}
$$
$$
\int_0^{\nu}d\xi_1\dots \int_0^{\nu}d\xi_n \int_0^{\nu_1}d\tilde{\xi}_1\dots \int_0^{\nu_1}d\tilde{\xi}_n (2\pi)^4\delta^4(q-q’-k_1-\dots k_n)
$$
$$
(-i)\exp\left[-2i\sum_{m=1}^n p’\cdot k_m(\nu-\xi_m)\right]\exp\left[-2i\sum_{\tilde{m}=1}^n q’\cdot k_{\tilde{m}}(\nu_1-\tilde{\xi}_{\tilde{m}})\right]$$
$$
\bigg[\sum_{m, m_1=1}^n k_m\cdot k_{m_1}\left(\nu-\frac{\xi_m+\xi_{m_1}}{2}-\frac{|\xi_m-\xi_{m_1}|}{2} \right)+
$$
\begin{equation}\sum_{m_2, m_3=1}^n k_{m_2}\cdot k_{m_3}\left(\nu_1-\frac{\tilde{\xi}_{m_2}+\tilde{\xi}_{m_3}}{2}-\frac{|\tilde{\xi}_{m_2}-\tilde{\xi}_{m_3}|}{2} \right)\bigg].
\label{ordensigo}
\end{equation}
The main difference with the formulas of the previous section is that now the terms $ k_m\cdot k_{m_1}$ are not neglected. By following the formulas (57)-(67) of reference \cite{fazio}, it is obtained that the contribution  to the scattering matrix is given by three terms. The first, after integration in $\nu_1$ and by use of the eikonal identities \eqref{idento}-\eqref{idento2}, can be written as
$$
iT^1(p,p’,q,q’)_{NE}=\lim\limits _{\substack{p^{2}\rightarrow0\\
p^{\prime2}\rightarrow0
}
}p^{2}p^{\prime 2}\sum_{n=1}^{+\infty}\frac{i(-1)^n\kappa^{2n}}{n!}\bigg(\frac{-2\pi i}{2M_\sigma}\bigg)^{n-1}
$$
$$
\prod_{m=1}^{n}\int\frac{d^3 k_m}{(2\pi)^4} iM_\sigma^2 p^{\prime\mu_{m}}p^{\prime\beta_{m}}D_{\mu_m \beta_m,00}\int_0^{+\infty}d\nu e^{i\nu(p’^2+i\epsilon)}
$$
$$
\int_0^{\nu}d\xi_1\dots \int_0^{\nu}d\xi_n \int_0^{\nu_1}d\tilde{\xi}_1\dots
(-i)\exp\left[-\sum_{m=1}^n 2i\vec{p}’\cdot \vec{k}_m(\nu-\xi_m)\right]
$$
\begin{equation}
\sum_{r=1}^n \vec{k}_r\cdot \vec{k}_r (\nu-\xi_r)(2\pi)^4\delta^3(\vec{\Delta}+\vec{k}_1+\dots +\vec{k}_n). \label{grups}
\end{equation}
In addition, the identity \eqref{cardio} described in the introduction, together with \eqref{idento}, leads to the following relation
$$
\int_0^{+\infty}d\nu e^{i\nu(p’^2+i\epsilon)}\int_0^{\nu}d\xi_1\dots \int_0^{\nu}d\xi_n \int_0^{\nu_1}d\tilde{\xi}_1\dots
\exp\left[-\sum_{m=1}^n 2i\vec{p}’\cdot \vec{k}_m(\nu-\xi_m)\right]
$$
$$
=\sum_{r=1}^n   \frac{\vec{k}_r\cdot \vec{k}_r}{2i\vec{p^{\prime}}\cdot \vec{k}_r}\sum_\pi \frac{1}{p^{\prime 2}+i\epsilon}\frac{1}{p^{\prime 2}-2i\vec{p^{\prime}}\cdot \vec{k}_{\pi(1)}+i\epsilon}\cdot\cdot\frac{1}{p^{\prime 2}-2i\vec{p^{\prime}}\cdot (\vec{k}_{\pi(1)}+..+\vec{k}_{\pi(n)})+i\epsilon}
$$
$$
-\frac{\nu}{p^{\prime 2}-2i\vec{p^{\prime}}\cdot (\vec{k}_{1}+..+\vec{k}_{n})+i\epsilon}\sum_{r=1}^n  \vec{k}_r\cdot \vec{k}_r\prod_{s=1}^n \frac{1}{2i\vec{p^{\prime}}\cdot \vec{k}_s}
$$
With this identity at hand, and by employing \eqref{idento2} it can be deduced that
$$
\lim\limits _{\substack{p^{2}\rightarrow0\\
p^{\prime2}\rightarrow0
}
}p^{2}p^{\prime 2}(2\pi)^4\delta(\vec{k}_{1z}+\dots +\vec{k}_{nz})\int_0^{+\infty}d\nu e^{i\nu(p’^2+i\epsilon)}
$$
$$
\int_0^{\nu}d\xi_1\dots \int_0^{\nu}d\xi_n \int_0^{\nu_1}d\tilde{\xi}_1\dots
\exp\left[-\sum_{m=1}^n 2i\vec{p}’\cdot \vec{k}_m(\nu-\xi_m)\right]
$$
\begin{equation}\label{fly}
=\frac{(-2\pi i)^{n-1}}{(2E_\phi)^{n-1}}\delta(k_{1z})..\delta(k_{nz})\sum_{r=1}^n     \frac{\vec{k}_r\cdot \vec{k}_r}{2iE_\phi k_{rz}}.
\end{equation}
The last formula was deduced by taking into account that $\vec{k}_{\pi(1)}+..+\vec{k}_{\pi(n)}=\vec{k}_{1}+..+\vec{k}_{n}=\vec{\Delta}$ with $\Delta$ a very small quantity, which was neglected. This implies that
$$
\lim_{p^2\to 0}\frac{p^{2}}{p^{\prime 2}-2i\vec{p^{\prime}}\cdot (\vec{k}_{\pi(1)}+..+\vec{k}_{\pi(n)})+i\epsilon}\sim 1,
$$
since $p^\prime\sim p$ in the eikonal limit. Even though the contribution of \eqref{fly}  seems to be divergent due to the presence of $k_z$ in the denominator, the integral
$$
\int_{-\infty}^\infty F(k_z) \frac{\delta(k_z)}{k_z}=0,
$$
for any  continuous  function $F(k_z)$ at $k_z=0$, as $k_z$ is odd and $\delta(k_z)$ is even.   The quantity $iM_\sigma^2 p^{\prime\mu_{m}}p^{\prime\beta_{m}}D_{\mu_m \beta_m,00}$ , as it will be seen below, is continuous at $k_z=0$.  This means that the contribution of the apparently divergent quantity \eqref{fly} is in fact vanishing.

The second  contribution that follows form the formulas (57)-(67) of  \cite{fazio} is given by
$$
iT^2(p,p’,q,q’)_{NE}\equiv\lim_{p_i^2\rightarrow m_i^2} p^2p’^2(2\pi)^4\sum_{n=1}^{+\infty}\frac{(-1)^n\kappa^{2n}}{n!}\sum\limits_{\substack{m=1\\m\neq \tilde{m}, m_1}}^n $$
$$
\int\frac{d^{3} k_1}{(2\pi)^4}\dots \frac{d^{3} k_n}{(2\pi)^4}\prod_{m=1}^n (iM_\sigma^2p^{\prime\mu_{m}}p^{\prime\beta_{m}}D_{\mu_m \beta_m,00})\delta^{3} (\vec{k}_1+\dots+ \vec{k}_n +\vec{\Delta}) \vec{k}_{\tilde{m}}\cdot \vec{k}_{m_1}\frac{i^n(-2\pi i)^{n-1}}{(2M_\sigma)^{n-1}}
$$
$$
\int_0^{+\infty}d\nu \prod\limits_{\substack{m=1\\m\neq \tilde{m}, m_1}}^n \frac{1-\exp(-2i\nu\vec{p}’\cdot \vec{k}_m)}{-2i\vec{p}’\cdot\vec{k}_m} 
\frac{[1-\exp(-2i\vec{p}’\cdot(\vec{k}_{\tilde{m}}+\vec{k}_{m_1})\nu)]}{(-2i\vec{p}’\cdot \vec{k}_{\tilde{m}})(-2i\vec{p}’\cdot \vec{k}_{m_1})(-2i\vec{p}’\cdot(\vec{k}_{\tilde{m}}+\vec{k}_{m_1}))}.
$$
This can be simplified by noticing that the  eikonal identities \eqref{idento} and \eqref{idento2} imply that
$$
\lim\limits_{\substack {p^2\rightarrow 0\\ p’^2\rightarrow 0}} p^2p^{\prime 2}\delta(k^z_1+\dots + k^z_n)
\int_0^{+\infty}d\nu \prod\limits_{\substack{m=1\\m\neq \tilde{m}, m_1}}^n \frac{1-\exp(-2i\nu\vec{p}’\cdot \vec{k}_m)}{-2i\vec{p}’\cdot\vec{k}_m}
$$
$$
\frac{[1-\exp(-2i\vec{p}’\cdot(\vec{k}_{\tilde{m}}+\vec{k}_{m_1})\nu)]}{-2i\vec{p}’\cdot(\vec{k}_{\tilde{m}}+\vec{k}_{m_1})}=
\frac{(-2\pi i)^{n-2}}{(2E_\phi)^{n-2}}\delta(k_1^z)\dots\delta(k^z_{\tilde{m}}+ k^z_{m_1})\dots \delta(k_{n}^z).
$$
The last formula allows to write the sough matrix element as
$$
iT^2_{NE}=i\sum_{n=2}^\infty\left(\frac{i\kappa^2 E_\phi M_\sigma}{8\pi^2}\right)^n\frac{1}{(n-2)!}\frac{M_\sigma}{2\pi} \int d^{2} k^\perp_3\dots \int d^{2} k^\perp_n
$$
$$
\prod_{m=3}^n \frac{1}{ k^{\perp 2}_m}\bigg[\frac{4}{3a(- k_m^{\perp 2})}+\frac{2a( k_m^{\perp 2})+d( -k_m^{\perp 2})-1}{6D( -k_m^{\perp 2})}\bigg]
$$
$$
\int\frac{d^{3} k_{1}}{k_{1}^2}\frac{d^{3} k_{2}}{k_{2}^2}\bigg[\frac{4}{3a(- k_1^{\perp 2})}+\frac{2a( k_1^{\perp 2})+d( -k_1^{\perp 2})-1}{6D( -k_1^{\perp 2})}
$$
$$
+\bigg(\frac{2}{ 3a(-k_1^2)}-\frac{2a(-k_1^2)+d(-k_1^2)-1}{6D(-k_1^2)}+\frac{1}{4D(-k_1^2)}\bigg)\frac{\eta_{00}(\vec{k}_1\cdot \vec{p}^\prime)^2}{E_\phi^2k^2_1}\bigg]  
$$
$$
\bigg[\frac{4}{3a(- k_2^{\perp 2})}+\frac{2a( k_2^{\perp 2})+d( -k_2^{\perp 2})-1}{6D( -k_2^{\perp 2})}
$$
$$
+\bigg(\frac{2}{ 3a(-k_2^2)}-\frac{2a(-k_2^2)+d(-k_2^2)-1}{6D(-k_2^2)}+\frac{1}{4D(-k_2^2)}\bigg)\frac{\eta_{00}(\vec{k}_2\cdot \vec{p}^\prime)^2}{E_\phi^2k^2_2}\bigg]  
$$
$$
 \frac{\vec{k}_{1}\cdot \vec{k}_{2}}{k_{1}^z k_{2}^z}\delta(k_{1}^z+k_{2}^z)(2\pi)^2\delta^2(\vec {k}^\perp_1+ \dots+\vec {k}^\perp_n +\vec{\Delta}).
$$
The last contribution is given by
$$
iT^3(p,p’,q,q’)_{NE}\equiv\lim_{p_i^2\rightarrow m_i^2} p^2p’^2(2\pi)^4\sum_{n=1}^{+\infty}\frac{(-1)^n\kappa^{2n}}{n!}\sum\limits_{\substack{m=1\\m\neq \tilde{m}, m_1}}^n\int\frac{d^{3} k_1}{(2\pi)^4}\dots \frac{d^{3} k_n}{(2\pi)^4} $$
$$
\prod_{m=1}^n (iM_\sigma^2p^{\prime\mu_{m}}p^{\prime\beta_{m}}D_{\mu_m \beta_m,00})\delta^{3} (\vec{k}_1+\dots+ \vec{k}_n +\vec{\Delta}) \frac{i^n(-2\pi i)^{n-1}}{(2M_\sigma)^{n-1}}\frac{(2i\vec{p}’\cdot(\vec{k}_m+\vec{k}_{m_1}))(\vec{k}_{\tilde{m}}\cdot \vec{k}_{m_1})}{(2ip’\cdot k_m) (2ip’\cdot k_{m_1})}
$$
$$
\int_0^{+\infty}d\nu\frac{(1-e^{-2i\vec{p}’\cdot \vec{k}_m \nu})}{(-2i\vec{p}’\cdot \vec{k}_m) }\frac{(1-e^{-2i\vec{p}’\cdot \vec{k}_{m_1} \nu})}{(-2i\vec{p}’\cdot \vec{k}_{m_1})}\prod\limits_{\substack{m=1\\m\neq \tilde{m}, m_1}}^n \frac{(1-e^{-2i\nu\vec{p}’\cdot \vec{k}_m})}{(-2i\vec{p}’\cdot\vec{k}_m)} .
$$
The eikonal identities \eqref{idento} and \eqref{idento2} converts the resulting contribution into
$$
iT^3(p,p’,q,q’)_{NE}\equiv\lim_{p_i^2\rightarrow m_i^2} p^2p’^2(2\pi)^4\sum_{n=1}^{+\infty}\frac{(-1)^n\kappa^{2n}}{n!}\sum\limits_{\substack{m=1\\m\neq \tilde{m}, m_1}}^n \int\frac{d^{2} k_1}{(2\pi)^4}\dots \frac{d^{2} k_n}{(2\pi)^4}\frac{(-2\pi i)^{n-2}}{(2E_\phi)^{n-2}}$$
$$
\prod_{m=1}^n \frac{-2iM_\sigma^2E_{\phi}^2}{ k_m^{\perp 2}}\bigg[\frac{4}{3a(- k_m^{\perp 2})}+\frac{2a( k_m^{\perp 2})+d( -k_m^{\perp 2})-1}{6D( -k_m^{\perp 2})}\bigg]\delta^{2} (\vec{k}^\perp_1+\dots+ \vec{k}^\perp_n +\vec{\Delta}) \frac{i^n(-2\pi i)^{n-1}}{(2M_\sigma)^{n-1}}
$$
$$
\int dk_{mz} dk_{m_1 z}\frac{(2i\vec{p}’\cdot(\vec{k}_m+\vec{k}_{m_1}))(\vec{k}_{\tilde{m}}\cdot \vec{k}_{m_1})}{(-2iE_\phi k_{mz}) (-2iE_\phi k_{m_1z})}\delta(k_{mz})\delta(k_{m_1 z}).
$$
The contribution of the integrals in $k_{mz}$ and $k_{m_1 z}$ vanish due to an argument analogous to the one given below \eqref{fly}.

Briefly, the unique contribution at next to leading order is simply $T^2(p,p’,q,q’)_{NE}$. Below, this quantity will be denoted aw $T(p,p’,q,q’)_{NE}$ without losing generality.

By taking the Fourier transform to the impact parameter space  of $T(p,p’,q,q’)_{NE}$ it is found that 
$$
i\widetilde{T}(\vec{b}^\perp)_{NE}=\int \frac{d^{2}\vec{\Delta}}{(2\pi)^2} e^{i\vec{\Delta}\cdot\vec{b}^\perp} iT_{NE}
$$
$$
=2i\int \frac{d^{2}\vec{\Delta}}{(2\pi)^2} e^{i\vec{\Delta}\cdot\vec{b}^\perp} \sum_{n=2}^\infty\left(\frac{i\kappa^2 E_\phi M_\sigma}{8\pi^2}\right)^n\frac{2}{(n-2)!}\frac{M_\sigma}{2\pi} \int d^{2} k^\perp_3\dots \int d^{2} k^\perp_n$$
$$
\prod_{m=3}^n \frac{1}{ k^{\perp 2}_m}\bigg[\frac{4}{3a(- k_m^{\perp 2})}+\frac{2a( -k_m^{\perp 2})+d( -k_m^{\perp 2})-1}{6D( -k_m^{\perp  2})}\bigg]
$$
$$
\int\frac{d^{3} k_{1}}{k_{1}^2}\frac{d^{3} k_{2}}{k_{2}^2}\bigg[\frac{4}{3a(- k_1^{2})}+\frac{2a(- k_1^{ 2})+d( -k_1^{2})-1}{6D( -k_1^{ 2})}
$$
$$
+\bigg(\frac{2}{ a(-k_1^2)}-\frac{2a(-k_1^2)+d(-k_1^2)-1}{6D(-k_1^2)}+\frac{\sqrt{3}}{4D(-k_1^2)}\bigg)\frac{\eta_{00}(\vec{k}_1\cdot \vec{p}^\prime)^2}{E_\phi^2k^2_1}\bigg]  
$$
$$
\bigg[\frac{4}{3a(- k_2^{ 2})}+\frac{2a(- k_2^{2})+d( -k_2^{ 2})-1}{6D( -k_2^{ 2})}
$$
$$
+\bigg(\frac{2}{ a(-k_2^2)}-\frac{2a(-k_2^2)+d(-k_2^2)-1}{6D(-k_2^2)}+\frac{\sqrt{3}}{4D(-k_2^2)}\bigg)\frac{\eta_{00}(\vec{k}_2\cdot \vec{p}^\prime)^2}{E_\phi^2k^2_2}\bigg]  
$$
$$
 \frac{\vec{k}_{1}\cdot \vec{k}_{2}}{k_{1}^z k_{2}^z}\delta(k_{1}^z+k_{2}^z)(2\pi)^2\delta^2(\vec {k}^\perp_1+ \dots+\vec {k}^\perp_n +\vec{\Delta}).
$$
The last formula can be written as
$$
i\widetilde{T}(\vec{b}^\perp)_{NE}=2i(s-M_\sigma^2)\tilde{\chi}_{NE}(\vec{b})\sum_{n=2}^\infty \frac{\tilde{\chi}^{n-2}_0(\vec{b})}{(n-2)!}
$$
Here $(s-M_\sigma^2)=2E_\phi M_\sigma$, the quantity $\chi_0(\vec{b}^\perp)$ is the exponentiated eikonal phase found in  (\ref{fase}) while
$$
\chi_{NE}(\vec{b}^\perp)=\frac{ \kappa^4 E_\phi M_\sigma^2}{256\pi^5}\int dk_z d^{2}k_1^\perp d^{2} k_2^\perp \frac{e^{-i\vec{b}^\perp\cdot \vec{k}_1^\perp}}{k_1^{\perp 2}+k_z^{ 2}} \bigg[\frac{4}{3a(-k_1^{\perp 2}-k_z^{2})}
$$
$$
+\frac{2a(-k_1^{\perp 2}-k_z^{ 2})+d( -k_1^{\perp 2}-k_z^{2})-1}{6D(-k_1^{\perp 2}-k_z^{2})}
+\bigg(\frac{2}{3 a(-k_1^{\perp 2}-k_z^{2})}
$$
$$
-\frac{2a(-k_1^{\perp 2}-k_z^2)+d(-k_1^{\perp 2}-k_z^{2})-1}{6D(-k_1^{\perp 2}-k_z^{2})}+\frac{1}{4D(-k_1^{\perp 2}-k_z^{2})}\bigg)\frac{\eta_{00}(k_1^\perp\cdot \Delta+E_\phi k_z)^2}{E_\phi^2(k_1^{\perp 2}+k_z^{2})}\bigg]  
$$
 $$
\frac{e^{-i\vec{b}^\perp\cdot \vec{k}_2^\perp}}{k_2^{\perp 2}+k_z^{2}} \bigg[\frac{4}{3a(-k_2^{\perp 2}-k_z^{2})}+\frac{2a(-k_2^{\perp 2}-k_z^{2})+d(-k_2^{\perp 2}-k_z^ 2)-1}{6D( -k_2^{\perp 2}-k_z^{ 2})}
$$
$$
+\bigg(\frac{2}{ 3a(-k_2^{\perp 2}-k_z^{ 2})}-\frac{2a(-k_2^{\perp 2}-k_z^{ 2})+d(-k_2^{\perp 2}-k_z^{2})-1}{6D(-k_2^{\perp 2}-k_z^{ 2})}+\frac{1}{4D(-k_2^{\perp 2}-k_z^{ 2})}\bigg)
$$
\begin{equation}\label{mylene}
\frac{\eta_{00}(k_2^\perp\cdot \Delta-E_\phi k_z)^2}{E_\phi^2(k_2^{\perp 2}+k_z^{ 2})}\bigg] \frac{\vec{k}_1^\perp\cdot \vec{k}_2^\perp -k_z^{ 2}}{k_z^{ 2}}.
 \end{equation}
The next task is to analyze the consequences of this apparently complicated expression.
\\

\emph{The GR limit}
\\

In the GR limit $a\to 1$, $2d\to -1$ and $4D\to -1$ the last quantity reduces to \cite{fazio}
 $$
\chi^{GR}_{NE}(\vec{b}^\perp)=\frac{ \kappa^4 E_\phi M_\sigma^2}{256\pi^5}\int dk_z d^{2}k_1^\perp d^{2} k_2^\perp \frac{e^{-i\vec{b}^\perp\cdot \vec{k}_1^\perp}}{k_1^{\perp 2}+k_z^{ 2}}\frac{e^{-i\vec{b}^\perp\cdot \vec{k}_2^\perp}}{k_2^{\perp 2}+k_z^{ 2}}\frac{\vec{k}_1^\perp\cdot \vec{k}_2^\perp -k_z^{2}}{k_z^{2}},
$$
which is an interesting consistency check.

 This  GR phase is divergent but can be regularized to zero by employing dimensional regularization techniques  \cite{nexteikonal2}.  In order to visualize this note that  the factor $k_1\cdot k_2$ can be found by taking derivatives respect to the impact parameter components $b_i$, these derivatives only affect the exponential factors. This of course is justified only if some assumptions about the integrands are made, and this is a key point.   These products  $k_1\cdot k_2$ can be obtained by taking derivatives of the impact parameter by employing the simple formula $k_i \exp(-ib\cdot k)=i\partial_{b_i}\exp(-ib\cdot k)$. 
 With this technique at hand and using the Bessel derivation formula\begin{equation}\label{besselder}\frac{d}{dx}x^\nu K_\nu(x)=-x^\nu K_{\nu-1}(x),\end{equation}
it follows, as in reference \cite{nexteikonal}, by defining the integrals by dimensional regularization with $\epsilon < 0$, $d^2 k_1^\perp \rightarrow d^{2-2\epsilon} k_1^\perp $ and $d^2 k_2^\perp \rightarrow d^{2-2\epsilon} k_2^\perp$, after integration, that

\begin{equation}
\begin{aligned}
\chi^{GR}_{NE}(\vec{b}^\perp)\sim \int_0^\infty \frac{dk_z}{k_z^2} [ &-4 \epsilon^2 (k_z b)^{-2\epsilon - 2} K_{-\epsilon}(k_z b) K_{-\epsilon}(k_z b)\\
&+ 4 \epsilon (k_z b)^{-2\epsilon - 1} K_{-\epsilon}(k_z b) K_{1+\epsilon}(k_z b)\\
&- (k_z b)^{-2\epsilon} K_{1+\epsilon}(k_z b) K_{1+\epsilon}(k_z b)\\
&-(k_z b)^{-2\epsilon+2} K_{-\epsilon}(k_z b)K_{-\epsilon}(k_z b)]
\end{aligned}
\end{equation}
Some of these integrals are in fact divergent when $\epsilon\to 0$ due to the bad behavior of some of the $K_\alpha(x)$ at the origin. The full result is divergent as well. There exist  integration formulas such as
\begin{equation}\label{toxic}\int_0^\infty dx x^{-\lambda} K_\mu(x)K_\nu(x)=\frac{2^{-\lambda-2}}{\Gamma(1-\lambda)}\prod_{c,d=\pm 1}\Gamma\bigg(\frac{1-\lambda +c \mu+ d\nu}{2}\bigg),\end{equation}
 however, they apply only if $\lambda <1-|\mu|-|\nu|$ when all these parameters are real. In other case, the integrand diverges equal or faster than $1/x$. The prescription to add an $\epsilon$ is to avoid that the Gamma functions take arguments at negative integers, where they have poles. These techniques are familiar in the context of dimensional regularization of gauge theories.
 
 There  are several ways to regularize the above introduced  integrals, even when $\epsilon\to 0$. For instance, the standard Hadamard regularization of the divergent integral $\int_{-1}^{1} x^{-2}dx $ replace it by
$$
\frac{d}{du}\bigg(P\int_{-1}^{1}\frac{1}{(x-u)}\bigg)\bigg|_{u=0}=\frac{d}{du}\log\frac{1-u}{1+u}\bigg|_{u=0}=-2.
$$

In any case, the regularization that is employed in the original reference \cite{nexteikonal2}
is to apply formula \eqref{toxic} by ignoring the fact that the condition  $\lambda <1-|\mu|-|\nu|$ are violated. The reason is that while the integral is divergent for values of $\lambda$ not satisfying this condition, the right hand side is perfectly defined. This is up to poles on the right hand side. And even there, the function can be regularized at those poles by subtracting the divergent part by the equivalent regularization methods of Hadamard or Riesz. This is called a meromorphic extension of the integral. The usual textbook method of dimensional regularization alone is not directly useful here and it is important to complement it with the more general renormalization techniques of meromorphic extension. See for example \cite{paycha}. Ultimately, the dimensional regularization parameter will be the one used for the meromorphic extension.

Going back to GR, if this prescription is applied to the problem, if some general formulas such as $\Gamma(x+1)=x\Gamma(x)$,  $\Gamma(\frac{1}{2})=\sqrt{\pi}$ and the Euler reflection formula $\Gamma(x)\Gamma(1-x)\sin(\pi x)=\pi$ are useful to obtain this cancellation, together with the specific values of the Gamma functions lead to \\
$$
\chi^{GR}_{NE}=0.
$$
This zero result was found already in  \cite{nexteikonal2}.
\\

\emph{The general regularization procedure}
\\

 The expression of \eqref{mylene} for general $a(k)$ and $d(k)$ coefficients turned on is more complicated. A quite explicit derivation for the Stelle model is presented in the appendix, just for a taste of the calculation involved.
 
 For a generic model, the correction \eqref{mylene}  is divergent and has to be regularized as well. In order to illustrate this fact, note that the terms $(k_1^\perp\cdot \Delta+E_\phi k_z)^2$ and $(k_2^\perp\cdot \Delta-E_\phi k_z)^2$ in \eqref{mylene} give rise to numerators which are sum of products of $k_{1i}$ or $k_{2i}$ up to third order in each variable. Therefore, the trick discussed in previous paragraphs of taking derivatives with respect to the components of the impact parameter $b$ now has to be employed up to order three. 

Furthermore, the quantities in parenthesis in \eqref{mylene} in several cases consist in division of polynomials with the denominators of higher order than the numerators. These quantities can be worked out by partial fractions, even when multiplied with factors proportional to $(k_1^{\perp 2}+k_z^{ 2})^{-1}$ or $(k_2^{\perp 2}+k_z^{2})^{-1}$.  By parametrization of the resulting integral \eqref{mylene} in cylindrical coordinates $d^2k_1=k_1 dk_1 d\theta_1$ and $d^2k_2=k_2 dk_2 d\theta_2$ and by performing the integrals in $\theta_1$ and $\theta_2$ it is obtained in \eqref{mylene} a cumbersome product of  linear combinations of integrals of the form
$$
I(\mu,\nu, a, b)=\int_0^\infty \frac{x^{1+\nu} J_\nu(bx) dx}{(x^2+a^2)^{1+\mu}}=\frac{a^{\nu-\mu} b^{\mu}}{2^\mu \Gamma(1+\mu)}K_{\nu-\mu}(ab),
$$
$$
L(\mu. \nu, a, b)=\int_0^\infty \frac{J_0(bx)x dx}{(x^2-a^2)}=\frac{\pi}{2}\bigg[H_0(ba)-Y_0(ba)\bigg],
$$
together with their derivatives with respect to the impact parameter components $b_i$, up to order three.

The last formulas are valid for $a>0$, $b>0$ and $-1<\text{Re}(\nu)<\text{Re}(2\mu+\frac{3}{2})$.  For the present case $\nu=0$ and for the Stelle gravity $\mu$ takes values 0 and 1. For other gravity models, higher orders may appear. The quantity $b$ will be identified with the impact parameter and $a=\sqrt{k_z^2\pm m^2_i}$ with $m_i$ being one of the induced masses in the gravity scenario in consideration. 

Up to the authors knowledge, there are no closed formulas such as \eqref{toxic} for the calculating the present integrals. The best we can do are simple estimations. In order to give an  example, after some calculation it is seen that the Bessel derivation formula \eqref{besselder}
 leads to the following derivatives of $I(\mu,\nu, a, b)$ with respect to the impact parameter
$$
\frac{\partial I}{\partial b_c}=-\frac{2\pi b^{2\mu-\nu-2}}{2^\mu \Gamma(1+\mu)}\bigg[(ba)^{\nu-\mu-1}K_{\nu-\mu-1}(ba) (ba)^2-(2\mu-\nu)(ba)^{\nu-\mu}K_{\nu-\mu}(ba)\bigg]b_c,
$$
$$
\frac{\partial I}{\partial b_b\partial b_c}=\frac{2\pi}{2^\mu \Gamma(1+\mu)}b^{2\mu-\nu-2}\bigg\{\bigg[\delta_{bc}+(2\mu-\nu-2)\frac{b_bb_c}{b^2}\bigg][(ba)^{\nu-\mu-1}K_{\nu-\mu-1}(ba) b^2a^2$$
$$
-(2\mu-\nu)(ba)^{\nu-\mu}K_{\nu-\mu}(ba)]+(2\mu-\nu+2)(ba)^{\nu-\mu-1}K_{\nu-\mu-1}(ba)b_b b_c a^2
$$
$$
-(ba)^{\nu-\mu-2}K_{\nu-\mu-2}(ba)b_b b_c b^2 a^4\bigg\},
$$
$$
\frac{\partial I}{\partial b_a\partial b_b\partial b_c}=\frac{2\pi}{2^\mu \Gamma(1+\mu)}\bigg\{b^{2\mu-\nu-4}(2\mu-\nu-2)\bigg[b_a\delta_{bc}+b_b\delta_{ac}+b_c \delta_{ab}
$$
$$
+(2\mu-\nu-2)(2\mu-\nu)\frac{b_ab_bb_c}{b^2}\bigg][(ba)^{\nu-\mu-1}K_{\nu-\mu-1}(ba) b^2a^2$$
$$
-(2\mu-\nu)(ba)^{\nu-\mu}K_{\nu-\mu}(ba)]+(2\mu-\nu-2)(ba)^{2\mu-\nu-6}b_ab_b b_c\bigg( \frac{a}{b}+1\bigg)
$$
$$
[(2\mu-\nu+2)(ba)^{\nu-\mu-1} K_{\nu-\mu-1}(ba) 
-(ba)^{\nu-\mu-2}K_{\nu-\mu-2}(ba) b^2 a^2]
\bigg\}.
$$
$$
+\frac{2\pi b^{2\mu-\nu-2}}{2^\mu \Gamma(1+\mu)}\bigg\{[(2\mu-\nu+2)(ba)^{\nu-\mu-1} K_{\nu-\mu-1}(ba) 
-(ba)^{\nu-\mu-2}K_{\nu-\mu-2}(ba)  b^2 a^2]
$$
$$
[b_a\delta_{bc}+b_b\delta_{ac}+b_c \delta_{ab}]a-b_a b_b  b_ca^3[(2\mu-\nu+4)(ba)^{\nu-\mu-2} K_{\nu-\mu-2}(ba) 
-(ba)^{\nu-\mu-3}K_{\nu-\mu-3}(ba) b^2 a^2]\bigg\}.
$$
Similar formulas can be found for  $L(\mu,\nu, a, b)$, as the Struve functions $H_\mu(x)$ have derivation formulas similar to the ones just quoted. Note that the last expressions is full of terms of the form $$(ba)^\alpha K_\alpha(ba).$$ When calculating the integrals \eqref{mylene} with respect to  $k_1$ and $k_2$ the quantities $a^2$ become identified with $k_z^2\pm m_i^2$ where $m_i$ are mass scales that arise after partial fractions.  The values of $\nu$ and $\mu$ are positive integers in general, which implies that the values of $\alpha$ are zero or negative. As the Bessel  functions $K_{\alpha}(x)=K_{-\alpha}(x)$ it is clear that $\alpha$ can be considered positive, and the resulting terms are of the form $K_\alpha(x)/(ba)^\alpha$ after switching $\alpha\to -\alpha$.  In other words, the powers of  $ba$ will appear exclusively at the denominators. Note however, that there are further powers of $a$ that multiply these terms, which do  not arise as a power of $ba$.

The considerations given above imply that resulting integral in $k_z$ has a form
$$
\chi_{NE}(\vec{b}^\perp)=\frac{ \kappa^4 E_\phi M_\sigma^2}{256\pi^5}\int_0^\infty dk_z C_1(k_z)C_2(k_z),
$$
where the functions $C_i(k_a)$ consist in cumbersome combinations of terms such as 
$I$, $\partial_{b_a}I$,  $\partial_{b_a}\partial_{b_b}I$, $\partial_{b_a}\partial_{b_b}\partial_{b_c}I$,
$L$,  $\partial_{b_a}L$,  $\partial_{b_a}\partial_{b_b}L$, $\partial_{b_a}\partial_{b_b}\partial_{b_c}L$ with coefficients proportional to powers of $E_\phi$,  $\Delta$ and $k_z$. In the appendix, the explicit derivation of these quantities for the model \cite{stelle1}-\cite{stelle2}, and they are quite involved.
However, by  looking the expressions for the derivatives found above it is clear that this integral is a combination of terms of the form
\begin{equation}\label{melancolica}
T(q, \alpha,\beta,\mu, \nu)=\int_0^\infty dk_z k_z^q\frac{K_\mu(b\sqrt{k_z^2+m_i^2})}{(b\sqrt{k_z^2+m_i^2})^{\mu-\alpha}}  \frac{K_\nu(b\sqrt{k_z^2+m_j^2})}{(b\sqrt{k_z^2+m_j^2})^{\nu-\beta}},
\end{equation}
with $\mu$ and $\nu$ positive and $q=-2,-1,0, 1, 2, 3$.   In addition, combinations involving the Struve functions can appear as well, The  last ones has to be separated in cases $k_z^2<m^2_i$ or $k_z^2>m^2_i$. It is clear that the explicit calculation of the full integral is rather challenging.

The integrals   $T(q,\alpha,\beta,\mu,\nu)$ defined in \eqref{melancolica} is generically divergent for $q=-2$ and  $q=-1$. For instance for $q=-2$ there is a divergence of the form
$$
T\sim  \lim_{k_z\to 0} \frac{ K_\mu(bm_i) K_\nu(bm_j)(bm_i)^{\mu-\alpha}(bm_j)^{\nu-\beta} m_k^\gamma}{k_z}.
$$
These integrals can be regularized by a method due to Hadamard, usually named ``Hadamard finite parts". Given a divergent integral of the form
$$
\int_0^\infty \frac{\phi(k)}{k^q}dk,
$$
where $\phi(k)$ is derivable to order $q$ and of compact support, the Hadamard regularization consist separating the integral as follows
$$
\int_\epsilon^\infty \frac{\phi(k)}{k^q}dk=\int_a^\infty \frac{\phi(k)}{k^q}dk+\int_\epsilon^a \frac{\phi(k)}{k^q}dk
$$
$$
=\int_a^\infty \frac{\phi(k)}{k^q}dk+\int_\epsilon^a  \frac{1}{k^q}\bigg[\phi(k)-\sum_{j=0}^{q-1}\frac{1}{j!}\frac{d^j \phi}{dk}\bigg|_{k=0} k^j\bigg]dk+\int_\epsilon^a \sum_{j=0}^{q-1}\frac{1}{j!}\frac{d^j \phi}{dk}\bigg|_{k=0} k^{j-k}dk
$$
$$
=\int_a^\infty \frac{\phi(k)}{k^q}dk+\int_\epsilon^a  \frac{1}{k^q}\bigg[\phi(k)-\sum_{j=0}^{q-1}\frac{1}{j!}\frac{d^j \phi}{dk}\bigg|_{k=0} k^j\bigg]dk
$$
$$
+ \sum_{j=0}^{q-2}\frac{1}{j! (q-j-1)}\frac{d^j \phi}{dk}\bigg|_{k=0} \bigg[\frac{1}{a^{q-j-1}}-\frac{1}{\epsilon^{q-j-1}}\bigg]+\frac{1}{(k-1)!}\frac{d^{q-1} \phi}{dk}\bigg|_{k=0}(\log a-\log \epsilon),
$$
and by deleting the terms that diverge in the limit $\epsilon\to 0$. In the case of Schwarz functions instead of compact support see reference \cite{paycha}. The result is independent on the choice of $a$ and one may chose $a=1$ in order to eliminate the logarithmic terms. The result is the Hadamard regularized integral 
$$
H\int_0^\infty \frac{\phi(k)}{k^q}dk=\int_1^\infty \frac{\phi(k)}{k^q}dk+\int_0^1  \frac{1}{k^q}\bigg[\phi(k)-\sum_{j=0}^{q-1}\frac{1}{j!}\frac{d^j \phi}{dk}\bigg|_{k=0} k^j\bigg]dk
$$
$$
+ \sum_{j=0}^{q-2}\frac{1}{j! (j-q+1)}\frac{d^j \phi}{dk}\bigg|_{k=0}.
$$
Note that the integral $\int_0^a$ is not divergent at $k=0$ due to the L´Hopital rule. 

The Hadamard regularization, applied for instance to the case $q=-2$  discussed above gives that the divergent integral
$$
T(-2, \alpha,\beta,\mu, \nu)=\int_0^\infty dk_z\frac{1}{k_z^2}\frac{K_\mu(b\sqrt{k_z^2+m_i^2})}{(b\sqrt{k_z^2+m_i^2})^{\mu-\alpha}}  \frac{K_\nu(b\sqrt{k_z^2+m_j^2})}{(b\sqrt{k_z^2+m_j^2})^{\nu-\beta}},
$$
becomes, by defining
$$
\phi(k_z, \alpha, \beta,\mu, \nu)=\frac{K_\mu(b\sqrt{k_z^2+m_i^2})}{(b\sqrt{k_z^2+m_i^2})^{\mu-\alpha}}  \frac{K_\nu(b\sqrt{k_z^2+m_j^2})}{(b\sqrt{k_z^2+m_j^2})^{\nu-\beta}},
$$
the Hadamard regularized expression
$$
T_R(-2, \alpha,\beta,\mu, \nu)=\int_1^\infty dk_z\frac{\phi(k_z, \alpha, \beta,\gamma)}{k_z^2}+\int_0^1  dk_z\frac{1}{k_z^2}\bigg[\phi-\phi|_{k_z=0}-\frac{d\phi}{dk_z}\bigg|_{k_z=0}k_z\bigg]+\phi|_{k_z=0}
$$
where
$$
\phi|_{k_z=0}= K_\mu(bm_i) K_\nu(bm_j)(bm_i)^{\alpha-\mu}(bm_j)^{\beta-\nu}.
$$
Similarly the case $q=-1$ is given by
$$
T_R(-1, \alpha,\beta, \nu, \mu)=\int_1^\infty dk_z\frac{\phi(k_z, \alpha, \beta,\gamma)}{k_z}+\int_0^1  dk_z\frac{1}{k_z}[\phi-\phi|_{k_z=0}].
$$
The other integrals are finite, although difficult to be calculated explicitly. 

Some further comments are desirable. First, in Hadamard renormalization, it is usually stated in the mathematical literature that it consist in replace the integrand by the Hadamard regularized integrand
$$
H\frac{\phi(k)}{k^q}=\lim_{\epsilon\to 0}\bigg[\frac{\phi(k-\epsilon)}{k^q}-\sum_{j=0}^{q-2}\frac{1}{j! (q-j-1)}\frac{\delta^j(k)}{\epsilon^{q-j-1}}+\frac{(-1)^{q-1}}{(k-1)!}\delta^{q-1}(k)\log \epsilon\bigg].
$$
With this definition it can be shown, see \cite{estrada} and references therein, that 
$$
\frac{d}{dk}\bigg(H\frac{\phi(k)}{k^q}\bigg)=-q\frac{d}{dk}\bigg(H\frac{\phi(k)}{k^q}\bigg)+\frac{(-1)^q\delta^q(k)}{q!}
$$
$$
k^lH\frac{\phi(k)}{k^q}=H\frac{\phi(k)}{k^{q-l}},\qquad l=0,..,q-1.
$$

The reader may consult  formulas (2.14)-(2.16) of \cite{estrada} for further details. The point is that the formulas just given have algebraic properties and behave under derivation  similar to powers $k^\lambda$.  This is the basis for another regularization method by Riesz, which turns out to be equivalent to the Hadamard method. Consider a parameter\footnote{For example it may be the parameter introduced when generalizing the integrals in ``dimensional regularization".} $\lambda$ such that $\text{Re}(\lambda)>-1$. Then the function 
$\phi(k) k^{\lambda}$ is locally integrable near $k=0$ and defines by the last formulas a distribution $k^\lambda_+$. This notation does not imply that it is given as a simple power, but it is indicating that its behavior under derivation and multiplication imitates the usual powers $k^n$ due to the last formulas. Given such function, it admits an analytic continuation to the punctured complex plane $C/\{-1, -2, -3,....\}$ by the definition
$$
k_+^\lambda=\frac{1}{(\lambda+n)..(\lambda-1)}\frac{d^n}{dk^n_+}(k_+^{\lambda+n})
$$
with $n$ a positive integer with $n+\text{Re}(\lambda)>-1$. This leads to a regularization of $\phi(k)k^\lambda$ if $\lambda\neq -1,-2,...$. In terms of these formulas a regularization of $\phi(k)/k^q$ can be found.

Note that this regularization can not be applied in the present form to  negative integer values of $q$ because $k_+^\lambda$ is not defined for these values. Nevertheless, its singular value at these limiting points may be subtracted. The result is equivalent to the Hadamard finite part \cite{estrada}
$$
H(\frac{\phi}{k^q})=\lim_{\lambda\to-q}\bigg[k_+^\lambda-\frac{(-1)^{q-1}\delta^{q-1}(k)}{q-1! (\lambda+q)} \bigg].
$$

Based on this discussion, we interpret our renormalization method, called Riesz regularization \cite{Gelfand} presented here with a generalization of the analytic continuation method presented in \cite{nexteikonal2}, adapted to the case where other masses besides the Planck mass are turned on in the gravity model. Unfortunately, the regularized integrals are more difficult to be found explicitly, but it is unlikely, in the authors opinion, that they vanish for generic masses turned on. 
\\

\emph{Further details about the integral calculations}
\\

The integrals described above, even though they may be regularized, are hard to be computed explicitly. However, they can be represented in series and can be regularized by meromorphic extension of some of the terms of the resulting series, thus generalizing the results of \cite{nexteikonal2} to the present situation. In order to see how this generalization follows,  consider again the generic integral
\begin{equation}\label{basica}I = \int_0^{+\infty}
\frac{K_\mu\left( \sqrt{t^2+z^2}\right)}{\left(t^2+z^2\right)^{{\mu / 2}-{\alpha / 2}}}
\frac{K_\nu\left( \sqrt{t^2+Z^2}\right)}{\left(t^2+Z^2\right)^{{\nu / 2}-{\beta / 2}}}\, \, t^{\eta}\, \, d t.
\end{equation}

The desired series expansion can be achieved by using the binomial expansion for the $\alpha$ and $\beta$ factors of the
denominators and by use the modified Gegenbauer's addition theorem \cite{watson}, as it will be explained now.

In order to give an example, consider  the case with $z,Z \in \mathbb{R}$ with $0<z<Z$, leading to the decomposition
$$
\int_0^{+\infty}= \int_0^{z} + \int_z^{Z}  + \int_Z^{+\infty}.
$$
Rewrite the denominators of the above integral $I$ as follows
$$\frac{1}{\left(t^2 + z^2\right)^{\mu/2 - \alpha/2}} = \frac{1}{\left(t^2 + z^2\right)^{\mu/2}} \cdot \left(t^2 + z^2\right)^{\alpha/2}.$$
$$\frac{1}{\left(t^2 + Z^2\right)^{\nu/2 - \beta/2}} = \frac{1}{\left(t^2 + Z^2\right)^{\nu/2}} \cdot \left(t^2 + Z^2\right)^{\beta/2}.$$
Then for $|t| > |z|$ and $|t| > |Z|$ expand the last factors as
\begin{equation*}
    \begin{aligned}
    \left(t^2 + z^2\right)^{\alpha/2} &= t^\alpha \left(1 + \frac{z^2}{t^2}\right)^{\alpha/2}  \, = t^\alpha \sum_{k=0}^\infty \binom{\alpha/2}{k} \left(\frac{z^2}{t^2}\right)^k,\\
    \left(t^2 + Z^2\right)^{\beta/2} &= t^\beta \left(1 + \frac{Z^2}{t^2}\right)^{\beta/2} = t^\beta \sum_{l=0}^\infty \binom{\beta/2}{l} \left(\frac{Z^2}{t^2}\right)^l.
\end{aligned}
\end{equation*}
Otherwise, in case $|t| < |z|$ and $|t| < |Z|$, the expansion that should be considered is instead
\begin{equation*}
    \begin{aligned}
    \left(t^2 + z^2\right)^{\alpha/2} &= z^\alpha \left(1 + \frac{t^2}{z^2}\right)^{\alpha/2}\, \, = z^\alpha \sum_{k=0}^\infty \binom{\alpha/2}{k} \left(\frac{t^2}{z^2}\right)^k,\\
    \left(t^2 + Z^2\right)^{\beta/2} &= Z^\beta \left(1 + \frac{t^2}{Z^2}\right)^{\beta/2} = Z^\beta \sum_{l=0}^\infty \binom{\beta/2}{l} \left(\frac{t^2}{Z^2}\right)^l.
\end{aligned}
\end{equation*}
The intermediate case is a combination of those situations. The Gegenbauer addition theorem for Bessel functions leads, for  $|t| > |z|$ and $|t| > |Z|$, to
the following expansion
\begin{equation*}
\frac{K_\mu(\sqrt{t^2 + z^2})}{\left(t^2 + z^2\right)^{\mu/2}} = 2^\mu \sum_{m=0}^\infty (-1)^m (\mu + 2m) \frac{K_{\mu+2m}(t)}{t^\mu} \frac{I_{\mu+2m}(z)}{z^\mu} \frac{\Gamma(\mu + m)}{m!},
\end{equation*}
\begin{equation*}
\frac{K_\nu(\sqrt{t^2 + Z^2})}{\left(t^2 + Z^2\right)^{\nu/2}} = 2^\nu \sum_{n=0}^\infty (-1)^n (\nu + 2n) \frac{K_{\nu+2n}(t)}{t^\nu} \frac{I_{\nu+2n}(Z)}{Z^\nu} \frac{\Gamma(\nu + n)}{n!}.    
\end{equation*}
Otherwise, in case $|t| < |z|$ and $|t| < |Z|$, then by interchanging $t \leftrightarrow z$ and $t \leftrightarrow Z$\\ respectively
\begin{equation*}
\frac{K_\mu(\sqrt{t^2 + z^2})}{\left(t^2 + z^2\right)^{\mu/2}} = 2^\mu \sum_{m=0}^\infty (-1)^m (\mu + 2m) \frac{K_{\mu+2m}(z)}{z^\mu} \frac{I_{\mu+2m}(t)}{t^\mu} \frac{\Gamma(\mu + m)}{m!},
\end{equation*}
\begin{equation*}
\frac{K_\nu(\sqrt{t^2 + Z^2})}{\left(t^2 + Z^2\right)^{\nu/2}} = 2^\nu \sum_{n=0}^\infty (-1)^n (\nu + 2n) \frac{K_{\nu+2n}(Z)}{Z^\nu} \frac{I_{\nu+2n}(t)}{t^\nu} \frac{\Gamma(\nu + n)}{n!}.    
\end{equation*}

So, in the first region, the integral  \eqref{basica} becomes
\begin{equation*}
\begin{aligned}
I_{0 z} = \frac{z^\alpha Z^\beta}{z^\mu Z^\nu}& \,  \,2^{\mu + \nu}\sum_{k,l,m,n=0}^\infty \frac{(-1)^{m+n}}{m!\, n!} \binom{\alpha/2}{k}
\binom{\beta/2}{l} {(\mu + 2m)(\nu + 2n)} \Gamma(\mu + m)\Gamma(\nu + n)\\
&\times z^{-2k} Z^{-2l}
K_{\mu+2m}(z)
K_{\nu+2n}(Z)
\int_0^z t^{\eta + 2(k + l) - \mu - \nu}   I_{\mu+2m}(t)
I_{\nu+2n}(t) dt.
\end{aligned}
\end{equation*}

For the integral between $z$ and $Z$ instead 
\begin{equation*}
\begin{aligned}
I_{z Z} = \frac{Z^\beta}{z^\mu Z^\nu}& \,  \,2^{\mu + \nu}\sum_{k,l,m,n=0}^\infty \frac{(-1)^{m+n}}{m!\, n!} \binom{\alpha/2}{k}
\binom{\beta/2}{l} {(\mu + 2m)(\nu + 2n)} \Gamma(\mu + m)\Gamma(\nu + n)\\
&\times z^{2k} Z^{-2l}
I_{\mu+2m}(z)
K_{\nu+2n}(Z)
\int_z^Z t^{\eta + \alpha - 2k + 2l - \mu - \nu}   K_{\mu+2m}(t)
I_{\nu+2n}(t) dt.
\end{aligned}
\end{equation*}

Finally for the integral from $Z$ to $+\infty$ the expression is
\begin{equation*}
\begin{aligned}
I_{Z \infty} = \frac{1}{z^\mu Z^\nu}& \,  \,2^{\mu + \nu}\sum_{k,l,m,n=0}^\infty \frac{(-1)^{m+n}}{m!\, n!} \binom{\alpha/2}{k}
\binom{\beta/2}{l} {(\mu + 2m)(\nu + 2n)} \Gamma(\mu + m)\Gamma(\nu + n)\\
&\times z^{2k} Z^{2l}
I_{\mu+2m}(z)
I_{\nu+2n}(Z)
\int_Z^{+\infty} t^{\eta + \alpha + \beta - 2(k + l) - \mu - \nu}    K_{\mu+2m}(t)
K_{\nu+2n}(t) dt.
\end{aligned}
\end{equation*}

By further employing the formulas in pages 400-401 of reference  \cite{bateman} it is seen that
\begin{equation}\label{IntKI}
\begin{aligned}
 x^\sigma &K_\mu(x) I_\nu(x) = 2^{-1} \pi^{-\frac{1}{2}}\\
&\times G_{2,4}^{2,2} \left[ x^2 \middle| \begin{array}{c}
\frac{1}{2}\sigma,\, \frac{1}{2}(\sigma+1) \\
\frac{1}{2}(\nu + \mu + \sigma),\, \frac{1}{2}(\nu - \mu + \sigma),\, \frac{1}{2}(-\nu + \mu + \sigma),\, \frac{1}{2}(-\nu - \mu + \sigma)
\end{array} \right],
\end{aligned}
\end{equation}

\begin{equation}\label{IntKK}
\begin{aligned}
 x^\sigma &K_\mu(x) K_\nu(x) = 2^{-1} \pi^{\frac{1}{2}}\\
&\times G_{2,4}^{4,0} \left[ x^2 \middle| \begin{array}{c}
\frac{1}{2}\sigma,\, \frac{1}{2}(\sigma+1) \\
\frac{1}{2}(\nu + \mu + \sigma),\, \frac{1}{2}(\nu - \mu + \sigma),\, \frac{1}{2}(-\nu + \mu + \sigma),\, \frac{1}{2}(-\nu - \mu + \sigma)
\end{array} \right].
\end{aligned}
\end{equation}
\\
By taking into account that $J_\mu (ix) = i^\mu I_\mu(x)$ it is further arrived to
\begin{equation}\label{IntII}
\begin{aligned}
 x^\sigma &I_\mu(x) I_\nu(x) = i^{-\mu - \nu} \pi^{-\frac{1}{2}}\\
&\times G_{2,4}^{1,2} \left[ -x^2 \middle| \begin{array}{c}
\frac{1}{2}(\sigma+1),\, \frac{1}{2}\sigma \\
\frac{1}{2}(\nu + \mu + \sigma),\, \frac{1}{2}(\nu - \mu + \sigma),\, \frac{1}{2}(-\nu + \mu + \sigma),\, \frac{1}{2}(-\nu - \mu + \sigma)
\end{array} \right]
\end{aligned}
\end{equation}
The utility  of writing everything in terms of the Meijer functions is that the following integration formulas, which can be found in page 348 of \cite{prudnikov}, can be employed in order to find the integrals expressed in the above series
\begin{equation}\label{definiteG_0a}
\begin{array}{c} \int_{0}^{a} x^{\alpha-1} (a-x)^{\beta-1} G_{pq}^{mn} \left( \omega x^{l/k} \middle| \begin{array}{c} (a_p) \\ (b_q) \end{array} \right) dx =\\
\\
\frac{k^{\mu^*} l^{-\beta} \Gamma(\beta)}{(2\pi)^{c^* (k-1)} a^{1-\alpha-\beta}} G_{kp+l,kq+l}^{km,kn+l} \left( \frac{\omega^k a^l}{k^{k(q-p)}} \middle| \begin{array}{c} \Delta(l, 1-\alpha), \Delta(k, (a_p)) \\ \Delta(k, (b_q)), \Delta(l, 1-\alpha-\beta) \end{array} \right) \end{array}
\end{equation}
\begin{equation}\label{definiteG_0inf}
\begin{array}{c} \int_{a}^{\infty} x^{\alpha-1} (x-a)^{\beta-1} G_{pq}^{mn} \left( \omega x^{l/k} \middle| \begin{array}{c} (a_p) \\ (b_q) \end{array} \right) dx =\\
\\
\frac{k^{\mu^*} l^{-\beta} \Gamma(\beta)}{(2\pi)^{c^* (k-1)} a^{1-\alpha-\beta}} G_{kp+l,kq+l}^{km+l,kn} \left( \frac{\omega^k a^l}{k^{k(q-p)}} \middle| \begin{array}{c} \Delta(k, (a_p)), \Delta(l, 1-\alpha) \\ \Delta(l, 1-\alpha-\beta), \Delta(k, (b_q)) \end{array} \right) \end{array}
\end{equation}
\\
Here ${\mu^*} = \sum_{j=1}^q b_j - \sum_{j=1}^p a_j + \frac{p-q}{2} + 1$, $c^* = m + n - \frac{p+q}{2}$ and in general,
$$\Delta(k, a)\equiv\frac{a}{k}, \frac{a+1}{k}, \ldots, \frac{a+k-1}{k}$$
$$\Delta(k, (a_p)) \equiv \left( \frac{a_1}{k}, \frac{a_1+1}{k}, \ldots, \frac{a_1+k-1}{k}, \ldots, \frac{a_p}{k}, \frac{a_p+1}{k}, \ldots, \frac{a_p+k-1}{k} \right)$$

For example, in order to calculate $I_{zZ}$, using that $\int_z^Z=\int_0^Z - \int_0^z$ (or $\int_z^Z=\int_z^{+\infty} - \int_Z^{+\infty}$), the  representation (\ref{IntKI}) may be employed  and  the further subtraction of the two integrals of the type (\ref{definiteG_0a}) (or (\ref{definiteG_0inf})).

The integrals \eqref{IntKI}, \eqref{IntKK}, \eqref{IntII} have a very long list of conditions required for convergence, only when those conditions are satisfied the integral in the LHS equals the stated result in the RHS. In any case, even if the LHS of the integral is non convergent, one may employ a meromorphic continuation of the integral as dictated by the RHS. In this way the expression is regularized using the RHS in symbolic way, regardless of convergence conditions. The expression can be extended, if necessary, even to regions where the RHS has poles by subtracting the divergent terms. This is the essence of the so called Riesz regularization. The resulting G-function in the RHS, may also be re expressed as a finite sum (with $km$, $kn$, $km+l$ or $kn+l$ terms) of generalized hypergeometric functions. For this and more properties of {G-functions} the reader may consult for example section 8.2 of the book \cite{prudnikov}. The point is that the above procedure  generalizes the results of \cite{nexteikonal2} to the case where additional masses in the gravity model are turned on. 

It may be interesting to find the explicit form of the integrals outlined above. They are found as follows.
 For the integral in $I_{0z}$ the formula (\ref{IntII}) applies by replacing $\mu \rightarrow \mu+2m$; $\nu \rightarrow \nu+2n$, $\sigma = \eta + 2(k + l) - \mu - \nu$ and in (\ref{definiteG_0a}) $a=z, \alpha=1, \beta=1, m=1,n=2, p=2, q=4, w=-1,l=2, k=1$. This leads to
    $$
\begin{array}{c} 
\begin{aligned}
\int_0^z t^{\eta + 2(k + l) - \mu - \nu} &I_{\mu+2m}(t) I_{\nu+2n}(t) \, dt = \frac{1}{2} \pi^{-1/2}\, z\, i^{-\mu - \nu - 2m - 2n} \\
&\times G_{4,6}^{1,4} \left( -z^2 \middle| \begin{array}{c} 
0, \frac{1}{2}, \frac{\eta + 2(k + l) - \mu - \nu + 1}{2}, \frac{\eta + 2(k + l) - \mu - \nu}{2}\\
\\
\frac{\eta + 2(k + l) + \mu + \nu}{2} + m + n, \frac{\eta + 2(k + l) - \mu + \nu}{2} - m + n, \\ 
\frac{\eta + 2(k + l) + \mu - \nu}{2} + m - n, \frac{\eta + 2(k + l) - \mu - \nu}{2} - m - n, -\frac{1}{2}, 0 
\end{array} \right)
\end{aligned}
\end{array}$$

For the integral in $I_{Z\infty}$ the formula (\ref{IntKK})  with the replacement: $\mu \rightarrow \mu+2m$, $\nu \rightarrow \nu+2n$, $\sigma = \eta + \alpha + \beta - 2(k + l) - \mu - \nu$ and in (\ref{definiteG_0inf}) $a=Z, \alpha=1, \beta=1,m=4, n=0, p=2, q=4, w=1,l=2, k=1$. This leads to
$$\begin{array}{c} 
\begin{aligned}
\int_Z^\infty t^{\eta + \alpha + \beta - 2(k + l) - \mu - \nu} &K_{\mu+2m}(t) K_{\nu+2n}(t) \, dt = \frac{1}{4} \pi^{1/2} Z \\
&\times G_{4,6}^{6,0} \left( Z^2 \middle| \begin{array}{c} 
\frac{\eta + \alpha + \beta - 2(k + l) - \mu - \nu}{2}, \frac{\eta + \alpha + \beta - 2(k + l) - \mu - \nu + 1}{2}, 0, \frac{1}{2}\\
\\
-\frac{1}{2}, 0, \frac{\eta + \alpha + \beta - 2(k + l) + \mu + \nu}{2} + m + n, \\ 
\frac{\eta + \alpha + \beta - 2(k + l) - \mu + \nu}{2} - m + n, \\ 
\frac{\eta + \alpha + \beta - 2(k + l) + \mu - \nu}{2} + m - n, \\ 
\frac{\eta + \alpha + \beta - 2(k + l) - \mu - \nu}{2} - m - n 
\end{array} \right)
\end{aligned}
 \end{array}$$

For the integral in $I_{zZ}$  the formula  (\ref{IntKI})  is the one to be employed, with the identifications $\mu \rightarrow \mu+2m$, $\nu \rightarrow \nu+2n$, $\sigma = \eta + \alpha - 2k + 2l - \mu - \nu$ and for each of the integrals (\ref{definiteG_0a}) $\alpha=1, \beta=1, m=2, n=2, p=2, q=4, w=1,l=2,k=1$ and $a=Z$ or $a=z$ depending on the integral. This gives
$$\begin{array}{c} 
\begin{aligned}
\int_z^Z &t^{\eta + \alpha - 2k + 2l - \mu - \nu} K_{\mu+2m}(t) I_{\nu+2n}(t) \, dt =\frac{1}{4} \pi^{-1/2}\\
&\times \left[ Z G_{4,6}^{2,4} \left( Z^2 \middle| \begin{array}{c} 
0, \frac{1}{2}, \frac{\eta + \alpha - 2k + 2l - \mu - \nu}{2}, \frac{\eta + \alpha - 2k + 2l - \mu - \nu + 1}{2} \\
\\
\frac{\eta + \alpha - 2k + 2l + \mu + \nu}{2} + m + n, \frac{\eta + \alpha - 2k + 2l - \mu + \nu}{2} - m + n, \\ 
\frac{\eta + \alpha - 2k + 2l + \mu - \nu}{2} + m - n, \frac{\eta + \alpha - 2k + 2l - \mu - \nu}{2} - m - n, -\frac{1}{2}, 0 
\end{array} \right) \right. \\
\\
&- \left. z G_{4,6}^{2,4} \left( z^2 \middle| \begin{array}{c} 
0, \frac{1}{2}, \frac{\eta + \alpha - 2k + 2l - \mu - \nu}{2}, \frac{\eta + \alpha - 2k + 2l - \mu - \nu + 1}{2} \\
\\
\frac{\eta + \alpha - 2k + 2l + \mu + \nu}{2} + m + n, \frac{\eta + \alpha - 2k + 2l - \mu + \nu}{2} - m + n, \\ 
\frac{\eta + \alpha - 2k + 2l + \mu - \nu}{2} + m - n, \frac{\eta + \alpha - 2k + 2l - \mu - \nu}{2} - m - n, -\frac{1}{2}, 0 
\end{array} \right) \right]
\end{aligned}
 \end{array}
 $$
 
In any of the cases, all the G-functions $G_{pq}^{mn} \left( x \middle| \begin{array}{c} (a_p) \\ (b_p) \end{array} \right)$ that appear in the integrand and on the RHS result of each integral, are well defined for all $x\neq 0$. This is because $p<q$. This condition ensures that the $G$-functions are well defined for every non zero $x$ value  and we can choose a proper contour of integration of the Mellin–Barnes integral defining G.\footnote{See condition for convergence 3) of \cite{prudnikov} page 617.}

Therefore, it is concluded that the RHS functions $G$'s are well defined for al $z,Z \neq 0$, being then these the regularizations of the LHS integrals arising for the problem.

\subsubsection{Seagull terms}

The other corrections that  references \cite{nexteikonal}-\cite{nexteikonal2} and \cite{fazio} work out  are two seagull contributions for GR, one involving a third order graviton-graviton interaction and one that does not. 
These will be generalized to higher order gravitational models. However, only the second contribution will be considered, as the three-graviton vertex in those references corresponds to GR. For instance, in GR, the bare  three graviton interaction vertex is given by
$$
\Gamma_{\alpha_1\beta_1\alpha_2\beta_2\alpha_3\beta_3}(k_1, k_2, k_3)=\frac{\delta S_{IGR}}{\delta g^{\alpha_1\beta_1}\delta g^{\alpha_2\beta_2}\delta g^{\alpha_3\beta_3}}\bigg|_{g^{\mu\mu}=\eta^{\mu\nu}}=-\text{sym} P_6(-4\eta_{\alpha_2 \alpha_3}\eta_{\beta_2 \alpha_1} \eta_{\beta_3 \beta_1} k_2\cdot k_3
$$
$$
+2\eta_{\alpha_2 \beta_2}\eta_{\alpha_3 \alpha_1} \eta_{\beta_3 \beta_1} k_2\cdot k_3-\eta_{\alpha_2 \beta_2}\eta_{\alpha_3 \beta_3} k_{2\alpha_1}k_{3\beta_|}+2\eta_{\alpha_2 \alpha_3}\eta_{\beta_2 \beta_3} k_{2\alpha_1} k_{3\beta_|}+4\eta_{\alpha_2 \alpha_1}\eta_{\beta_1 \beta_3} k_{2\alpha_3} k_{3\beta_2})
$$
Here $S_{IGR}$ is the interacting part of the lagrangian of GR, the term sym indicates symmetrization in every pair of indices $\alpha_i\beta_i$ with $i=1,2,3$ and $P_6$ indicates the sum over the permutations of the indices $\alpha_i\beta_i k_i$.  The above expression omits the delta that enforce momentum conservation. It is clear with this example that 
for other gravity theories, these vertices will acquire new terms, due to the polynomial terms appearing in \eqref{tat}. These terms have to be worked out independently and the study of their effect is laborious. This will be considered in a separate work.

In  the seagull approximation, the results of \cite{fazio}  show that Green function of the light particle \eqref{lighparticle} has to be replaced by the following
\begin{equation}\label{mastercard}
<p’|G^c(x,y|h)|p>=-\frac{i\kappa^2}{2} \lim\limits_{\substack{p^2\rightarrow 0 \\ p^{\prime 2}\rightarrow 0}}p^2 p^{‘2}(2\pi)^4 \sum_{n=0}^{+\infty}\frac{(-i)^n\kappa^n}{n!}\int_0^{+\infty}d\nu\, e^{i\nu(p^{’2}+i\epsilon)}\int\frac{d^4 \bar{k} d^4\underline{k}}{(2\pi)^8}
\end{equation}
$$
\int\frac{d^4 k_1\dots d^4k_n}{(2\pi)^{4n}}\int_0^\nu d\nu’ \int_0^\nu d\xi_1\nonumber\dots d\xi_n \delta^4(p’-p-k_1-\dots -k_n-\bar{k}-\underline{k})
$$
$$
 e^{-\sum\limits_{m=1}^n 2ip’\cdot k_m(\nu-\xi_m)} e^{-2ip’\cdot (\bar{k}+\underline{k})(\nu-\nu’)} p^{\prime\mu_1}p^{\prime\beta_1}\hat{h}_{\mu_1\beta_1}(k_1)\dots p^{\prime\mu_n}p^{\prime\beta_n}
 $$
 $$
 \hat{h}_{\mu_n\beta_n}(k_n)\hat{h}_{\gamma\delta}(\bar{k})\left(p^{\prime\alpha}p^{\prime\beta}\eta^{\gamma\delta}-2p^{\prime\gamma}p^{\prime\beta}\eta^{\delta\alpha}\right)\hat{h}_{\alpha\beta}(\underline{k}).
$$
The only difference between the last matrix element and the one in \eqref{lighparticle} is the insertion of the factor $\hat{h}_{\gamma\delta}(\bar{k})\left(p^{\prime\alpha}p^{\prime\beta}\eta^{\gamma\delta}-2p^{\prime\gamma}p^{\prime\beta}\eta^{\delta\alpha}\right)\hat{h}_{\alpha\beta}(\underline{k})$. The heavy particle Green function \eqref{heavyparticle}
instead,  is unchanged under this approximation.  By following the same procedure than the one sketched in \eqref{fisura1}-\eqref{matrix}
together with the use of the formula \eqref{rodnis}
it is arrived to 
$$
iT(p,p’,q,q’)_{SG}=\frac{i}{2} \lim_{p_i^2\rightarrow m_i^2} p^2 p’^2 (q^2-M_\sigma^2)(q’^2-M_\sigma^2)\sum_{r=2}^{+\infty}\frac{(-i)^{2r}\kappa^{2r}}{(r-2)!}\int_0^{+\infty}d\nu e^{i\nu(p’^2+i\epsilon)}$$
$$
\int_0^{+\infty} \frac{d\nu_1 e^{i\nu_1(q’^2-M_\sigma^2 +i\epsilon)}}{(2\pi)^{4r}}\int d^4 \bar{k} d^4 \underline{k}  \bigg[M_\sigma^4 p^{\prime \alpha}p^{\prime \beta} D_{\alpha\beta,00}(\bar{k})\eta^{\epsilon \phi} D_{\epsilon\phi, 00}(\underline{k})-2M_\sigma^4 p^{\prime \alpha}p^{\prime \epsilon} D_{\alpha\beta,00}(\bar{k})\eta^{\beta \phi} D_{\epsilon\phi, 00}(\underline{k})\bigg]
$$
$$
\bigg(\prod_{m=3}^{r}\int d^4 k_ m p^{\prime\mu_{m}}p^{\prime\beta_{m}}D_{\mu_m \beta_m,00}(k_m)\bigg)
\int_0^\nu d\nu’ e^{-2ip’\cdot (\bar{k}+\underline{k})(\nu-\nu’)}\int_0^\nu e^{-\sum_{m=1}^{r-2}2ip’\cdot k_m(\nu-\xi_m)}d\xi_1\dots d\xi_{r-2}$$

\begin{equation}\label{seaguli}
\int_0^{\nu_1}\exp\left[\sum_{\tilde{m}=1}^{r}2iq’\cdot k_{\tilde{m}}(\nu_1-\tilde{\xi}_{\tilde{m}})\right]d\tilde{\xi}_1\dots d\tilde{\xi}_{r}\delta^4(p’-p-k_1-\dots-k_n-\bar{k}-\underline{k}).
\end{equation}
The fact that the sum starts at $r=2$ reflects  the fact that the factor $\hat{h}_{\gamma\delta}(\bar{k})\left(p^{\prime\alpha}p^{\prime\beta}\eta^{\gamma\delta}-2p^{\prime\gamma}p^{\prime\beta}\eta^{\delta\alpha}\right)\hat{h}_{\alpha\beta}(\underline{k})$ involving two metrics  is always present in this approximation.  In addition, $n=r-2$.
By applying the eikonal identities \eqref{idento} -\eqref{idento2} the last expression becomes
$$
i\mathcal{T}^{NL}(p,p’,q,q’)_{SG}=-\frac{1}{2} \lim_{p_i^2\rightarrow m_i^2} p^2 p’^2\sum_{r=2}^{+\infty}\frac{(-i)^{2r}\kappa^{2r}}{(r-2)!}\int_0^{+\infty}d\nu\,\frac{e^{i\nu(p’^2+i\epsilon)}}{(2\pi)^{4r}}
$$
$$
\int d^4 \bar{k} d^4 \underline{k}\bigg[M_\sigma^4 p^{\prime \alpha}p^{\prime \beta} D_{\alpha\beta,00}(\bar{k})\eta^{\epsilon \phi} D_{\epsilon\phi, 00}(\underline{k})-2M_\sigma^4 p^{\prime \alpha}p^{\prime \epsilon} D_{\alpha\beta,00}(\bar{k})\eta^{\beta \phi} D_{\epsilon\phi, 00}(\underline{k})\bigg]
$$
$$
\bigg(\prod_{m=3}^{r}\int d^4 k_ m p^{\prime\mu_{m}}p^{\prime\beta_{m}}D_{\mu_m \beta_m,00}(k_m)\bigg)\int_0^\nu\exp\left[-\sum_{m=1}^{r-2}2ip’\cdot k_m(\nu-\xi_m)\right]d\xi_1\dots d\xi_{r-2}$$
\begin{equation}\label{laburando}
\int_0^\nu d\nu’ e^{-2ip’\cdot (\bar{k}+\underline{k})(\nu-\nu’)}
i^r\frac{(-2\pi i)^{r-1}}{(2M_\sigma)^{r-1}}
\delta(k^0_1)\dots \delta(k^0_r)
(2\pi)^4\delta^3(\vec{k}_1+\dots+\vec{k}_n+\vec{\bar{k}}+\vec{\underline{k}}+\vec{\Delta}).
\end{equation}
In addition, the eikonal identity at the z axis \eqref{idento}, applied to this case is
$$
  \lim_{p_{i}^{2}\rightarrow m_{i}^{2}}p^{2}p^{2}\delta(k_{1}^{z}+\dots k_{r-2}^{z}+\bar{k}^{z}+\underline{k}^{z})\int_{0}^{+\infty}d\nu e^{i\nu(p^{2}+i\epsilon)}\frac{1-e^{2i\nu E_{\phi}(\bar{k}^{z}+\underline{k}_{z})}}{-2iE_{\phi}(\bar{k}^{z}+\underline{k}_{z})}\prod_{m=3}^{r}\frac{1-e^{2i\nu E_{\phi}k_{m}^{z}}}{-2iE_{\phi}k_{m}^{z}}
  $$
  \begin{equation}\label{identos}
 = i^{r}\frac{(-2\pi i)^{r-2}}{(2E_{\phi})^{r-2}}\delta(k_{1}^{z})\dots\delta(k_{r-2}^{z})\delta(\bar{k}^{z}+\underline{k}^{z}),
\end{equation}
and turns the amplitude into the following
$$
iT(p,p’,q,q’)_{SG}=\frac{k^4}{32\pi^3 M_\sigma }\sum_{r=2}^{+\infty}\frac{1}{(r-2)!}\bigg(\frac{i\kappa^2 M_\sigma E_\phi}{(4\pi)^2}\bigg)^{r-2}
$$
$$
\int d^3 \bar{k} d^3 \underline{k}\bigg[M_\sigma^4 p^{\prime \alpha}p^{\prime \beta} D_{\alpha\beta,00}(\bar{k})\eta^{\epsilon \phi} D_{\epsilon\phi, 00}(\underline{k})-2M_\sigma^4 p^{\prime \alpha}p^{\prime \epsilon} D_{\alpha\beta,00}(\bar{k})\eta^{\beta \phi} D_{\epsilon\phi, 00}(\underline{k})\bigg]\delta(\bar{k}^k+\underline{k}^z)
$$
\begin{equation}\label{laburandoafull}
\prod_{m=3}^{r}\int d^2 k^\perp_ m p^{\prime\mu_{m}}p^{\prime\beta_{m}}D_{\mu_m \beta_m,00}(k^\perp_m)\delta^2(\vec{k}^\perp_1+\dots+\vec{k}^\perp_n+\vec{\bar{k}}^\perp+\vec{\underline{k}}^\perp+\vec{\Delta}^\perp).
\end{equation}
By taking  the formulas \eqref{propagador} and \eqref{evaluando1} into account, after a simple calculation, it is found that
$$
M_\sigma^4 p^{\prime \alpha}p^{\prime \beta} D_{\alpha\beta,00}(\bar{k})\eta^{\epsilon \phi} D_{\epsilon\phi, 00}(\underline{k})-2M_\sigma^4 p^{\prime \alpha}p^{\prime \epsilon} D_{\alpha\beta,00}(\bar{k})\eta^{\beta \phi} D_{\epsilon\phi, 00}(\underline{k})
$$
$$
=\frac{M_\sigma^4E_\phi^2}{\bar{k}^2\underline{k}^2}\bigg[\bigg(\frac{4}{3\overline{a}}+\frac{2\overline{a}+\overline{d}-1}{6\overline{D}}\bigg)\eta_{00}
$$
$$
+\bigg(\frac{2}{3 \overline{a}}-\frac{2\overline{a}+\overline{d}-1}{6\overline{D}}+\frac{1}{4\overline{D}}\bigg)\frac{(\overline{k}\cdot \vec{p}^\prime)^2}{E_\phi^2\overline{k}^2}\bigg]
\bigg[\frac{2\underline{a}+\underline{d}-1}{6\underline{D}}
+\frac{1}{4\underline{D}}\bigg]
$$
$$
-\frac{2M_\sigma^4E_\phi^2}{\bar{k}^2\underline{k}^2}\bigg[\frac{2}{\bar{a} E_\phi} p^{\prime\alpha} \eta_{\alpha 0}\eta_{00}\delta_0^\gamma+\bigg(\frac{2\bar{a}+\bar{d}-1}{2 \bar{D}}-\frac{2}{\bar{a}}\bigg) \frac{p^{\prime\gamma}}{3E_\phi} +\bigg(\frac{2}{3\bar{a}}-\frac{2\bar{a}+\bar{d}-1}{6\bar{D}}+\frac{1}{4\bar{D}}\bigg)\frac{(\bar{k}\cdot p^\prime)\bar{k}^\gamma}{E_\phi\bar{k}^2}\bigg]
$$
$$
\times \bigg[\frac{2}{\underline{a}E_\phi} p^{\prime\alpha} \eta_{\alpha 0}\eta_{0\gamma}\eta_{00}+\bigg(\frac{2\underline{a}+\underline{d}-1}{2\underline{D}}-\frac{2}{\underline{a}}\bigg)  \frac{p^\prime_\gamma}{3E_\phi}+\bigg(\frac{2}{3\underline{a}}-\frac{2\underline{a}+\underline{d}-1}{6\underline{D}}+\frac{1}{4\underline{D}}\bigg) \frac{(\underline{k}\cdot p^\prime)\underline{k}_\gamma}{E_\phi\underline{k}^2}\bigg],
$$
where the quantities such as $\bar{a}$ denote the dependence in $\bar{k}$, for instance $\bar{a}=a(\bar{k})$ and so on. The same logic follows for other quantities such as $\underline{a}$. In addition
$$
\bar{D}=\frac{1}{4}(\bar{a}+3\bar{d})(2\bar{a}+\bar{d}-1)-\frac{3}{16},
$$
and the analogous definition follows for $\underline{D}$. From here it is deduced that
$$
iT(p,p’,q,q’)_{SG}\equiv -\sum\limits_{r=2}^{+\infty}\frac{i}{(r-2)!}\left(\frac{i\kappa^2 M_\sigma E_\phi}{8\pi^2}\right)^{r-2}\left(\frac{\kappa^4 M_\sigma^3 E_\phi^2}{32\pi^3}\right)$$
$$
\int\prod\limits_{m=3}^r\frac{ d^2\tilde{k}^\perp_m}{\tilde{k}^{\perp2}_m }\bigg[\frac{4}{3a(- k_m^{\perp 2})}+\frac{2a( k_m^{\perp 2})+d( -k_m^{\perp 2})-1}{6D( -k_m^{\perp  2})}\bigg]
$$
$$
\int \frac{d^3 \bar{k} d^3\overline{k}}{\bar{k}^2\overline{k}^2}\bigg\{\frac{M_\sigma^4E_\phi^2}{\bar{k}^2\overline{k}^2}\bigg[\bigg(\frac{4}{3\overline{a}}+\frac{2\overline{a}+\overline{d}-1}{6\overline{D}}\bigg)\eta_{00}
$$
$$
+\bigg(\frac{2}{ \overline{a}}-\frac{2\overline{a}+\overline{d}-1}{6\overline{D}}+\frac{\sqrt{3}}{4\overline{D}}\bigg)\frac{(\overline{k}\cdot \vec{p}^\prime)^2}{E_\phi^2\overline{k}^2}\bigg]
\bigg[\frac{2\underline{a}+\underline{d}-1}{6\underline{D}}
+\frac{1}{4\underline{D}}\bigg]
$$
$$
-\frac{2M_\sigma^4E_\phi^2}{\bar{k}^2\underline{k}^2}\bigg[\frac{2}{\bar{a} E_\phi} p^{\prime\alpha} \eta_{\alpha 0}\eta_{00}\delta_0^\gamma+\bigg(\frac{2\bar{a}+\bar{d}-1}{2 \bar{D}}-\frac{2}{\bar{a}}\bigg) \frac{p^{\prime\gamma}}{3E_\phi} +\bigg(\frac{2}{3\bar{a}}-\frac{2\bar{a}+\bar{d}-1}{6\bar{D}}+\frac{1}{4\bar{D}}\bigg)\frac{(\bar{k}\cdot p^\prime)\bar{k}^\gamma}{E_\phi\bar{k}^2}\bigg]
$$
$$
\times \bigg[\frac{2}{\underline{a}E_\phi} p^{\prime\alpha} \eta_{\alpha 0}\eta_{0\gamma}\eta_{00}+\bigg(\frac{2\underline{a}+\underline{d}-1}{2\underline{D}}-\frac{2}{\underline{a}}\bigg)  \frac{p^\prime_\gamma}{3E_\phi}+\bigg(\frac{2}{3\underline{a}}-\frac{2\underline{a}+\underline{d}-1}{6\underline{D}}+\frac{1}{4\underline{D}}\bigg) \frac{(\underline{k}\cdot p^\prime)\underline{k}_\gamma}{E_\phi\underline{k}^2}\bigg]\bigg\}
$$
$$
\delta(\bar{k}^z+\underline{k}^z)\delta^2(\vec{\Delta}+\vec{k}_1+\dots+\vec{k}_n+\vec{\bar{k}}+\vec{\underline{k}}).
$$
By taking the Fourier transform in the impact parameter $\vec{b}$ as before, it follows that the resulting contribution is
$$
i\widetilde{T}(\vec{b}^\perp)_{SG}=2i(s-M_\sigma^2)\tilde{\chi}_{SG}(\vec{b})\sum_{n=2}^\infty \frac{\tilde{\chi}^{n-2}_0(\vec{b})}{(n-2)!}
$$
with  $\chi_0(\vec{b}^\perp)$ is the ´phase  (\ref{fase}) while
$$
\chi_{SG}(\vec{b}^\perp)=
\frac{ \kappa^4 E_\phi M_\sigma^2}{256\pi^5}\int dk_1^z d^{2}k_1^\perp d^{2} k_2^\perp \frac{e^{-i\vec{b}^\perp\cdot \vec{k}_1^\perp}}{k_1^{\perp 2}+k_1^{z 2}}\frac{e^{-i\vec{b}^\perp\cdot \vec{k}_2^\perp}}{k_2^{\perp 2}+k_1^{z 2}}\bigg\{\bigg[\bigg(\frac{4}{3a(-k_1^{\perp 2}-k_1^{z 2})}
$$
$$
+\frac{2a(-k_1^{\perp 2}-k_1^{z 2})
+d(-k_1^{\perp 2}-k_1^{z 2})-1}{6D(-k_1^{\perp 2}-k_1^{z 2})}\bigg)\eta_{00}
$$
$$
+\bigg(\frac{2}{ a(-k_1^{\perp 2}-k_1^{z 2})}-\frac{2a(-k_1^{\perp 2}-k_1^{z 2})+d(-k_1^{\perp 2}-k_1^{z 2})-1}{6D(-k_1^{\perp 2}-k_1^{z 2})}
$$
$$
+\frac{\sqrt{3}}{4D(-k_1^{\perp 2}-k_1^{z 2})}\bigg)\frac{(k_1^\perp\cdot \Delta+E_\phi k_1^{z})^2}{E_\phi^2(k_1^{\perp 2}+k_1^{z 2})}\bigg]
$$
$$
\bigg[\frac{2a(-k_2^{\perp 2}-k_1^{z 2})+d(-k_2^{\perp 2}-k_1^{z 2})-1}{6D(-k_2^{\perp 2}-k_1^{z 2})}
+\frac{1}{4D(-k_2^{\perp 2}-k_1^{z 2})}\bigg]
$$
$$
-2\bigg[\frac{2 p^{\prime\alpha} \eta_{\alpha 0}\eta_{00}\delta_0^\gamma}{a(-k_1^{\perp 2}-k_1^{z 2}) E_\phi}
+\bigg(\frac{2a(-k_1^{\perp 2}-k_1^{z 2})+d(-k_1^{\perp 2}-k_1^{z 2})-1}{2 D(-k_1^{\perp 2}-k_1^{z 2})}-\frac{2}{a(-k_1^{\perp 2}-k_1^{z 2})}\bigg) \frac{p^{\prime\gamma}}{3E_\phi} 
$$
$$
+\bigg(\frac{2}{3a(-k_1^{\perp 2}-k_1^{z 2})}
-\frac{2a(-k_1^{\perp 2}-k_1^{z 2})+d(-k_1^{\perp 2}-k_1^{z 2})-1}{6D(-k_1^{\perp 2}-k_1^{z 2})}
$$
$$
+\frac{1}{4D(-k_1^{\perp 2}-k_1^{z 2})}\bigg)\frac{(k_1^\perp\cdot \Delta+E_\phi k_1^{z})k_1^{+\gamma}}{E_\phi(k_1^{\perp 2}+k_1^{z 2})}\bigg]
$$
$$
\times \bigg[\frac{2 p^{\prime\alpha} \eta_{\alpha 0}\eta_{0\gamma}\eta_{00}}{a(-k_2^{\perp 2}-k_1^{z 2})E_\phi}+\bigg(\frac{2a(-k_2^{\perp 2}-k_1^{z 2})+d(-k_2^{\perp 2}-k_1^{z 2})-1}{2D(-k_2^{\perp 2}-k_1^{z 2})}-\frac{2}{a(-k_2^{\perp 2}-k_1^{z 2})}\bigg)  \frac{p^\prime_\gamma}{3E_\phi}
$$
$$
+\bigg(\frac{2}{3a(-k_2^{\perp 2}-k_1^{z 2})}-\frac{2a(-k_2^{\perp 2}-k_1^{z 2})+d(-k_2^{\perp 2}-k_1^{z 2})-1}{6D(-k_2^{\perp 2}-k_1^{z 2})}
$$
\begin{equation}\label{shake}
+\frac{1}{4D(-k_2^{\perp 2}-k_1^{z 2})}\bigg) \frac{(k_2^\perp\cdot \Delta-E_\phi k_1^{z})k_{2\gamma}^{-}}{E_\phi(k_2^{\perp 2}+k_1^{z 2})}\bigg]\bigg\}.
\end{equation}
Here $\vec{k}_1^+=(k_1^\perp, k_1^z)$ and  $\vec{k}_2^-=(k_2^\perp, -k_1^z)$.

The integrals \eqref{shake} can be studied with the same techniques as for the next to leading order. The problem of divergences is softened as there is no $k_z^2$ in the denominators. It  is difficult in general to calculate the integrals in $k_z$ both for the  next to leading order and the seagull approximation. However, as the Bessel functions involved have dependence of the form $K_\mu(b\sqrt{k_z^2+m_i^2})$ then, if the impact parameter $b$ is large enough, these expressions may be replaced by their asymptotic expansions which may convert the integrals in something more tractable. 

\section{Discussion of the results}
In the present work, the eikonal approximation for the renormalizable models introduced in \cite{modesto1}-\cite{modesto4} was studied.  In particular, for the scattering between  one almost massless and one largely massive particle.  The results presented along the text suggest that if one of the mass scales of these generalized theories is lower than the Planck scale, then the deviation angle and time delay are proportional to the GR one, but corrected by a numerical  factor. The analysis of the validity of the eikonal approximation however is changed by the introduction of  this new scale.  It has been shown that if it is of order of the GUT this  puts stronger lower bounds on the impact parameter $b$, which are absent in GR. This makes sense for the authors, since the presence of a scale smaller than the Planck scale is equivalent of a larger length scale, and a semi classical description in this case requires a larger impact parameter.  We have found that the eikonal approximation is not necessarily reliable in this case, as the number of interchanged gravitons may be not so large, which is one of the main eikonal conditions.

If instead all the new mass scales are larger than the Planck mass, the result is  GR result plus subleading corrections. For certain limiting values of the impact parameter these corrections may be noticeable. In addition, the next to leading order and the seagull diagrams not containing three graviton vertices were estimated. The first vanishes in GR, after employing a  convenient regularization. For the models studied here  it is non zero instead. 

A regularization scheme for dealing with the next to eikonal order was developed. This scheme relies on certain integrals that are divergent when some parameters tend to specific value, by analytically continuing the resulting primitive to the forbidden region of parameters. This leads to a  regularization which  involves the Hadamard or Riesz method, and the result may be employed in series by use of Meijer functions. This scheme reduces to the one presented in \cite{nexteikonal} for the GR case.

The presented results are dependent on the functional form of the effective gravitational coupling for the problem. However, the validity of the eikonal results are based on some assumptions about the Schwarzschild radius, which is not known for these higher derivative theories. In any case, if the mass of the massive particle is large enough, we believe that our estimations are correct as the resulting radius should be close to GR. In other words, outside and far from the largest horizon, the deviation from the GR black hole solution may not be a leading order effect

The calculation of the next to eikonal corrections were partially estimated. It remains to understand which of the two corrections is leading at the next order, those of the seagull and next lo leading order or the GR corrections in $\Delta/m_i$ of the eikonal approximation, obtained by the non trivial curvature terms. We hope to fill this hole in the future.

The  calculations performed along the text are all related to scalar fields. However, it may be interesting to include in the picture gauge fields and to study causality issues, in the same fashion as in \cite{edelstein}-\cite{zhiboedov}, and the next to eikonal terms by use of \cite{nexteikonal}-\cite{nexteikonal2}, \cite{vietnam}.  General applications, related to the open problems cited in the introduction section, will be considered elsewhere.

\section*{Acknowledgements}
The authors are supported by CONICET, Argentina and O. S is supported by the Grant
PICT 2020-02181. A discussion with J. Serra is acknowledged.

\appendix

\section{The next to eikonal expressions corresponding to the Stelle model}

For the Stelle model, the phase \eqref{mylene} becomes
$$
\chi_{NE}(\vec{b}^\perp) = \frac{\kappa^4 E_\phi M_\sigma^2}{256\pi^5} \int dk_1^z \, d^2k_1^\perp \, d^2k_2^\perp \, \mathcal{I}_1 \, \mathcal{I}_2 \, \frac{\vec{k}_1^\perp \cdot \vec{k}_2^\perp - k_1^{z 2}}{k_1^{z 2}},
$$
where
$$
\mathcal{I}_1 = \frac{e^{-i\vec{b}^\perp \cdot \vec{k}_1^\perp}}{k_1^{\perp 2} + k_1^{z 2}} \mathcal{F}_1 \, ; \quad \mathcal{I}_2 = \frac{e^{-i\vec{b}^\perp \cdot \vec{k}_2^\perp}}{k_2^{\perp 2} + k_1^{z 2}} \mathcal{F}_2.
$$
Here the notation \(x = k_1^{\perp 2} + k_1^{z 2}\) was introduced, together with the following quantities 
$$
a(-x) = 1 + \beta x, \quad d(-x) = -\frac{1}{2} + 2\alpha x,
$$
$$D(-x)=\frac{1}{4}\bigg[-\frac{1}{2}+(6\alpha+\beta) x\bigg]\bigg[\frac{1}{2}+2(\alpha+\beta) x\bigg]-\frac{3}{16},$$
The last expression is factorized as
$$
D(-x) = c(x + \tilde{x}_1)(x + \tilde{x}_2), \quad c = \frac{1}{2}(6\alpha + \beta)(\alpha + \beta),$$
the last expression is given in terms of the roots
$$
\tilde{x}_{1,2} = \frac{x_1 + x_2}{2} \pm \sqrt{\frac{3}{16c} + \left(\frac{x_2 - x_1}{2}\right)^2}, \quad x_1 = \frac{-1}{2(6\alpha + \beta)}, \quad x_2 = \frac{1}{4(\alpha + \beta)}.
$$
The  following notation
$$
\mathcal{F}_1 = \bigg[\mathcal{F}_{11} + \mathcal{F}_{12} \frac{\eta_{00}(\vec{k}_1^\perp \cdot \Delta + E_\phi k_1^{z})^2}{E_\phi^2}\bigg],
$$
has been introduced, where
$$
\mathcal{F}_{11} = \frac{4}{3a(-x)} + \frac{2a(-x) + d(-x) - 1}{6D(-x)}, \quad x = k_1^{\perp 2} + k_1^{z 2},
$$
$$
\mathcal{F}_{12} = \left(\frac{2}{3a(-x)} - \frac{2a(-x) + d(-x) - 1}{6D(-x)} + \frac{1}{4D(-x)}\right)\frac{1}{x}.
$$
$$
\mathcal{F}_2 = \bigg[\mathcal{F}_{21} + \mathcal{F}_{22} \frac{\eta_{00}(\vec{k}_2^\perp \cdot \Delta - E_\phi k_1^{z})^2}{E_\phi^2}\bigg]
$$
with
$$
\mathcal{F}_{21} = \frac{4}{3a(-y)} + \frac{2a(-y) + d(-y) - 1}{6D(-y)}, \quad y = k_2^{\perp 2} + k_1^{z 2},
$$
$$
\mathcal{F}_{22} = \left(\frac{2}{3a(-y)} - \frac{2a(-y) + d(-y) - 1}{6D(-y)} + \frac{1}{4D(-y)}\right)\frac{1}{y}.
$$
By use of partial fractions, it is seen after substitution   \(a(-x) = 1 + \beta x\), \(d(-x) = -\frac{1}{2} + 2\alpha x\), and \(D(-x) = c(x + \tilde{x}_1)(x + \tilde{x}_2)\) that
$$
\frac{\mathcal{F}_{11}}{x} = \frac{4}{3x(1 + \beta x)} + \frac{\frac{1}{2} + 2(\alpha + \beta)x}{6c x(x + \tilde{x}_1)(x + \tilde{x}_2)}.
$$
From the decomposition
$$
\frac{4}{3x(1 + \beta x)} = \frac{4}{3}\left(\frac{1}{x} - \frac{\beta}{1 + \beta x}\right),
$$
$$
\frac{\frac{1}{2} + 2(\alpha + \beta)x}{6c x(x + \tilde{x}_1)(x + \tilde{x}_2)} = \frac{A}{x} + \frac{B}{x + \tilde{x}_1} + \frac{C}{x + \tilde{x}_2},
$$
with the coefficients
$$
A = \frac{1}{12c\tilde{x}_1 \tilde{x}_2}, \quad B = \frac{\frac{1}{2} - 2(\alpha + \beta)\tilde{x}_1}{-6c\tilde{x}_1(\tilde{x}_2 - \tilde{x}_1)}, \quad C = \frac{\frac{1}{2} - 2(\alpha + \beta)\tilde{x}_2}{-6c\tilde{x}_2(\tilde{x}_1 - \tilde{x}_2)}.
$$
it is arrived to
\begin{equation*}
{\frac{\mathcal{F}_{11}}{x} = \left(\frac{4}{3} + \frac{1}{12c\tilde{x}_1 \tilde{x}_2}\right) \frac{1}{x} - \frac{4\beta}{3(1 + \beta x)} + \frac{B}{(x + \tilde{x}_1)} + \frac{C}{(x + \tilde{x}_2)}}
\end{equation*}
By  taking into account that $4c\tilde{x}_1 \tilde{x}_2=-1$ it is calculated that
\begin{equation}
\frac{\mathcal{F}_{11}}{x} = \frac{1}{x} - \frac{4/3}{(x + 1 / \beta)} + \frac{B}{(x + \tilde{x}_1)} + \frac{C}{(x + \tilde{x}_2)}
\end{equation}
Furthermore, $\frac{\mathcal{F}_{12}}{x}$ can be expressed as
\begin{equation*}
\frac{\mathcal{F}_{12}}{x} = \frac{8\alpha(\alpha + \beta)x^2 - \beta x}{4c \, x^2 (1 + \beta x)(x + \tilde{x}_1)(x + \tilde{x}_2)} =  \frac{8\alpha(\alpha + \beta)x - \beta }{4c \, x (1 + \beta x )(x + \tilde{x}_1)(x + \tilde{x}_2)}
\end{equation*}
in which the $x^2$ type divergences cancel out. That is
\begin{equation*}
{\frac{\mathcal{F}_{12}}{x} = -\frac{\beta }{4 c \tilde{x}_1 \tilde{x}_2 x} + \frac{D}{(1 + \beta x)} + \frac{E}{(x + \tilde{x}_1)} + \frac{F}{(x + \tilde{x}_2)}}
\end{equation*}
where
$$D = \frac{\beta^2 (\beta^2 + 8 \alpha( \alpha + \beta) )}{4 c (-1 + \tilde{x}_1 \beta) (-1 + \tilde{x}_2 \beta)} = \frac{2 \beta ^2}{3}, \quad E = \frac{\beta +8 \alpha  \tilde{x}_1 (\alpha +\beta )}{4 c \tilde{x}_1 (\beta  \tilde{x}_1-1) (\tilde{x}_1-\tilde{x}_2)},$$
and $$F = \frac{\beta +8 \alpha  \tilde{x}_2 (\alpha +\beta )}{4 c \tilde{x}_2 
(\beta  \tilde{x}_2-1) (\tilde{x}_2-\tilde{x}_1)},$$ Again, the relation $4c\tilde{x}_1 \tilde{x}_2=-1$ leads to
\begin{equation*}
\frac{\mathcal{F}_{12}}{x} = \frac{\beta }{x} + \frac{2\beta /3}{(x + 1 / \beta)} + \frac{E}{(x + \tilde{x}_1)} + \frac{F}{(x + \tilde{x}_2)}
\end{equation*}
The full phase can then expressed as 
$$\chi_{NE}\equiv  \chi_{+} + \chi_{-}$$
where the components are
$$\chi_+ = \frac{\kappa^4 E_\phi M_\sigma^2}{256\pi^5} \int dk_1^z \, d^2k_1^\perp \, d^2k_2^\perp \, \mathcal{I}_1 \, \mathcal{I}_2 \, \frac{\vec{k}_1^\perp \cdot \vec{k}_2^\perp}{k_1^{z 2}}$$
$$\chi_- = -\frac{\kappa^4 E_\phi M_\sigma^2}{256\pi^5} \int dk_1^z \, d^2k_1^\perp \, d^2k_2^\perp \, \mathcal{I}_1 \, \mathcal{I}_2$$
with
$$\mathcal{I}_1 = \frac{e^{-i\vec{b}^\perp \cdot \vec{k}_1^\perp}}{k_1^{\perp 2} + k_1^{z 2}} \left[\mathcal{F}_{11} + \mathcal{F}_{12} \frac{\eta_{00}(\vec{k}_1^\perp \cdot \Delta + E_\phi k_1^{z})^2}{E_\phi^2}\right]$$
$$\mathcal{I}_2 = \frac{e^{-i\vec{b}^\perp \cdot \vec{k}_2^\perp}}{k_2^{\perp 2} + k_1^{z 2}} \left[\mathcal{F}_{21} + \mathcal{F}_{22} \frac{\eta_{00}(\vec{k}_2^\perp \cdot \Delta - E_\phi k_1^{z})^2}{E_\phi^2}\right]$$
The phase $\chi_-$ can be regularized as \cite{nexteikonal2}
$$\chi_- = -\frac{\kappa^4 E_\phi M_\sigma^2}{256\pi^5} \int dk_1^z  \underbrace{\int d^{2-2\epsilon}\vec{k}_1^\perp \mathcal{I}_1}_{\equiv \mathcal{J}_1} \,\,\,
\underbrace{\int d^{2-2\epsilon}\vec{k}_2^\perp \mathcal{I}_2}_{\equiv \mathcal{J}_2} $$
The $\epsilon$ value is added in order to avoid the singularities of $\Gamma(x)$ functions that may appear at integer negative values $x=-1,-2,--$, see formula \eqref{toxic}.  By taking this regularization into account it is seen that
$$\mathcal{J}_1 \equiv \mathcal{J}^{11}_{\text{base}}+ \text{Derivatives of }\left[\mathcal{J}^{12}_{\text{base}}\right]$$
and
\begin{equation*}
\mathcal{J}^{11}_{\text{base}} = \int d^{2-2\epsilon}\vec{k}_1^\perp \frac{e^{-i\vec{b} \cdot \vec{k}_1^\perp}}{k_1^{\perp 2} + k_1^{z 2}} \mathcal{F}_{11}
\end{equation*}
\begin{equation*}
\mathcal{J}^{12}_{\text{base}} = \int d^{2-2\epsilon}\vec{k}_1^\perp \frac{e^{-i\vec{b} \cdot \vec{k}_1^\perp}}{k_1^{\perp 2} + k_1^{z 2}} \mathcal{F}_{12}
\end{equation*}
and analogously for $\vec{k}_{2}^\perp$. All the terms coming from the partial fraction decomposition are of the form
\begin{equation}\label{JM}
\begin{aligned}
\mathcal{J}_{M^2} (k_1^{z}) &\equiv \int d^{2-2\epsilon}\vec{k}^\perp e^{-i \vec{b} \cdot \vec{k}^\perp} \frac{1}{k^{\perp 2} + k_1^{z 2} + M^2}\\
&= 2\left(\frac{b}{2}\right)^\epsilon \pi^{1-\epsilon} \left(\sqrt{k_1^{z 2} + M^2}\right)^{-\epsilon} K_{-\epsilon}\left(b \sqrt{k_1^{z 2} + M^2}\right)  
\end{aligned}
\end{equation}
The terms accompanying $\mathcal{F}_{12}$ and $\mathcal{F}_{22}$, are given by
$$(\vec{k}^\perp \cdot {\Delta} \pm E_\phi k^z)^2 = (\vec{k}^\perp \cdot {\Delta})^2 \pm 2 E_\phi k_1^z (\vec{k}^\perp \cdot {\Delta}) + E_\phi^2 (k_1^z)^2,$$and can be obtained by taking derivatives. In fact, note that the terms like $(\vec{k}^\perp \cdot {\Delta})^2 $correspond to the action  of $$-\Delta^\mu \Delta^\nu \frac{\partial^2}{\partial b^\mu \partial b^\nu},$$ inside the integrand.  This leads to
$$\int d^{2-2\epsilon}\vec{k}^\perp e^{-i\vec{b} \cdot \vec{k}^\perp} \frac{(\vec{k}^\perp \cdot {\Delta})^2}{k^{\perp 2} + k_z^2 + M^2} = -\Delta^\mu \Delta^\nu \frac{\partial^2}{\partial b^\mu \partial b^\nu} \mathcal{J}_{M^2}$$
For $\vec{k}^\perp \cdot {\Delta} $ the corresponding operator is $ i {\Delta} \cdot \nabla_{\vec{b}} \,$ and the following formal identity
$$\int d^{2-2\epsilon}\vec{k}^\perp e^{-i\vec{b} \cdot \vec{k}^\perp} \frac{\vec{k}^\perp \cdot {\Delta}}{k^{\perp 2} + k_z^2 + M^2} =  i{\Delta} \cdot \nabla_{\vec{b}} \mathcal{J}_{M^2}$$
takes place (under some assumptions about the integrand). Similarly, to obtain $\vec{k}_1^\perp \cdot \vec{k}_2^\perp$ in $\chi_{+}$ we also have to make the derivative of each integral separately and make the dot product of them

$$\chi_+ = -\frac{\kappa^4 E_\phi M_\sigma^2}{256\pi^5} \int \frac{dk_1^z}{k_1^{z 2}} \, \nabla_{\vec{b}} \mathcal{J}_{1} \cdot \nabla_{\vec{b}} \mathcal{J}_{2}$$
So ultimately, in order to obtain $\chi_{NE}(\vec{b}^\perp) $ all the integrals of $\vec{k}^{\perp}$'s are done by calculating the integrals $\mathcal{J}^{11}_{\text{base}}$ and $\mathcal{J}^{12}_{\text{base}}$ that depends themselves on integrals of the type of $\mathcal{J}_{M^2}$ and its derivatives, present on each term (each with different $M^2$ value) of the partial fraction decomposition of $\mathcal{F}_{11}/x$ and $\mathcal{F}_{12}/x$.

Based on this, consider the integrals
\begin{equation*}
\begin{aligned}
\mathcal{J}^{11}_{\text{base}} &= \int d^{2-2\epsilon}\vec{k}_1^\perp e^{-i\vec{b} \cdot \vec{k}_1^\perp} \frac{\mathcal{F}_{11}(k_1^{\perp 2} + k_1^{z 2})}{k_1^{\perp 2} + k_1^{z 2}}
\\
&=\mathcal{J}_{M^2=0}-\frac{4}{3} \mathcal{J}_{1/\beta}+B \mathcal{J}_{\tilde{x}_1}+C \mathcal{J}_{\tilde{x}_2},
\end{aligned}
\end{equation*}
and
\begin{equation*}
\begin{aligned}
\mathcal{J}^{12}_{\text{base}} &= \int d^{2-2\epsilon}\vec{k}_1^\perp e^{-i\vec{b} \cdot \vec{k}_1^\perp} \frac{\mathcal{F}_{12}(k_1^{\perp 2} + k_1^{z 2})}{k_1^{\perp 2} + k_1^{z 2}}
\\
&=\beta \mathcal{J}_{0}+\frac{2 \beta}{3} \mathcal{J}_{1/\beta}+E \mathcal{J}_{\tilde{x}_1}+F \mathcal{J}_{\tilde{x}_2}.
\end{aligned}
\end{equation*}
The analogous quantities for the $\vec{k}_{2}^\perp$ integrals. will be denoted $\mathcal{J}^{21}_{\text{base}}$ and $\mathcal{J}^{22}_{\text{base}}$  So, as to obtain the integrals we have that
$$(\vec{k}^\perp \cdot {\Delta})^2 \pm 2 E_\phi k_1^z (\vec{k}^\perp \cdot {\Delta}) + E_\phi^2 (k_1^z)^2 \rightarrow -\Delta^\mu \Delta^\nu \frac{\partial^2}{\partial b^\mu \partial b^\nu} \pm 2 i E_\phi k_1^z  {\Delta} \cdot \nabla_{\vec{b}} + E_\phi^2 (k_1^z)^2.$$
We can define the differential operator acting on the impact parameter $\vec{b}$
$$\partial_{\pm}= \frac{\eta_{00}}{E_\phi^2} \left[ -\Delta^\mu \Delta^\nu \frac{\partial^2}{\partial b^\mu \partial b^\nu} \pm 2 i E_\phi k_1^z  {\Delta} \cdot \nabla_{\vec{b}} + E_\phi^2 (k_1^z)^2\right].$$
Therefore
$$\chi_- = -\frac{\kappa^4 E_\phi M_\sigma^2}{256\pi^5} \int dk_1^z  \left[ \mathcal{J}^{11}_{\text{base}} + \partial_{+}\, \mathcal{J}^{12}_{\text{base}} \right] 
\left[ \mathcal{J}^{11}_{\text{base}} + \partial_{-}\, \mathcal{J}^{12}_{\text{base}} \right],$$

$$\chi_+ = -\frac{\kappa^4 E_\phi M_\sigma^2}{256\pi^5} \int \frac{dk_1^z}{k_1^{z 2}} \, \nabla_{\vec{b}} \left[ \mathcal{J}^{11}_{\text{base}} + \partial_{+}\, \mathcal{J}^{12}_{\text{base}} \right] \cdot \nabla_{\vec{b}} \left[ \mathcal{J}^{11}_{\text{base}} + \partial_{-}\, \mathcal{J}^{12}_{\text{base}} \right].$$
This is the most simplified expression  that we were able to find for the Stelle phase. The explicit calculation of the integrals is a difficult task that we have not solved yet.

\end{document}